\documentclass{SCIS2025}
\usepackage{bm,amsmath,amssymb,amsfonts,graphicx,epsfig,amsthm,color, xcolor,longtable}
\usepackage[most]{tcolorbox}

\begin{document}
\ArticleType{Review}
\Year{2025}
\Month{}
\Vol{}
\No{}
\DOI{}
\ArtNo{}
\ReceiveDate{}
\ReviseDate{}
\AcceptDate{}
\OnlineDate{}
\AuthorMark{}
\AuthorCitation{}

\title{Data-driven control of network systems: Accounting for communication adaptivity and security}{Data-driven control of network systems: Accounting for communication adaptivity and security}

\author[1]{Gang Wang}{}
\author[1]{Wenjie Liu}{}
\author[1]{Yifei Li}{}
\author[2]{Xin Wang}{}
\author[1]{Jian Sun}{}
\author[3,1]{Jie Chen}{}


\address[1]{State Key Lab of Autonomous Intelligent Unmanned Systems and the School of Automation, \\Beijing
Institute of Technology, Beijing 100081, China}
\address[2]{{School of Artificial Intelligence}, Changsha University of Science and Technology, Changsha 410114, China}
\address[3]{Department of Control Science and Engineering, Harbin Institute of Technology, Harbin, 150001, China}

\abstract{Over the past decades, network systems have surged in significance, driven by merging technological advancements. These systems play pivotal roles in diverse applications ranging from autonomous driving to smart grids, yet they confront complexities arising from network imperfections and intricate interconnections, which challenge system identification, controller design, as well as stability and performance analysis. This survey provides an in-depth exploration of network systems from most recent data-driven perspective, across four key issues:  communication delay, aperiodic sampling, network security, and distributed configurations.  By doing so, this survey enhances our comprehension of the challenges and theoretical innovations within the realm of network systems.}

\keywords{Data-driven control, 
Networked control,
Aperiodic sampling,
Resilient control,
Distributed control}

\maketitle

\section{Introduction}
\label{sec:intro}

The last two decades have borne witness to an unprecedented surge of interest in network systems, driven by the rapid evolution and merge of computing, communication, networking, and data-centric learning technologies. This heightened enthusiasm finds its roots in the pivotal role that network systems now occupy within contemporary engineering systems. These systems span a vast spectrum of engineering applications, including autonomous driving, intelligent manufacturing, smart grids, and remote healthcare \cite{FB-LNS,hespanha2007survey,he2024analysis,vlacic2024automation,zou2024leader,wang2024social}. 
Concurrently, industries are intensifying their call for systematic methodologies capable of adeptly modeling, analyzing, designing, and enhancing network systems in a robust, resilient, and sample-efficient manner, where sensor and control data traverse intricate digital communication channels. This escalating demand underscores the pressing need for innovative solutions and continued research in the ever-evolving domain of network systems.

Yet, amidst this technological fervor, it is paramount to acknowledge that this type of system is not immune to several network-induced challenges, including communication imperfections, safety, security, and scalability.
Numerous efforts have been made addressing these challenges when an explicit system model is available; see 
 e.g., \cite{Paulo2007,mo2009secure,2013Attack,sandberg2015cyberphysical,wang2010event}.
However, as the scale of the modern networked systems growing, e.g., robotics \cite{Armanini2023robot}, biology \cite{kitano2002systems}, or human-in-the-loop systems \cite{zhang2023obstacle}, modeling using first principle or identifying an exact system model from data become difficult \cite{dorfler2023data}, which highly restricts the practical implementation of the aforementioned control patterns.

Very recently, a convergence of a hundred researchers from diverse backgrounds within the IEEE Control Systems Society gathered to collectively contemplate a scientific roadmap for the future of our control discipline \cite{dorfler2023data}. To quote from Section $4$ of the ``Control for Societal-Scale Challenges Roadmap $2030$'' \cite{annaswamy2023control} and Introduction of \cite{dorfler2023data} report: ``One of the major developments in control over the past decade—and one of the most important moving forward—is the interaction of machine learning and control systems.'' 
Since the key feature of machine learning is to learn a control law directly from data, this statement indicates the increasing importance of data-centric methods in control theory, referred to as data-driven control \cite{Hou2013,zhang2016unsupervised}.
  
Most recently, the fundamental lemma in \cite{WILLEMS2005} has attracted reviving interest in this literature due to its advantage in designing data-driven controllers with rigorous theoretical guarantees.
	To be specific, this lemma provides a non-parametric representation of a linear time-invariant (LTI) system using a trajectory of the system.
	Inspired by this work, a number of applications and generalizations have been developed, including model predictive control (MPC) \cite{berberich2019data,Liu2021data,pang2022comparison,hu2025robust,liu2025switchmpc,wang2024ddmpc,wolff2024robust}, optimal control 
\cite{Coulson2019,persis2020data,vanwaarde2023quadratic,liu2023learning,wang2025data}, robust control \cite{Zhao2022data,hu2024hinfty}, 
dissipativity analysis \cite{van2022diss,Anne2022diss}, nonlinear control \cite{depersis2021sos,martin2023guarantees}, among others. 

In the remainder of this review, we focus on the fundamental lemma-based data-driven method, which we will refer to simply as the data-driven control method, and embark on an expedition through the multifaceted landscape of network systems. Our exploration is categorized into four  themes of concern: communication delay, aperiodic transmission, network security, and distributed configuration. 
These themes are arguably the most representative  issues of network systems and have been considered in, e.g., \cite{wang2008networked,zhang2019networked}.

\subsection{Communication delay}
Communication delay is a prevalent factor in almost all practical engineering systems, including power systems, vehicle suspension systems, and communication networks. 
This temporal discrepancy can lead to performance degradation and even instability.
The primary challenge revolves around stability analysis, specifically determining the maximal allowable upper bound (MAUB) for the delay.

 \emph{Model-based approaches.} 
When the exact system model is known, several methodologies have been developed, among which the Lyapunov-Krasovskii stability theorem plays an important role due to its adaptability to  time-varying delay.
Guided by this theorem, extensive efforts have been made to reduce the conservatism of stability conditions by enlarging the 
MAUB.
To this aim, two aspects are often taken into account.
One is to develop a tighter estimation method of the derivative of the Lyapunov functional, while the other lies in designing a suitable Lyapunov functional.
More references can be found in \cite{Fridman2014,LIU2017BL,tcns2022wangxin,tcb2023wzc,chen2021mean,chen2024state,CHEN2025TS}.

\emph{Data-driven approaches.} 
So far, data-driven solutions to this problem have only been considered in \cite{Rueda2022datadelay, wang2021data}, where the delays are assumed to be known constants during offline data collection.
Under this assumption, data-based stability conditions under noise-free, measurement noise-corrupted, process noise-corrupted data were derived. Although the assumption on exact time-delay removes most of the difficulties, their results can be seen as a starting point in this area, leading to several possible research directions. 


\subsection{Aperiodic transmission}
\label{sec:intro:etc}
{Central to the investigation of network systems is the challenge of determining the optimal execution frequency for sensors, controllers, and/or actuators. 
Striking the right balance between communication cost and overall system performance is the crux of this matter. 
Conventional periodic or time-triggered schemes, characterized by fixed transmission rates, often result in inefficient utilization of communication resources. 
To tackle this issue, a resource-efficient scheduling
approach for data transmissions, known as event-based control,
has been widely studied.
There are two effective event-based approaches, namely,
event-triggered control (ETC) and self-triggered control (STC) \cite{Paulo2007,wang2010event,heemels2012introduction}.}

\emph{Model-based approaches.}  
An ETC periodically or continuously assesses the system's state/output to determine if predefined conditions warrant data transmission \cite{aaarzen1999simple,heemels2013model}. In contrast, an STC determines the next transmission time by sequentially comparing the predicted future state/output with the most recent triggered state/output.
Consequently, under STCs, sensors can be completely deactivated between sampling times, thereby conserving energy and extending the sensor’s lifespan, \cite{anta2010to,gommans2015resource,matsume2020resilient}.

\emph{Data-driven approaches.} 
In the literature of data-driven aperiodic transmission, numerous efforts have been made, including both data-driven ETCs and STCs. 
Specifically, various data-driven ETCs have been proposed with respect to noise-free and noisy data, different types of systems, and different event-triggered strategies (ETSs) \cite{Matsuda2022event,persis2022event,chen2023data,wang2021data,Wang2023Tc,Qi2023event,yang2025ddset,liu2023ddevent}.
On the other hand, since designing a data-driven STC requires predicting future trajectories from noisy data, making this problem rather challenging, only a few results have been derived.
Data-driven MPC-based STCs were proposed in \cite{Liu2023data,Liu2023self} under measurement noise-corrupted data.
In the presence of process noise, leveraging a switched system approach, a self-triggered strategy (STS) was designed in \cite{WangCDC} by solving data-based LMIs.

\subsection{Network security}
\label{sec:intro:security}
{In the realm of security for network systems, cyberattacks are categorized based on their objectives: disrupting availability or compromising data integrity, as discussed in the taxonomy in \cite{SminSecure}. Attacks aimed at disrupting availability manifest as communication interruptions orchestrated by malicious attackers or compromised network components, employing techniques like denial-of-service (DoS) \cite{hou2022deep}. Conversely, attacks targeting data integrity involve eavesdropping on authentic data and injecting into transmitted data packets false information through false data injection (FDI) attacks \cite{huo2022secure}. 
These attacks not only result in financial losses but also pose substantial risks to real-world infrastructure, e.g., the smart power grid. 
Therefore, effective mechanisms should be designed to defend against attacks and/or to mitigate the associated damages on physical entities.
To this aim, a possible remedy is to design control strategies such that satisfied performance can be maintained regardless of attack strategies, which is referred to as resilient control.}

\emph{Model-based approaches.} 
Model-based resilient controllers against DoS attacks have been developed since at least 2015, see \cite{depersis2015input}.  Contributions in this domain are comprehensively detailed in \cite{FengResilient, wakaiki2019stabilization,Liu2021resilient}, along with their associated references. 
For instance, \cite{chen2025tac} proposed an asynchronous sampling-and-holding countermeasure that can effectively realizes attack detection even under stealthy attacks. 
However, achieving a balance between enhanced sensitivity to cyberattacks and minimized false alarms during normal system operation remains a formidable challenge, as highlighted in \cite{bai2017data}. As cyberattack strategies become increasingly sophisticated and intelligent, the risk of undetectable attacks escalates, potentially leading to catastrophic performance degradation and even system collapse.
Consequently, the latest research in this domain has shifted its focus from addressing individual attack instances to enhancing the system's overall robustness against attacks, see e.g., \cite{hashemi2022co,wu2019optimalswitching}.

\emph{Data-driven approaches.}  
Although considering network security in the context of data-driven design is still an area of exploration, several noteworthy endeavors deserve mention, e.g., FDI detection in \cite{krishnan2021data}, resilient control against FDI attacks in \cite{Liu2023fdi,liu2024robustfdi}, and resilient control against DoS attacks in \cite{Liu2021data}.
To be specific, the work \cite{krishnan2021data} designed a data-driven FDI detector, which has demonstrated performance levels comparable to its model-based counterparts. 
Moreover, \cite{Liu2021data} developed a data-driven predictive controller achieving maximum resilience against DoS attacks. 
Furthermore, when noise-free input-state data is available, data-driven resilient control under FDI attacks was investigated in \cite{Liu2023fdi}, and this was extended to consider noisy data in \cite{liu2024robustfdi}.
These pioneering efforts highlight the potential of data-driven techniques in enhancing the cybersecurity of network systems.

\subsection{Distributed configuration}
Distributed network systems represent a specific configuration within the realm of network systems, characterized by two key features \cite{jiang2024distributed}. Firstly, information exchange between subsystems or individual agents is facilitated through a shared communication network, linking components like sensors, controllers, and actuators. Secondly, the plant itself consists of numerous simple interacting units, often distributed physically and interconnected to collaboratively perform complex tasks. Within this context, each controller has the capability to share its local information with neighboring controllers, providing additional insights into the plant's dynamics. This distributed configuration offers significant advantages, including modularity, scalability, and robustness, all of which are essential in practical engineering systems, such as transportation networks, electrical power grids, smart manufacturing systems, and flocking systems.

On the other hand, multi-agent systems (MASs) are commonly used to model distributed network systems, where local information gathered from neighboring agents is utilized to control the system's global behavior. Consensus control is a fundamental challenge in MASs, attracting substantial attention from both academia and industry over the past two decades. This is due to the widespread applications of MASs in diverse domains, including satellite attitude alignment, formation control of multiple robots, estimation over sensor networks, and power management in electrical grids.

\emph{Model-based approaches.} 
The core idea behind consensus in MASs is to design a networked control protocol that ensures all agents converge to a common point or state value. 
The theoretical foundations for posing and solving consensus problems in MASs were established by earlier works such as \cite{Jad2003, Olfati2004, Ren2005}. Subsequently, a plethora of research has been conducted, contributing to different aspects of the consensus problem, including leader-following consensus, formation control, output synchronization, event-triggered consensus, and distributed optimization. Notable works in this area include \cite{Lizk2010, Su2012, Dima2012,Li2021,Yang2021unmanned,Chen2020scis,xu2023reinforcement,li2023topology,li2024preserving}. 

\emph{Data-driven approaches.} 
The aforementioned advancements in data-driven control have led to the development of various distributed data-driven control techniques to address a wide range of control problems.
Notable research in this domain includes works such as \cite{Li2023data, Wang2023Distributed,wang2023event,Jiao2021,li2023poly,Tan2025Data,CHAO2025Data}, which delve into distributed controller design and stability analysis.
To be specific, articles \cite{Li2023data, Wang2023Distributed,li2023selftriggered} explore distributed data-driven event/self-triggered consensus control (ETC/STC) by leveraging agent-wise data-driven reformulations of input-state parametrizations.
Further, distributed data-driven controllers tailored for heterogeneous MASs were developed in \cite{Jiao2021}, assuming perfect knowledge of process noise during offline data collection. 
To remove this strict assumption, the study in \cite{li2023poly} proposed a distributed data-driven polytopic approach for output synchronization in heterogeneous MASs.
Most recently, leveraging the internal model principle, \cite{liu2025OutReg} proposed data-driven methods for both linear and nonlinear systems, and achieved zero tracking error for LTI systems.

\subsection{Paper structure}
In this paper, our primary objective is to offer a comprehensive overview of data-driven control and stabilization analysis methodologies as applied in recent studies investigating unknown network systems. We delve into key aspects, including time delay, variable sampling/transmission intervals, security concerns, and distributed control problems. Our paper unfolds across four structured sections, guiding readers through this multifaceted exploration.

\section{Preliminaries on data-driven control}
In the context of data-driven control, it is often assumed that noisy data can be collected beforehand, resulting in a multitude of systems consistent with these data. Therefore, rather than designing a stabilizing controller for a single system as in the model-based approach, the task typically involves designing a stabilizing controller for a set of systems.
One of the important factors deciding whether such a control law is feasible is the size of the set, which depends on the noise assumption.
In this section, we review two data-based system representations under two commonly used noise assumptions.

\subsection{Notation}
Throughout the paper, we adhere to a standard set of notations, which are detailed in Table \ref{tab:notation} for clarity and reference.
\begin{longtable}{lp{12cm}}
    \caption{Table of notation} \label{tab:notation} \\
    \toprule 
    Variable & Definition \\
    \midrule
    \endfirsthead
    \toprule 
    Variable & Definition \\
    \midrule
    \endhead
    \bottomrule
    \endfoot
    \bottomrule
    \endlastfoot
    $\mathbb{N}$ ($\mathbb{N}_+$) & Set of non-negative (positive)
    integers \\
    $\mathbb{N}_{[a,b]}$ & $\mathbb{N}\cap [a,b]$, $a,b\in \mathbb{N}$\\
    $\mathbb{N}_{a}$ & Set of all integers greater than or equal to $a$, with $a \in \mathbb{N}$\\
    $\mathbb{R}$ & Set of all real numbers\\
    $\mathbb{R}_{>a}~(\mathbb{R}_{\ge a}$) & Set of all real numbers greater
    than (greater than or equal to) $a$, with $a \in \mathbb{R}$\\
    $\mathbb{R}^n$ & Set of all $n$-dimensional vectors\\
    $\mathbb{R}^{n\times m}$ & Set of all $n \times m$-dimensional matrices\\
    $a!$ &  $0! = 1$ and $a! := a(a-1)\cdots 1$ where $a \in \mathbb{N}_+$\\
    $C_{m}^i$ &  $C_{m}^i := m!/i!(m-i)!$ with $m,i \in \mathbb{N}$ and $m \ge i$\\
    $P\succ 0$ ($P\succeq 0$) &$P$ is a symmetric positive (semi)definite matrix\\
    ${\rm diag}\{\cdots\}$ & A block-diagonal matrix\\
    ${\rm Sym}\{P\}$&  The sum of $P^{\top}$ and $P$\\
    $[\cdot]$& If symmetry elements in the matrix\\
    $0$ ($I$) & zero (identity) matrices
    of appropriate
    dimensions\\
    $\ast$ & The symmetric term in block matrices\\
    $\underline{\lambda}_{P}$ ($\overline{\lambda}_{P}$) & The smallest (largest) eigenvalue of a square matrix $P$\\
    $M^\dag$& The left pseudo-inverse of matrix $M$\\
    $\Vert x\Vert_1$, $\Vert x\Vert$, $\Vert x\Vert_{\infty}$& The $\ell_1$-, $\ell_2$- (a.k.a., Euclidean), and $\ell_\infty$-norm, respectively\\
    $\Vert x \Vert_P$ & The weighted norm $\sqrt{x^\top P x}$ with $P = P^\top \succ 0$\\
    $\Vert M\Vert$& The spectral norm of matrix $M$\\
    $\mathbb{B}_{\delta}$ & $\{x \in \mathbb{R}^{n}~|~\Vert x\Vert \le \delta \}$\\
    ${\rm Tr} (M)$& The trace of  matrix $M$\\
    $x_{[t_1, t_1 + T - 1]}$ &$[x^\top(t_1) \cdots x^\top(t_1 + T - 1)]^\top$\\
    $x^{pre}$ &Pre-collected data $x^{pre} := x_{[0,T-1]}$\\
    $\mathcal{L}_2[0,~\infty]$ & The space of square-summable vector sequences over $[0,~\infty]$\\
    $x(t)\in \mathcal{L}_2[0,~\infty]$ &{$\|x(t)\|_{\mathcal{L}_2}=(\sum_{t=0}^\infty x^\top(t) x(t) )^{1/2}$}  \\
\end{longtable}
In addition, an important definition indicating the richness of the collected data is given below.
\begin{definition}[{Persistency of excitation}]\label{def:pe}
 Given any $L \in \mathbb{N}_+$, a signal $x_{[0,T-1]} \in \mathbb{R}^{n}$ with $T \in \mathbb{N}_{(n + 1)L - 1}$ is called persistently exciting of order $L$ if ${\rm {rank}}(H_{L}(x_{[0,T-1]})) = n L$ where {$H_{L}(x_{[0,T-1]})$ is the Hankel matrix of signal $x_{[0,T-1]}$ defined by}
{ \begin{equation*}
				  H_{L}(x_{[0,T-1]}) := 
      \left[ \begin{smallmatrix}
x(0)&x(1)&\cdots&x(N-L)\\
            x(1) & x(2) & \cdots &x(N - L + 1)\\
            \vdots & \vdots & \ddots & \vdots\\
            x(L - 1) & x(L) & \cdots & x(N - 1)
				    \end{smallmatrix}\right].
				\end{equation*}}
\end{definition}

\subsection{QMI-formed data-driven representation}\label{representatioin:noise}
Consider the following perturbed dynamic system
\begin{equation}\label{sys:data:perturbed}
x(t+1)=A x(t)+B u(t)+B_w w(t),
\end{equation}
where $x(t )\in \mathbb{R}^n$ is the state, $u(t) \in \mathbb{R}^m$ is the control input and $w(t)\in \mathbb{R}^{n_w}$ is the unknown noise accounting for various factors,  e.g., process noise or unmodeled system dynamics.
Matrices $A$, $B$ are unknown matrices but assumed to be controllable, and matrix $B_w\in \mathbb{R}^{n\times n_w}$ is a known matrix with full column rank.
For a given $T \in \mathbb{N}_+$, suppose  input-state data $(u_{[0,T-1]}, x_{[0,T]})$ can be obtained from offline experiments.
Define data matrices $U$, $X$, $X_+$, and $W$ by
\begin{subequations}\label{eq:data:all}
    \begin{align}
	U &:= [{u}(0)~ {u}(1)~ \cdots~ {u}(T - 1)],\label{eq:data:U}\\
	X &:= [{x}(0)~ {x}(1)~ \cdots~ {x}(T- 1)],\label{eq:data:X}\\
	X_+ &:= [{x}(1)~ {x}(2)~ \cdots~ {x}(T)],\label{eq:data:X+}\\
 W &:=[w(0)~w(1)~\cdots~ w(T-1) ],
\end{align}
\end{subequations}
where $W$ represents the unknown noise sequence $\{w(t)\}^{{T}-1}_{t=0}$. 
It becomes evident from \eqref{sys:data:perturbed} that
\begin{align}\label{formu:data}
X_+=AX+BU+B_wW.
\end{align}
Although exact knowledge on matrix $W$ is unavailable, in practical scenarios, it is often reasonable to assume that this noise remains bounded.

\begin{assumption}[Quadratic noise]\label{Ass:disturbance}
The noise sequence $\{w(t)\}^{{T}-1}_{t=0}$ assembled within the matrix $W$, satisfies $W \in \mathcal{W}$ with
\begin{equation}\label{data:disturbance}
\mathcal{W}=\bigg\{\bar{W}\in\mathbb{R}^{n_w\times T} \Big |
\left[\begin{array}{cc}\begin{smallmatrix}\bar{W}^{\top}\\I_{n_w} \\ \end{smallmatrix}\end{array}\right]^{\top}
  P_d
  \left[\begin{array}{cc}\begin{smallmatrix}\bar{W}^{\top}\\I_{n_w} \\ \end{smallmatrix}\end{array}\right]\succeq0 \bigg\},
\end{equation}
where $P_d$ is a predetermined symmetric matrix satisfying the condition $\left[\begin{array}{cc}\begin{smallmatrix}I \\0 \\ \end{smallmatrix}\end{array} \right]^\top P_d\left[\begin{array}{cc}\begin{smallmatrix}I \\0 \\ \end{smallmatrix}\end{array} \right]\prec 0$.
\end{assumption}

\begin{remark}\label{rmk:pd}
Assumption \ref{Ass:disturbance}, dubbed as the quadratic noise assumption, provides a versatile framework for characterizing bounded additive noise. This framework has been instrumental and has found applications in related studies, e.g., \cite{persis2020data,van2020noisy,wildhagen2021datadriven}. 
For instance, Assumption \ref{Ass:disturbance} can be specialized to represent quadratic full-block noise bounds by
\begin{equation}\label{noise:full}
\left[\begin{array}{cc}\bar{W}^{\top}\\I \\\end{array}\right]^{\top}
  \left[\begin{array}{cc}Q_d& S_d\\\ast & R_d \\\end{array}\right]
  \left[\begin{array}{cc}\bar{W}^{\top}\\I \\\end{array}\right]\succeq0,
  \end{equation}
  for some known matrices $Q_d \prec 0 \in \mathbb{R}^{T\times T}$, $S_d \in \mathbb{R}^{T\times n_w}$, and $R_d=R_d^{\top} \in \mathbb{R}^{n_w\times n_w}$.
  Considering pointwise bounded disturbances of the form $\|w(t)\|_2\leq \bar{w}$ for all $t\in \mathbb{N}$, the multiplier $P_d$ of the constraint \eqref{data:disturbance} is chosen as 
  \begin{equation}\label{multiplier:single}
      P_d= \left[\begin{array}{cc}-\epsilon I & 0\\\ast & \epsilon {T} \bar{w}^2 I \\\end{array}\right],~ \epsilon>0 ,
  \end{equation}
  which is a special case of \eqref{noise:full} with $Q_d=-\epsilon I$, $R_d=\epsilon T\bar{w}^2 I$, and $S_d=0$, where $\epsilon>0$ is a free scalar to be designed.
  Alternatively, considering the noise bound of each data separately, $P_d$ can be given as \begin{equation}\label{multiplier:diagonal}
			{P_d=
			\left[ \begin{array}{cc}-{\rm diag}\{\epsilon_i\}^{T-1}_{i=0} & 0\\\ast & \sum_{i=0}^{T-1} \epsilon_i \bar{w}^2 I\\\end{array} \right]
			, ~\epsilon_i>0.}
		\end{equation}
{Choosing {$P_d$} in \eqref{multiplier:diagonal} is less conservative compared to the form in \eqref{multiplier:single}, since $T$ free scalars $\{\epsilon_i\}^{T-1}_{i=0}$ are used to enhance the flexibility of the multiplier.}
\end{remark}

Under Assumption \ref{Ass:disturbance},
although an equivalent data-driven representation of the actual system matrices $[A~B]$ cannot be established, a set of systems contains $[A~B]$ can be constructed from data, presented in the following lemma. 

\begin{lemma}[{{QMI-formed data-based representation}}] \label{lem:sys_rep:qmi}  If the matrix $\left[\begin{smallmatrix}U\\X\end{smallmatrix}\right]$ has full row rank, the set $\Sigma_{AB}=\{[\bar{A}~ \bar{B}]\in\mathbb{R}^{n\times (n+m)}  |
\left[\begin{array}{cc}\begin{smallmatrix}[\bar{A}~ \bar{B}]^{\top}\\I \\ \end{smallmatrix}\end{array}\right]^{\top}
  \Theta_{AB}[\cdot]^{\top}\succeq 0
\}$
where $\Theta_{AB}:=
\left[\begin{array}{cc}\begin{smallmatrix}-X & 0 \\ -U & 0 \\ \hline X_+ & B_w \\ \end{smallmatrix}\end{array}\right]
  P_d
  [\cdot]^{\top}$
  characterizes all matrices $(\bar{A}, \bar{B})$  consistent with the data $(U, X, X_+)$.
\end{lemma}
Noting that the actual system $[A~B] \in \Sigma_{AB}$, if a controller is designed stabilizing all systems within this set, the actual system $[A~B]$ can be stabilized.
This is accomplished using robust control methods, such as the S-lemma in \cite{van2020noisy} and Petersen's lemma in \cite{bisoffi2022data}.

In this setting, it has been shown that reducing the size of the system set can improve the feasibility of the controller.
Noting that the system representation in Lemma \ref{lem:sys_rep:qmi} is constructed based on an ellipsoid noise bound, one possible way to reduce the size of this set is to consider a more accurate noise model, such as using a polytope. 
This will be introduced in the next section.

\subsection{Data-driven polytopic representation}\label{sec:representatioin:dncs}

In this subsection, we consider distributed systems and review a data-driven polytopic representation, as presented in \cite{li2023poly}. 
Compared with the QMI formulation in above, this representation not only reduces the size of the system set, but also provides a flexible way to account for uncertainty in system behavior.

Assume that we can gather a finite set of state-input-output data $\{x_i(t),u_i(t),y_i(t)\}_{t=0}^{T}$ for each agent $i\in\mathbb{N}_{[1,N]}$. These data points are collected offline by imposing a sequence of persistently exciting control inputs to the following {perturbed} system 
\begin{equation}
\label{masnoisy}
	\begin{split}
		x_i(t+1)&={A}_{i}x_i(t)+{B}_{i}u_i(t)+w_i(t)\\
  y_i(t)&={C}_ix_i(t)+v_i(t).
	\end{split}
\end{equation}
Here, $x_i(t)\in \mathbb{R}^n$, $u_i(t)\in \mathbb{R}^m$, $w_i(t)\in\mathbb{R}^{n_{w_i}}$  and $v_i(t)\in\mathbb{R}^{n_{v_i}}$ represent the state, control input, unknown process noise, and unknown measurement noise of $i$th agent, respectively.
The unknown noises $w_i(t)$, $v_i(t)$ satisfy the following assumption.
\begin{assumption}[Polytopic noise]
	\label{as:noise:poly}
 For every time step $t$ and agent $i\in\mathbb{N}_{[1,N]}$, the process noise $w_i(t)$ and measurement noise $v_i(t)$ are bounded by well-defined polytopic sets. These sets are defined as 
\begin{equation*}
	\begin{split}
	\mathcal{P}_{w_i}&=\Big\{\bar{w}_i|\bar{w}_i=\sum_{k=1}^{\gamma_{w_i}}\beta^{(k)}_{w,i}\hat{w}_{i}^{(k)},\beta^{(k)}_{w,i}\geq0, \sum_{k=1}^{\gamma_{w_i}}\beta^{(k)}_{w,i}=1 \Big\},\\
\mathcal{P}_{\bar{v}_i}&=\Big\{\bar{v}_i|\bar{v}_i=\sum_{k=1}^{\gamma_{v_i}}\beta^{(k)}_{v,i}\hat{v}_{i}^{(k)},\beta^{(k)}_{v,i}\geq0, \sum_{k=1}^{\gamma_{v_i}}\beta^{(k)}_{v,i}=1 \Big\},
	\end{split}
\end{equation*}
where $\hat{w}_{i}^{(k)}$ and $\hat{v}_{i}^{(k)}$ represent the $k$-th vertices of polytopes $\mathcal{P}_{w_i}$ and $\mathcal{P}_{v_i}$, respectively, and $\gamma_{w_i}$ and $\gamma_{v_i}$ denote the number of vertices.
\end{assumption}


The unknown process noise of length $T$ is denoted by $\{w_i(t)\}_{t=0}^{T-1}$. Consequently, for each agent $i$, the stacked matrix $W_i=[w_i(0) ~w_i(1) ~ \cdots ~ w_i(T-1)]$ belongs to a set denoted as $\mathcal{M}_{W_i}$, formally expressed as $W_i \in \mathcal{M}_{W_i}$. It is worth noting that $\mathcal{M}_{W_i}$ is a matrix polytope, described by the following formulation
\begin{equation}
	\label{M_poly}
	\mathcal{M}_{W_i}=\Big\{\bar{W}_i\Big|\bar{W}_i=\sum_{k=1}^{\gamma_{w_i} T}\beta_{W,i}^{(k)}\hat{W}_{i}^{(k)},\beta_{W,i}^{(k)}\geq0, \sum_{k=1}^{\gamma_{w_i}T}\beta_{W,i}^{(k)}=1 \Big\}.
\end{equation} 
This matrix polytope $\mathcal{M}_{W_i}$ results from the concatenation of multiple disturbance polytopes $\mathcal{P}_{w_i}$ and is constructed by
\begin{equation*}
	\begin{split}
		\hat{W}_i^{(1+(k-1){T})}&=\big[\hat{w}_{i}^{(k)} \quad {0}_{n_i\times ({T}-1)} \big],\\
	\hat{W}_i^{(l +(k-1){T})}&=\big[{0}_{n_{w_i}\times (l-1)}\quad \hat{w}_{i}^{(k)}\quad {0}_{n_{w_i}\times ({T}-l)} \big],	\\
	\hat{W}_i^{({T}+(k-1){T})}&=\big[{0}_{n_{w_i}\times ({T}-1)}\quad \hat{w}_{i}^{(k)} \big],
	\end{split}
\end{equation*}
for all $k\in\mathbb{N}_{[1,\gamma_{w_i}]}$, $l\in\mathbb{N}_{[2,T-1]}$, and $i \in\mathbb{N}_{[1,N]}$.
It should be emphasized that the noise is represented using a predefined polytope at each instant, and then these polytopes are concatenated to form a comprehensive matrix polytope $\mathcal{M}_{W_i}$ across time length $T$. In this way, the noise can be adequately characterized  at each instant rather than over a period of time $T$.

Similarly, the sequence of measurement noise $\{v_i(t)\}_{t=0}^{T-1}$ is collected in the matrix $V_i=[v_i(0) \; v_i(1) \; \cdots \; v_i(T-1)]$, which also belongs to a matrix polytope $\mathcal{M}_{V_i}$, defined as
\begin{equation}
	\mathcal{M}_{V_i}=\Big\{\bar{V}_i\big|\bar{V}_i=\sum_{k=1}^{\gamma_{v_i}{T}}\beta_{V,i}^{(k)}\hat{V}_{i}^{(k)},\beta_{V,i}^{(k)}\geq 0, \sum_{k=1}^{\gamma_{v_i}{T}}\beta_{V,i}^{(k)}=1 \Big\}.
\end{equation}
Upon introducing these notation, a data-based polytopic representation of MASs is given as follows.

\begin{lemma}[Data-based polytopic representation {\cite{li2023poly}}]\label{lem:sys_rep:poly}
$\\$Consider input-state-output data generated per agent from \eqref{masnoisy} gathered in matrices
\begin{equation*}
	\begin{split}
		U_i&:=\left[u_i(0) \quad u_i(1) \quad \dots \quad u_i(T-1)\right],	\\
		X_i&:=\left[x_i(0) \quad x_i(1) \quad \dots \quad x_i(T-1)\right],\\
  X_{i+}&:=\left[x_i(1) \quad x_i(2) \quad \dots \quad x_i(T)\right],\\
  Y_i&:=\left[y_i(0) \quad y_i(1) \quad \dots \quad y_i(T-1)\right].
	\end{split}
\end{equation*}
The set of system matrices $(\bar{A}_i,\bar{B}_i,\bar{C}_i)$ that can explain the data $(U_i,X_i,X_{i+},Y_i)$ is given as follows
\begin{equation}
	\begin{split}\label{eq:sigma}
		\Sigma_{i}:=\big\{(\bar{A}_i&,\, \bar{B}_i,\,\bar{C}_i)| X_{i+}=\bar{A}_iX_i+\bar{B}_iU_i+\bar{W}_i ,~Y_i=\bar{C}_iX_i+\bar{V}_i, \bar{W}_i\in\mathcal{M}_{W_i},\bar{V}_i\in\mathcal{M}_{V_i}\big\}.\\
		\end{split}
\end{equation}

Suppose Assumption~\ref{as:noise:poly} holds. If  {the data matrices $\left[\begin{smallmatrix}U_i \\X_i\end{smallmatrix}\right]$ and $X_{i}$} for $i\in\mathbb{N}_{[1,N]}$ have full row rank, the matrix polytope
		\begin{equation}
		\label{M_AB}
		\begin{split}
			\mathcal{M}_{i}=\big\{
			(\mathcal{M}_{{Z}_i},\mathcal{M}_{{C}_i})\big|
			\mathcal{M}_{{Z}_i}&=(X_{i+}-\mathcal{M}_{W_i})\left[\begin{smallmatrix}U_i \\X_i\end{smallmatrix}\right]^{\dagger},
			\mathcal{M}_{{C}_i}=(Y_{i}-\mathcal{M}_{V_i})X_i^{\dagger}\big\}.
		\end{split}
	\end{equation}
 characterizes all matrices $(\bar{A}_i, \bar{B}_i,\bar{C}_i)$ consistent with the data $(U_i,X_i,X_{i+},Y_i)$, i.e., $\mathcal{M}_{i} \supseteq \Sigma_i$.
\end{lemma}

\begin{remark}[QMI-formed vs polytopic representations] 
Ellipsoids (
{cf.} Assumption \ref{Ass:disturbance}) and polytopes ({cf.} Assumption \ref{as:noise:poly}) are two standard approaches to modeling unknown yet bounded noise. 
 {According to \cite{Blanchini1999}, the shape of polytopes is, in some sense, more flexible than that of ellipsoids (i.e., typically in terms of a quadratic full-block bound). 
Specifically, a polytope can adjust its facets independently in each direction, whereas an ellipsoid  is described by a single quadratic form and must remain centrally symmetric.
As such, a polytopic noise set can match the data more closely and yields less conservative data-driven stability conditions, albeit at the price of higher computational effort. }
Conversely, the QMI-formed representation generally requires fewer computational resources, particularly when dealing with large-scale systems or extensive data, where its advantages are more pronounced. 
\end{remark}

\section{Communication delay}\label{sec:time-delay}

This section briefly reviews the main results in \cite{Rueda2022datadelay,wang2021data}, i.e., data-driven design for time-delay systems under measurement noise corrupted data in \cite{Rueda2022datadelay} and under process noise corrupted data in \cite{wang2021data}.

\subsection{Data-based design under measurement noise}

Consider a discrete-time LTI system with unknown time-varying delays described by
\begin{subequations}\label{model:delay:noisefree}
    \begin{align}
        x(t+1) & =Ax(t)+Bu(t-d(t)) \label{model:delay:noisefree:x},\\
        u(t) &= Kx(t),
    \end{align}
\end{subequations}
where $d(t)$ represents input delays with an upper bound of $\bar{d}$. 
The objective is twofold: i) for a given $\bar{d}$, find a stabilizing matrix $K$, and ii) enlarge the given bound $\bar{d}$ to find a maximum tolerable upper bound of the delay, i.e., an MAUB, and its corresponding gain $K$. 

To derive stability conditions in the absence of the system model, a data-based system representation as in Lemma \ref{lem:sys_rep:qmi}, should be derived.
This is challenging due to the presence of unknown time-varying time delay $d(t)$.
  To address this problem, the work in \cite{Rueda2022datadelay} considered a simplified situation by assuming that offline-collected data is generated from system \eqref{model:delay:noisefree}  with a known constant delay.

Specifically, for a time horizon $[0,T]$, suppose the input-state data $(x^{pre}, u^{pre}) := \{x(t),u(t-d_0)\}_{t=0}^T$ is generated from the system $x(t+1) = Ax(t) + Bu(t - d_0)$, where $d_0 \in \mathbb{N}_+$ is known.
Assume that a sequence of measurement noise-corrupted input-state data is available and collected in the matrix $\Phi_d := \Phi_d^{\rm nom} + \Phi_d^{\rm noise}$, where $\Phi_d^{\rm nom} = [U_d^\top~X^\top]^\top$ represents the noise-free input-state data, and $\Phi_d^{\rm noise}$ represents the noise matrix.
Let $\mathcal{U}\in \mathbb{R}^{(m+n)\times (m+n)}$ and $\mathcal{V} \in \mathbb{R}^{T\times T}$ be orthonormal matrices such that $\mathcal{U}$ and $\Phi_d$ share the same row space, and $\mathcal{V}$ and {$\Phi_d^{\top}$} share the same row space, i.e., $Range(\mathcal{U})=Range(\Phi_d)$ and 
 $Range(\mathcal{V})=Range(\Phi_d^{\top})$. 
Decompose matrices $\Phi_d$ and $\Phi_d^{\rm{nom}}$ by $\Phi_d=\mathcal{U}[\Phi_d^{11}~0]\mathcal{V}^{\top}$ and $\Phi_d^{\rm{nom}}=\mathcal{U}[\Phi_d^{\rm{nom},11}~\Phi_d^{\rm{nom},12}]\mathcal{V}^{\top}$.
 To implement the data-driven design, the following assumption is imposed.
 
 \begin{assumption}[{\cite[Assumption V.1]{Rueda2022datadelay}}]
 \label{measurement:noise}
 Matrices $\Phi_d$ and $\Phi_d^{\rm{nom}}$ satisfy 
\begin{enumerate}
		\item $rank(\Phi_d)=rank(\Phi_d^{\rm{nom}})=m+n$,
\item $\|\Phi_d^{\rm{nom},11}\|_2 \|\Phi_d^\dag\|_2=\|\Phi_d^{\rm{nom},11}\|_2\|(\Phi_d^{11})^{-1}\|_2<1$.
\end{enumerate}
 \end{assumption}
 Under this assumption, a stabilizing gain $K$ can be designed following \cite[Theorem V.3]{Rueda2022datadelay}.
However, it can be observed that  Assumption \ref{measurement:noise} relies on the unknown noise-free data matrix $\Phi_d^{\rm nom}$.
 This renders the condition rather challenging to be verified, and hence limits the practical use of the design method.
 To address this concern, two possible research directions include verifying these conditions using data and finding other data-driven methods that relax these conditions.

\subsection{Data-based design under process noise}
Due to strict conditions in Assumption \ref{measurement:noise}, in the context of data-driven design using input-state data, 
it is often assumed that offline collected data are corrupted by process noise rather than measurement noise. 
In the following, we review the method in \cite{wang2021data}, which extends Lemma \ref{lem:sys_rep:qmi} to constant time-delay systems and presents a data-based control method.


Upon collecting input-state data from the perturbed time-delay system 
\begin{equation}\label{model:delay:noise}
	x(t+1)=Ax(t)+Bu(t-d_0) + B_w w(t),
\end{equation}
we get data matrices $X_+$, $U_d$, and $X$, satisfying $X_+ \!= A X +\! BU_d+ B_wW$ with $W$ obeying Assumption \ref{Ass:disturbance}.


Similar to Lemma \ref{lem:sys_rep:qmi}, a set $\Sigma^d_{AB}$ containing all the systems consistent with collected data can be formulated by
\begin{align*}%
	\Sigma^d_{AB}=\bigg\{[\bar{A}~\bar{B}]\in\mathbb{R}^{n\times (n+m)} \Big |
	\left[\begin{array}{cc}[\bar{A}~\bar{B}]^{\top}\\I \\\end{array}\right]^{\top}
	\Theta^d_{AB} {\left[\begin{array}{cc}[\bar{A}~\bar{B}]^{\top}\\I \\\end{array}\right]}\succeq0
	\bigg\},
\end{align*}
where 
\begin{align*}
	\Theta^d_{AB}&:=
	\left[\begin{array}{cc}-X & 0 \\ -U_d & 0 \\ \hline X_+ & B_w \\\end{array}\right]
	\left[\begin{array}{cc}Q_d& S_d\\\ast & R_d \\\end{array}\right]
	 {\left[\begin{array}{cc}-X & 0 \\ -U_d & 0 \\ \hline X_+ & B_w \\\end{array}\right]^{\top}}.
\end{align*}
Subsequently, by applying the Lyapunov function $V(t)$ in \cite{Fridman2014} in conjunction with Jensen-type inequality and reciprocally convex approaches, a data-based  stability condition with time delays is derived as follows.

\begin{theorem}\label{delay:stability:processnoise}
For a given scalar $\bar{d} \geq 0$, the system \eqref{model:delay:noisefree} remains stable for any $[\bar{A}~\bar{B}] \in \Sigma^d_{AB}$ with $d(t) \in[0,\bar{d}]$, provided there exists a scalar $\varepsilon>0$, along with matrices $P\succ0$, $Q\succ0$, $R\succ0$, $S$, and $F$, {such that} the following LMI is satisfied
	\begin{align}
		\left[
		\begin{array}{ccc}
			\mathcal{G}_3& \mathcal{G}_2+F^{\top}\\
			\ast &  \mathcal{G}_1+\Xi+\hat{\Psi}
		\end{array}
		\right]\prec0,
	\end{align}
	where 
	\begin{align*}
		\hat{\Psi} &:=- FL_4-(FL_4)^{\top},\\
		\mathcal{G}_1&:=\varepsilon\mathcal{Y}_1^{\top}\bar{\Theta}^d_{AB}\mathcal{Y}_1,\mathcal{G}_2:=\varepsilon\mathcal{Y}_2^{\top}\bar{\Theta}^d_{AB}\mathcal{Y}_1,\mathcal{G}_3:=\varepsilon\mathcal{Y}_2^{\top}\bar{\Theta}^d_{AB}\mathcal{Y}_2,\\
		\bar{\Theta}_{AB}^{d}&:=\left[\begin{array}{cc}-\bar{R}_d & {\bar{S}_d^{\top}}\\\ast &  -\bar{Q}_d \\\end{array}\right],\left[\begin{array}{cc}\bar{Q}_d & \bar{S}_d\\\ast & \bar{R}_d \\\end{array}\right]:=
		\left[\begin{array}{cc}Q_d & S_d\\ \ast & R_d \\\end{array}\right]^{-1},\\
		\mathcal{Y}_1&:=
		\left[\begin{array}{ccc}0 & L_1^{\top}& (KL_2)^{\top}\\\end{array}\right]^{\top},~
		\mathcal{Y}_2:=
		\left[\begin{array}{ccc}I& 0&0\\\end{array}\right].
	\end{align*}
\end{theorem}
Furthermore, to achieve better performance under different time delays, the following theorem provides a data-based LMI for designing controller gain $K$.

\begin{theorem}\label{delay:design:processnoise}
	For a given scalar $\bar{d} \geq 0$, there exists a controller gain $K$ that ensures the stability of system \eqref{model:delay:noisefree} for any $[\bar{A}~\bar{B}] \in \Sigma^d_{AB}$ with $d(t) \in[0,\bar{d}]$. This is achievable when the following conditions are met: a scalar $\varepsilon>0$, matrices $P\succ0$, $Q\succ0$, $R\succ0$, $S$, $G$, and $K_c$, satisfying the subsequent LMI
	\begin{align}\label{Th2:LMI1}
		\left[
		\begin{array}{ccc}
			\mathcal{B}_1& \mathcal{B}_2+[GL_1^{\top}~K_cL_2^{\top}]^{\top}\\
			\ast &  \mathcal{B}_3+\Xi+\bar{\Psi}
		\end{array}
		\right]\prec0.
	\end{align}
	Here, the terms are defined as follows
	\begin{align*}
		\bar{\Psi}&:=-(L_1^{\top}+ L_4^{\top})GL_{4}-((L_1^{\top}+ L_4^{\top})GL_{4})^{\top},\\
		\mathcal{B}_1&:=\varepsilon\mathcal{V}_1\Theta^d_{AB}\mathcal{V}_1^{\top},~
		\mathcal{B}_2:=\varepsilon\mathcal{V}_1\Theta^d_{AB}\mathcal{V}_2^{\top},~
		\mathcal{B}_3:=\varepsilon\mathcal{V}_2\Theta^d_{AB}\mathcal{V}_2^{\top},\\
		\mathcal{V}_1&:=
		\left[\begin{array}{ccc}\begin{smallmatrix}I & 0& 0\\
			0 & 1 & 0\\\end{smallmatrix}\end{array}\right],~
		\mathcal{V}_2:=
		\left[\begin{array}{ccc}0 & 0& (L_1^{\top}+\epsilon L_2^{\top})\\\end{array}\right].
	\end{align*}
	Moreover, the controller gain $K$ is designed as $K=K_cG^{-1}$.
\end{theorem}

{By iteratively enlarging $\bar{d}$ until LMI \eqref{Th2:LMI1} no longer holds, one can derive an MAUB.}
{Theorems \ref{delay:stability:processnoise} and \ref{delay:design:processnoise} are extensions of \cite[Theorems 3 and 4]{wang2021data} in the discrete-time domain.}
The proofs of these theorems can be deduced from those of \cite[Theorems 3 and 4]{wang2021data}.
{It is known that the matrix set $\Sigma^d_{AB}$ for Theorems \ref{delay:stability:processnoise} and \ref{delay:design:processnoise}  is consistent with the noise bound in Assumption \ref{Ass:disturbance}, as established in Lemma \ref{lem:sys_rep:qmi}. }
Compared to  \cite{Rueda2022datadelay}, {Theorems \ref{delay:stability:processnoise} and \ref{delay:design:processnoise} deal with process noise under a more general assumption (cf. Assumption \ref{Ass:disturbance}), which is easier to verify.} However, the data-driven design presented above also relies on the assumption that the collected data are generated from a system with a known constant time delay, which limits its practical application.
 {Incorporating time-varying delays in control design, where the exact delays within bounded intervals remain uncertain, better aligns with practical engineering requirements. 
However, such delays fundamentally intensify the complexity of data-driven control systems. 
Under the Willems' fundamental lemma framework, the time-varying causality between inputs and outputs compromises the data's capability to accurately characterize system dynamics. 
This motivates a robust control methodology that models delayed states as bounded uncertainties with known boundary information (i.e., delay upper/lower limits), thereby reformulating the problem into convex-constrained control synthesis. 
The computational complexity escalates geometrically since each delayed state must be represented as a vertex of the convex hull. 
This limitation underscores the necessity for refined delay characterization techniques to achieve practically feasible solutions. 
A critical open challenge persists: how to systematically extract effective control strategies from temporally ordered data under incomplete delay information, which serves a direction for our future research.}

\section{Aperiodic transmission}


As discussed in Section \ref{sec:intro:etc}, ETC and STC are two effective event-based approaches reducing transmission frequency.
Hence, in the following sections, we start by reviewing a data-driven ETC presented in \cite{Wang2023Tc}, followed by data-driven STCs under measurement noise in \cite{Liu2023self,Liu2023data} and process corrupted noise in \cite{WangCDC}.

\subsection{Data-based event-triggered control}
\label{sec:etc}
To start with, we introduce a standard problem setup in the context of event-based control under a periodic dynamic ETS.
Consider the following discrete-time linear state-feedback aperiodically sampled system
\begin{subequations}\label{sec1:sys:disLTI}
    \begin{align}
    x(t+1) & =A x(t)+Bu(t),~x(0)=x_0\in \mathbb{R}^{n} \label{sec1:sys:disLTI:x},\\
    u(t) &= Kx(t_k),~~t \in \mathbb{N}_{[t_k, t_{k + 1}-1]},
\end{align}
\end{subequations}
where $t_k \in \mathbb{N}$ is the sampling time generated by 
a periodic dynamic ETS introduced as follows
\begin{align}\label{sys:trigger}
	t_{k+1}=t_k+h\min\Big\{j \in \mathbb{N}_+\Big|\eta({\tau_j^k})+\theta\rho({\tau_j^k})<0\Big\},
\end{align}
where $h\in \mathbb{N}_+$ is the sampling interval obeying $1\leq \underline{h} \leq h \leq \bar{h}$ for given lower and upper bounds $\underline{h},\bar{h} \in \mathbb{N}_+$; $\theta>0$ is to be designed;$\tau_j^k:=t_k+jh, j \in \mathbb{N}_+$;
and $\rho({\tau_j^k})$ is given by
\begin{equation*}\label{trigger:function}
	\rho({\tau_j^k}):=\sigma_1 x^{\top}({\tau_j^k})\Omega x({\tau_j^k})+\sigma_2 x^{\top}(t_k)\Omega x(t_k)-e^{\top}({\tau_j^k})\Omega e({\tau_j^k}),
\end{equation*}
where $\Omega\succ0$ is some weight matrix; $\sigma_1\geq0$ and $\sigma_2\geq0$ are triggering parameters to be designed; $e({\tau_j^k}):=x({\tau_j^k})-x(t_k)$ denotes the error between the sampled signals $x({\tau_j^k})$ at the current sampling instant and $x(t_k)$ at the latest transmission instant; and, $\eta(t)$ is a dynamic variable, satisfying
$\eta(t)=\eta({\tau_j^k})$ for $t \in \mathbb{N}_{[{\tau_j^k}, {\tau_{j+1}^k}-1]}$ and
\begin{equation}\label{sys:dynamic}
	\eta({\tau_{j+1}^k})-\eta({\tau_j^k})=-\lambda\eta({\tau_j^k})+\rho({\tau_j^k}),
\end{equation}
where $\eta(0)\geq0$ and $\lambda>0$ are given.

The objective of designing a data-driven ETC is that given noisy input-state data $(u^{pre}, x^{pre})$ generated from system \eqref{sys:data:perturbed},  design a matrix $K$ and the matrix $\Omega$ in the ETS \eqref{sys:trigger} guaranteeing the stability of the system \eqref{sec1:sys:disLTI}.
This can be achieved by the following theorem.

\begin{theorem}[{\cite[Theorem 2]{Wang2023Tc}}]
\label{Th2}
	{For a given set of scalars $\bar{h}>\underline{h}>1$, $ \sigma_{1}\geq0$, $\sigma_{2}\geq0$, $\epsilon$, $\lambda>0$, and $\theta>0$ satisfying $1-\lambda-\frac{1}{\theta}\geq0$, the existence of a controller gain $K$ is guaranteed, leading to asymptotic stability for the system \eqref{sec1:sys:disLTI} under the triggering condition \eqref{sys:trigger} for any $[\bar{A} ~\bar{B}]\in \Sigma_{AB}$. Additionally, it is shown that $\eta({\tau_j^k})$ converges to the origin, provided there exists a scalar $\varepsilon>0$, and matrices $P\succ0$, $R_1\succ0$, $R_2\succ0$, $\Omega_z\succ0$, $S$, $N_1$, $N_2$, $G$, and $K_c$, such that the following LMIs hold for all $h\in[\underline{h},  \bar{h}]$
 \begin{equation}\label{inqua:th3}
\left[
  \begin{array}{ccc}
     \mathcal{T}_1& \mathcal{T}_2+\mathcal{F}& 0\\
    \ast &  \mathcal{T}_3+\Xi_0+h\Xi_\varsigma+\bar{\Psi}+\bar{\mathcal{O}}  & hN_\varsigma\\
     \ast &  \ast & -h\mathcal{R}_\varsigma
  \end{array}
\right]\prec0,
\end{equation}
where $\varsigma=1,2$, $\bar{\Psi}:={\rm Sym}\big\{-\mathcal{D}GL_{2}\big\}$ 
\begin{align*}
\bar{\mathcal{O}}&:=\sigma_1 L_3^{\top} \Omega_z L_3+
\sigma_2 L_{7}^{\top}\Omega_z L_{7}- (L_3-L_{7})^{\top} \Omega_z(L_3-L_{7}),\\
\mathcal{D}&:=(L_1+\epsilon L_2)^{\top},~\mathcal{F}:=\big[L_1^{\top}G^{\top},\,L_{7}^{\top}K_c^{\top} \big]^{\top},\\
\mathcal{T}_1&:=\varepsilon\mathcal{V}_1\Theta_{AB}\mathcal{V}_1^{\top},~
\mathcal{T}_2:=\varepsilon\mathcal{V}_1\Theta_{AB}\mathcal{V}_2^{\top},~
\mathcal{T}_3:=\varepsilon\mathcal{V}_2\Theta_{AB}\mathcal{V}_2^{\top},\\
\mathcal{V}_1&:=
\left[\begin{array}{ccc}I & 0\\\end{array}\right],~
\mathcal{V}_2:=
\left[\begin{array}{ccc} 0& \mathcal{D}\\\end{array}\right],
\end{align*}
and other matrices are outlined in  {\cite[Theorem 2]{Wang2023Tc}}.
 Furthermore, the controller gain $K$ is determined as $K=K_cG^{-1}$, and the triggering matrix $\Omega$ is characterized as ${G^{-1}}^{\top} \Omega_z G^{-1}$.}
\end{theorem}
The theorem is built on the looped-function approach \cite{Wang2023Tc}, which provides less conservative stability conditions  compared with other Lyapunov functions.
 {Theorem \ref{Th2} follows the co-design methodology established in \cite{Peng2013,Yue2013}, with its core contribution
being the joint optimization framework for the control gain under the proposed event-triggered mechanism in \eqref{sys:trigger}. This framework inherently generates a unified convex-constrained synthesis problem
that simultaneously addresses stability and resource efficiency. 
The proposed formulation achieves
superior engineering applicability through structural parsimony and systematic constraint handling.}

 {The data-driven inequalities in Theorem \ref{Th2} are derived via S-procedure \cite{Sche2001} through a systematic integration of Lyapunov stability theory and the QMI-formed data-driven representation. We now outline the key steps of this derivation. Chose the functional $V(z,t)$ where $x(t)=Gz(t)$ and $G\in \mathbb{R}^{n \times n}$ is nonsingular. The forward difference of $V(z,t)$ satisfies 
\begin{equation*}
\Delta V(z,t)
\leq \xi_z^{\top}(t)\left[\frac{t-{\tau_j^k}}{h}\bar{\Upsilon}_1(h) +\frac{{\tau_{j+1}^k}-t}{h}\bar{\Upsilon}_2(h) \right]\xi_z(t),
\end{equation*}
where the terms $\bar{\Upsilon}_1(h)$ and $\bar{\Upsilon}_2(h)$ are restructured as follows
\begin{equation*}
\bar{\Upsilon}_\varsigma(h):=
\left[\begin{array}{cc}[\mathcal{D}\bar{A}~\mathcal{D}\bar{B}]^{\top}\\I \\\end{array}\right]^{\top}\left[\begin{array}{cc}0 & \mathcal{F}\\\ast & \Xi_0+h\Xi_\varsigma+\bar{\Psi}+\bar{\mathcal{O}}+
hN_\varsigma \mathcal{R}_\varsigma^{-1} N_\varsigma^{\top} \\\end{array}\right]
\left[\cdot\right], ~\varsigma=1,2.
\end{equation*}
According to the data-based representation in Lemma \ref{lem:sys_rep:qmi}, it holds for any $[\bar{A} ~\bar{B}]\in \Sigma_{AB}$ that
\begin{equation*}
\left[\begin{array}{cc}[\bar{A}~\bar{B}]^{\top}\\I \\\end{array}\right]^{\top}
  \Theta_{AB}
  \left[\begin{array}{cc}[\bar{A}~\bar{B}]^{\top} \\
   I\\\end{array}\right]\succeq0.
\end{equation*}
By the full-block S-procedure \cite{Sche2001}, we have $\bar{\Upsilon}_1(h)\prec0$ and $\bar{\Upsilon}_2(h)\prec0$ for any $[\bar{A} ~\bar{B}]\in \Sigma_{AB}$ if there exists a scalar $\varepsilon>0$ such that for $\varsigma=1,2$
 \begin{equation}\label{Th2:fullblock1}
\left[\begin{array}{cc}0 & \mathcal{F}\\\ast & \Xi_0+h\Xi_\varsigma+\bar{\Psi}+\bar{\mathcal{O}}
+hN_\varsigma \mathcal{R}_\varsigma^{-1} N_\varsigma^{\top}\\\end{array}\right]+
\varepsilon \left[\begin{array}{cc}\mathcal{V}_1\Theta_{AB}\mathcal{V}_1^{\top} & \mathcal{V}_1\Theta_{AB}\mathcal{V}_2^{\top}\\\ast &  \mathcal{V}_2\Theta_{AB}\mathcal{V}_2^{\top} \\\end{array}\right] \prec0.
\end{equation}
Finally, the Schur Complement Lemma ensures that the inequalities in \eqref{Th2:fullblock1} are equivalent to \eqref{inqua:th3}.
This derivation systematically bridges classical Lyapunov stability criteria with data-driven formulations through the S-Procedure, thereby establishing data-driven verifiable conditions for guaranteeing the negative definiteness of the Lyapunov function difference. Such theoretical integration, achieved by constructing data-driven counterparts to model-based stability conditions through S-procedure, serves as the theoretical cornerstone of the mentioned methodologies in the paper.
}

\subsection{Data-driven self-triggered control}\label{sec:self}


This section briefly reviews the main results in \cite{Liu2023data,Liu2023self,wang2021data}, i.e., data-driven STC under measurement noise corrupted data in \cite{Liu2023data,Liu2023self}, and under process noise corrupted data in \cite{WangCDC}.

\subsubsection{Data-driven STC under measurement noise}
\label{sec:sts:mpc}
For the design of an STC, we investigate the same system with the ETC section, i.e., system \eqref{sec1:sys:disLTI}, but replace the control input by $u(t) = K \zeta(t_k)$ with
\begin{equation}
	\zeta(t_k) = x(t_k) + v(t_k),
\end{equation}
where $\zeta(t_k)$ represents the observed state and $v(t_k)$ is the noise in the measurement. 
For distinction, let $v^{p}(t)$ represent the noise in the offline collected data and $v(t)$ the noise during online operation. 
We assume that $v(t_k)$ belongs to a known bounded set $\mathbb{B}_{\bar{v}}$ where $\bar{v} \ge 0$ is a constant, such that $\|v^{p}(t)\|_{\infty} \le \bar{v}$ for all $t \in \mathbb{N}_{[0,T-1]}$ and $\|v(t)\|_{\infty} \le \bar{v}$ for all $t \in \mathbb{N}$.
The triggering time $t_k$ is generated by 
\begin{equation}\label{eq:selftri:noise:setup}
	t_{k} := t_{k-1} + s_{k-1},\quad t_0 := 0, k \in \mathbb{N}_+,\\
\end{equation}
where $s_k$ is the inter-triggering time between two consecutive transmissions.
The objective here is that, given a sequence of measurement noise corrupted input-state data $\{u(t), \zeta(t)\}_{t = 0}^T$, design a matrix $K$ and an STS generating $s_k$ such that system \eqref{sec1:sys:disLTI:x} in closed-loop with the controller $u(t) = K \zeta(t_k)$ achieves input-to-state stability (ISS).
Different from Theorem \ref{Th2}, the following method designs matrix $K$ and inter-triggering time $s_k$ separately.

To be specific, a matrix $K$ rendering $A + BK$ Schur-stable is first designed from $\{u(t), \zeta(t)\}_{t = 0}^T$ using any data-driven method, and then $s_k$ is generated by a MPC-based method below.
For a given prediction horizon $L \in \mathbb{N}_+$, at each triggering time $t_k$, the self-triggering module solves the following optimization problem based on $\zeta(t_k)$, to predict the states for the subsequent $L$ time instances
\begin{subequations}\label{eq:ddmpc:noise}
	\begin{align}
		J^*_L(\zeta(t_k)) :=&\!\!\!\underset{g(t_k), h(t_k)\atop
			\bar{x}_i(t_k)}{\min}
		\sum_{i = 0}^{L - 1}\Vert \bar{x}_i(t_{k})\Vert_{Q}+ (\lambda_h/\bar{v}) \Vert h(t_{k})\Vert^2   + {\lambda_g \bar{v}} \Vert g(t_{k})\Vert^2 \label{eq:ddmpc:noise:1}\\
		&~~{\rm s.t.} \quad
		\left[
		\begin{matrix}
			u(t_k)\\
			\bar{x}(t_k) + h(t_k)
		\end{matrix}
		\right] \!=\! 
		\left[
		\begin{matrix}
			H_{L}(u^{pre})\\
			H_{L}(\zeta^{pre})
		\end{matrix}
		\right] g(t_k) \label{eq:ddmpc:noise:2},\\
		& \qquad\quad
		\bar{x}_{0}(t_k) = \zeta(t_k) ,\label{eq:ddmpc:noise:3}
	\end{align}
\end{subequations}
where $u(t_k)$ is replicated $L$ times as $\bar{u}(t_k)$, and $\bar{x}(t_k)$, $h(t_k)$, $g(t_k)$ are the optimization variables; $\bar{x}(t_k)$ predicts the future $L-1$ states from $t_k$, $g(t_k)$ is defined as in the fundamental lemma in \cite{WILLEMS2005}, and $h(t_k)$ is a slack vector compensating for the influence of measurement noise.

The self-triggering time in \eqref{eq:selftri:noise:setup} is determined based on the optimal solution $(\bar{x}^*(t_k), g^*(t_k), h^*(t_k))$ of Problem \eqref{eq:ddmpc:noise}, i.e., 
\begin{equation}\label{eq:selftri:noise}
	t_{k + 1} := t_{k} + \min\{L - 1, s_k\},\quad t_0 := 0,\\
\end{equation}
with 
\small
\begin{align}\label{eq:phi:noise}
	s_k &:= \sup\big\{s_k\in \ \mathbb{N}_+\,|\Vert \bar{x}^*_{\tau}(t_{k}) - \zeta(t_{k})\Vert _{\infty}
	+ \Vert h^*_{s_k}(t_{k})\Vert_{\infty} \nonumber\\
	&\qquad\quad+ \rho^{s_k}\big(\bar{v} + \bar{v}\Vert g^*(t_{k})\Vert_1 + \Vert h^*_0(t_{k})\Vert_{\infty}\big)+  \bar{v} \Vert g^*(t_{k})\Vert_1  \le \sigma\Vert \zeta(t_{k})\Vert_{\infty}\big\},
\end{align}
\normalsize
where the constant $\sigma \in (0,1)$ balances between the system performance and the inter-triggering time, and $\rho^{s_k}:=\Vert A^{s_k}\Vert_{\infty}$ is the system divergence rate when $u(t) = 0$; see \cite[Algorithm 1]{Liu2023data} for details on a data-driven over-approximation method for $\rho^{s_k}$.
Recursive feasibility of problem \eqref{eq:ddmpc:noise} and the stability analysis of the closed-loop system can be found in
\cite[Theorem 3.1]{Liu2023data}.

When only noisy input-output data are available, the aforementioned MPC-based STC can be extended in two ways.
First, one can construct an output feedback controller as in \cite{li2024controller,van2023behavioral} and extend both Problem \eqref{eq:ddmpc:noise} and the STS in \eqref{eq:selftri:noise} directly.
Alternatively, noting that MPCs are capable of providing a sequence of optimal inputs at each time, instead of applying the output feedback controller, one can design a data-driven MPC and use its optimal solution for both system control and STS design.
In the following, we briefly review the latter approach, which was extended from \cite{Liu2023self} by considering noise also in the offline collected data.

Consider the following dynamics
\begin{subequations}\label{eq:sys:self:out}
	\begin{align}
		x(t + 1) &= A x(t) + B u(t), \quad t \in \mathbb{N},\\
		y(t) &= Cx(t) + Du(t) + v(t),
	\end{align}
\end{subequations}
where $u(t)$ and $y(t)$ are available.
The equilibrium point is denoted by $(u^e, y^e)$.
Assume that the pair $(A,C)$ is observable with a known observability index $\eta$.
Unlike the state feedback case presented above, we only consider triggering on the output side, i.e., the input $u(t)$ is transmitted to the plant at each time $t$, while the output $y(t)$ is transmitted only at the triggering times $t_k$.
The objective here is that given noisy input-output data $(u^{pre},y^{pre})$ generated from \eqref{eq:sys:self:out},  design $u(t)$ and an STS generating $s_k$ rendering the closed-loop system \eqref{eq:sys:self:out} ISS.

Before preceding, some notations are introduced.
For $t \in \mathbb{N}_\eta$, define the extended state  $\xi(t)$ by
\begin{equation}\label{eq:extx}
	\xi(t):= \left[
	\begin{matrix}
		u_{[t - \eta, t - 1]}\\
		y_{[t - \eta, t - 1]}
	\end{matrix}
	\right]\in \mathbb{R}^{n_\xi},
\end{equation}
where $n_\xi := (m + p)\eta$.
It follows iteratively from \eqref{eq:sys:self:out} that 
\begin{align*}\label{eq:yu}
	y(t) &= CA^{\eta}x(t - \eta) + CA^{\eta - 1}Bu(t - \eta) +\cdots  \\
	& ~~~+ CBu(t- 1)+ Du(t) + E v_{[t - \eta, t]}.
\end{align*}
where $E$ is related to matrices $A$, $B$, $C$, and $D$.
Since the pair $(A, C)$ is observable, there exist some matrices $\Upsilon$,  $\Upsilon_1$, $\Upsilon_2$, and $\Upsilon_v$ such that $x(t - \eta)= \Upsilon_1 u_{[t - \eta, t- 1]} + \Upsilon_2 y_{[t - \eta, t- 1]} + \Upsilon_v v_{[t- \eta, t - 1]} := \Upsilon \xi(t) + \Upsilon_v v_{[t - \eta, t - 1]}$.
Building on this result, system \eqref{eq:sys:self:out} can be transformed into 
\begin{subequations}\label{eq:extsys}
	\begin{align}
		\xi(t+1) &= \tilde{A} \xi(t) + \tilde{B} u(t) + \tilde{E} v_{[t - \eta, t]}, \quad t \in \mathbb{N}_\eta,\\
		y(t) &= \tilde{C}\xi(t) + \tilde{D} u(t) + v(t),
	\end{align}
\end{subequations}
for suitable matrices $(\tilde{A},\tilde{B},\tilde{C},\tilde{D}, \tilde{E})$ depending on $(A,B,C,D)$.
For subsequent analysis, let us denote the equilibrium point of the new system \eqref{eq:extsys} by 
\begin{equation*}
	\xi^e := \big[
		\underbrace{{u^e}^\top ~ \cdots~ {u^e}^\top}_{\eta~{\rm times}} ~\underbrace{{y^e}^\top ~\cdots ~{y^e}^\top}_{\eta~{\rm times}} 
	\big]^\top,
\end{equation*} 

An assumption regarding this extended system is imposed.

\begin{assumption}\label{as:terminal}
	There exist matrices $P = P^\top \succ 0$, $R = R^\top \succ 0$, $Q = Q^\top \succ 0$, $K \in \mathbb{R}^{m \times n_\xi}$, and a set $\Xi_r :=\{\xi \in \mathbb{R}^{n_\xi} | \Vert \xi - \xi^e\Vert_{P} \le r \} \subseteq \mathbb{U}^\eta \times \mathbb{R}^\eta$ such that for all $\xi \in \Xi_r$, $u = u^e + K(\xi - \xi^e)$, and $y = (\tilde{C} + \tilde{D }K)\xi$, the following statements hold true:
	\begin{enumerate}
		\item $u \in \mathbb{U}$, $\tilde{A}\xi + \tilde{B} u \in \Xi_r$, and,
		\item the following inequality holds
		\begin{equation}\label{eq:a+bkxi}
			\Vert (\tilde{A} + \tilde{B} K)\xi\Vert_{P}^2 \le \Vert \xi\Vert_{P}^2 - \Vert K\xi\Vert_{R}^2 -\Vert y\Vert_Q^2.
		\end{equation}
	\end{enumerate}
\end{assumption}

Building upon Assumption \ref{as:terminal}, a data-driven MPC-based STC is specified as follows. 
Set $\xi^e = 0$ for simplicity.
At each triggering time $t_k$,
the STS solves the following optimization problem based on an output packet containing the past $\eta$ measurements, i.e., $y_{[t_k - \eta, t_k - 1]}$ to predict the trajectory for the subsequent $L$ time instants.
\begin{subequations}\label{eq:ddmpc}
	\begin{align}
		J^*_L(u_{t_k},&  y_{t_k}) :=\nonumber 	\\
		\underset{\{g(t_k), h(t_k)\atop
			\bar{y}(t_k), \bar{u}(t_k)\}}{\min}
		&\sum_{i = -\eta}^{L - 1} \Vert \bar{u}_i(t_{k})\Vert^2_{R}  + \Vert \bar{y}_i(t_{k})\Vert^2_{Q} + \frac{\lambda_h}{\bar{v}}\Vert h(t_k) \Vert^2 \nonumber\\
		&+ \lambda_g \bar{v}\Vert g(t_k) \Vert^2  + \Vert \bar{\xi}_L(t_{k})\Vert^2_{P} \label{eq:ddmpc0}\\
		{\rm s.t.}\quad 
		&\left[
		\begin{matrix}
			\bar{u}(t_k)\\
			\bar{y}(t_k) + h(t_k)
		\end{matrix}
		\right] = 
		\left[
		\begin{matrix}
			H_{L + \eta}({u}^{pre})\\
			H_{L + \eta}({y}^{pre})
		\end{matrix}
		\right] g(t_k)\label{eq:ddmpc1},\\
		&	\left[
		\begin{matrix}
			\bar{u}_{[-\eta, -1]}(t_k)\\
			\bar{y}_{[-\eta, -1]}(t_k)
		\end{matrix}
		\right] = 
		\left[
		\begin{matrix}
			u_{[t_k - \eta, t_k - 1]}\\
			y_{[t_k - \eta, t_k - 1]}
		\end{matrix}
		\right] ,\label{eq:ddmpc2}\\
		& \bar{\xi}_L(t_k) \in \Xi_\epsilon
		\label{eq:ddmpc3} ,\\
		& \bar{u}_i(t_k) \in \mathbb{U},\quad i \in \mathbb{N}_{[1,L - 1]}\label{eq:ddmpc4}.
	\end{align}
\end{subequations}

The following theorem outlines the MPC-based STC and the stability conditions, which is extended from \cite[Theorem 1]{Liu2023self} by considering noisy offline collected data.

\begin{theorem}\label{lem:tau}
	Assume that \eqref{eq:ddmpc} is feasible at $t_0$.
	For appropriate $r > 0$ and $P \succ 0$ satisfying Assumption \ref{as:terminal}, there exist constants $\bar{\lambda}_g > \underline{\lambda}_g > 0$, $\bar{\lambda}_h > \underline{\lambda}_h > 0$, and $\bar{v}_0 >0$ such that for all $\underline{\lambda}_g \le \lambda_g \le \bar{\lambda}_g$, $\underline{\lambda}_h \le\lambda_h \le \bar{\lambda}_h$, and $0 \le \bar{v} < \bar{v}_0$,  Problem \eqref{eq:ddmpc} is feasible at all $t_{k}\in \mathbb{N}_+$, whose optimal solutions are $\bar{u}^*(t_k)$, $\bar{y}^*(t_k)$, $g^*(t_k)$, and $h^*(t_K)$.
 System \eqref{eq:sys:self:out} applying control inputs $u(t) = \bar{u}^*_{t - t_k}(t_k)$ with $t \in \mathbb{N}_{[t_k, t_{k + 1}-1]}$ is ISS, 
	if i) $\epsilon$ in \eqref{eq:ddmpc3} obeys
	\begin{align}\label{eq:rleepsilon}
		\Big(1 - \frac{\underline{\lambda}_{K^\top R K}}{\bar{\lambda}_{P}}\Big) r^2 \le \epsilon^2 \le r^2,
	\end{align}
	and ii) the inter-triggering time satisfies 
	\begin{align}\label{eq:tau}
		s_k = \min\big\{ \hat{s}_k,\check{s}_k,\, L - 1 \big\},
	\end{align}
	with integers $\hat{s}_k$ and $\check{s}_k$ defined by
	 {\begin{align}
		&\hat{s}_k:= \sup\bigg\{s_k \in \mathbb{N}_+ \Big| (\sqrt{\eta}\bar{n} + \Vert h^*_{[-\eta, -1]}(t_k)\Vert)\sqrt{\sum_{i = \tau_k}^{s_k + \eta - 1} \rho^i} \nonumber\\
			&~\quad \quad\quad+ \Vert h^*_{[s_k - \eta, s_k - 1]}(t_k)\Vert  \le \frac{r}{\bar{\lambda}_{P}} - \Vert \bar{\xi}^*_{s_k}(t_k)\Vert \bigg\},\label{eq:tautau1}
	\end{align}}
	and 
	\begin{align}
		\check{s}_k  &:=\sup\bigg\{s_k\in \mathbb{N}_+\Big| 2\bar{\lambda}_Q\!\!\sum_{i = 0}^{s_{k} - 1}\Vert h^*_i(t_k)\Vert^2
		\!+\!\lambda_h\Big(\!\eta\bar{v} \!-\! \frac{\Vert h^*(t_k) \Vert^2}{\bar{v}}\Big)\nonumber\\
		&+ \sqrt{\bar{\lambda}_{P}}(r + \epsilon)\Big(\big(\bar{v} \Vert g^*(t_k)\Vert_1 + \Vert h_{s_k}^*(t_k)\Vert_\infty){\sum_{i = s_k}^{s_k + \eta - 1} \rho^i} \nonumber\\
		&+ \bar{v} \Vert g^*(t_k)\Vert_1 + \Vert h^*_{[s_k - \eta, s_k - 1]}(t_k)\Vert_{\infty} \Big) + \lambda_g\bar{v}\big(\Vert H_{u\xi}^{\ddag}\Vert^2(2 \nonumber\\
		& + \bar{\lambda}_{P}/\underline{\lambda}_{R})r^2 - \Vert g^*(t_k)\Vert^2\big)\! +\! \bar{\lambda}_Q\big(2 \eta\bar{v}^2\! +\! 2 \Vert h^*_{[-\eta, -1]}(t_k)\Vert^2\big)\sum_{i = 0}^{s_k - 1}\rho^{i + \eta}\le \sigma \sum_{i = 0}^{s_k - 1} \Vert \bar{\xi}^*_i(t_k)\Vert^2  \bigg\},\label{eq:tautau2}
	\end{align}
	where $\rho^i := \Vert CA^i\Phi^\dag\Vert$ can be over-approximated using the set-based method as in \cite[Algorithm 1]{Liu2023data}, $H_{u\xi}$ is the right pseudo-inverse of matrix $H_{u\xi}$, and $H_{u\xi}$ is defined by $H_{u \xi} = \left[
		\begin{smallmatrix}
			H_{L + \eta}({u}^{pre})\\
			H_1(\xi^{pre}_{[\eta,T-L]})
		\end{smallmatrix}
		\right]$, 
	and the constant $0<\sigma<\bar{\sigma}<1$ balances between the system performance and the inter-triggering time.
\end{theorem}

\subsubsection{Data-driven STC under process noise}
\label{sec:sts:swithced}
In addition to the measurement noise addressed above, this section reviews a data-driven STS design method using process noise corrupted data.
Specifically, consider system \eqref{sec1:sys:disLTI} with the triggering time $t_k$ generated by \eqref{eq:selftri:noise:setup}.
The objective here is that given noisy input-state data $(u^{pre}, x^{pre}$ generated from system \eqref{sys:data:perturbed}, design a matrix $K$ and an STS generating $s_k$ in \eqref{eq:selftri:noise:setup} ensuring the stability of system \eqref{sec1:sys:disLTI}.
To this aim, we introduce a switched system approach-based STC method.


We begin by reformulating the system \eqref{sec1:sys:disLTI} into a switched system.
Define for any $s \in \mathbb{N}_+$
\begin{equation*}
	\begin{aligned}
		\underline{B}^s:=
		\left[\begin{matrix}A^{s-1}B&A^{s-2}B&\cdots &B \\\end{matrix}\right],~~~
		\underline{K}^s:=\big[~\underbrace{K^{\top} ~~ K^{\top} ~~\cdots ~~K^{\top}}_{s~ \text{times}}\big]^{\top}.
	\end{aligned}
\end{equation*}
Recalling for any $k \in \mathbb{N}$ that the inter-triggering time $s_k = t_{k + 1} - t_k$ and satisfying $s_k\in \mathbb{N}_{[1, \,\bar{s}]}$ for some given constant $\bar{s} \in \mathbb{N}_{2}$, system \eqref{sec1:sys:disLTI} can be rewritten as 
\begin{equation}\label{sys:switch}
	x(t_k+s_k)=(A^{s_k}+\underline{B}^{s_k} \underline{K}^{s_k}) x(t_k).
\end{equation}
Since matrices $A$, $B$ are unknown, the system matrices in the switched system $A^{s_k}$, $\underline{B}^{s_k}$, $ \underline{K}^{s_k}$ are also unknown.
In addition, for $k \in \mathbb{N}$, determine the triggering time by
\begin{equation}\label{sys:self:function}
	t_{k+1}=t_k+\max\big\{s_k\in \mathbb{N}_{[1,\bar{s}]}\big|\mathcal{Q}(x(t_k),s_k)\geq 0\big\},
\end{equation}
with
\begin{align}\label{sys:QMI}
	\mathcal{Q}(x(t_k),s_k)= &
	\left[\!\!\begin{array}{cc}(A^{s_k}+\underline{B}^{s_k}\underline{K}^{s_k}) x(t_k)\\x(t_k) \\\end{array}\!\!\right]^{\top}  \left[\!\!\begin{array}{cc}(\sigma_1-1)\Omega & \!\!\Omega\\\ast & \!\!(\sigma_2-1)\Omega \\\end{array}\!\!\right][\cdot]\geq 0,
\end{align}
where parameters $\sigma_1 \geq0$, $\sigma_2 \geq0$ are given constants, and matrix $\Omega$ is to be designed. 

Instead of $U$ and $X_+$ in previous sections, we formulate the following data matrices for $s \in \mathbb{N}_{[1,\bar{s}]}$.
\begin{equation*}
	\begin{split}	
  X_{+}^s&:=\left[\begin{matrix}x(s) \quad x(s+1) \quad \dots \quad x(T+s-1)\end{matrix}\right],\\
		U^s&:=\left[\begin{matrix}
			u(0) & u(1) & \cdots &u(T-1) \\ \vdots & \vdots &\ddots &\vdots \\  u(s-1) & u(s) & \cdots &u(T+s-2)
		\end{matrix}\right],
	\end{split}
\end{equation*}
 and let
 \begin{equation*}
	\begin{split}	
		W^1 &:=W, \\
		W^s& :=\left[ A^{s-1}B_w ~ A^{s-2}B_w ~ \cdots ~ B_w \right]\underline{W}^s,
  \\
		\underline{W}^s&:=\left[\begin{matrix}
			w(0) & w(1) & \cdots &w(T-1) \\ \vdots & \vdots &\ddots &\vdots \\  w(s-1) & w(s) & \cdots &w(T+s-2)
		\end{matrix}\right],
	\end{split}
\end{equation*}
for $s \in \mathbb{N}_{[2,\bar{s}]}$.


Naturally, the noise bound in Assumption \ref{Ass:disturbance} is shifted into its switched system counterpart as follows.

\begin{assumption}[{\cite[Assumption 15]{wildhagen2021datadriven}}]\label{Ass:disturbance:self}
	The noise sequence $\{w(t)\}^{{T}+s-2}_{t=0}$ collected in the matrix $W^s$ satisfies $W^s\in\mathcal{W}^s$ with $\mathcal{W}^s=\{\bar{W}^s\in\mathbb{R}^{n_w^s\times{T}}  |
		\left[\begin{smallmatrix}\bar{W^s}^{\top}\\I \\\end{smallmatrix}\right]^{\top}
		P_d^s
		\left[\begin{smallmatrix}\bar{W^s}^{\top}\\I \\\end{smallmatrix}\right]\succeq0 \}$,
	where $P_d^s$ is a given symmetric matrix obeying $\left[\!\begin{array}{cc}\begin{smallmatrix}I \\0 \\\end{smallmatrix}\end{array} \!\right]^\top P_d^s\left[\!\begin{array}{cc}\begin{smallmatrix}I \\0 \\\end{smallmatrix}\end{array}\! \right]\!\prec \!0$. 
\end{assumption}
For simplicity, we consider a quadratic full-block form of the matrix $P_d^s$ as in \eqref{noise:full}, i.e.,
$P_d^s=\left[\begin{smallmatrix}Q_{d}^s & S_{d}^s \\ * & R_{d}^s\end{smallmatrix}\right]$, 
where matrices $Q_d^s \prec 0 \in \mathbb{R}^{T\times T}$, $S_d^s \in \mathbb{R}^{T\times n_w^s}$, and $R_d^s=R_d^{s\top} \in \mathbb{R}^{n_w^s\times n_w^s}$, $n_w^1:=n_w$ and $n_w^s:=n$ for $s\in \mathbb{N}_2$.
Other choices of matrix $P_d^s$ can be found in Remark \ref{rmk:pd}. 

Moreover, instead of the full row rank assumption of matrix $[X^\top~U^\top]$, a more strict assumption on the data richness is imposed below.
\begin{assumption}[Requirement of data]
	\label{as:Theta:single}
	The matrix 	
$\Theta_{AB}^s=\left[\begin{array}{cc}\begin{smallmatrix}Q_c^s & S_c^s\\\ast & R_c^s \\ \end{smallmatrix} \end{array}\right]:=\left[\begin{array}{cc}\begin{smallmatrix}-X & {0} \\ -U^s & {0} \\ \hline X_{+}^s & B_w^s\end{smallmatrix}\end{array}\right]\left[\begin{array}{cc}\begin{smallmatrix}Q_{d}^s & S_{d}^s \\ * & R_{d}^s\end{smallmatrix}\end{array}\right]\left[\begin{smallmatrix} \cdot\end{smallmatrix} \right]^\top$
	has full column rank.
\end{assumption}

Based on Assumptions \ref{Ass:disturbance:self} and \ref{as:Theta:single},
a data-based representation of system \eqref{sys:switch} is given by
\begin{align}
	\Sigma_{AB}^s=~&\bigg\{[\bar{A}^s~ \underline{\bar{B}}^s]\in \mathbb{R}^{n \times (n+sm)} \Big | \left[\begin{array}{cc}[\bar{A}^s~ \underline{\bar{B}}^s]^{\top}\\I \\\end{array}\right]^{\top}
	\Theta_{AB}^s
	\left[\begin{array}{cc}[\bar{A}^s~ \underline{\bar{B}}^s]^{\top} \\\label{data:represent:self}
		I\\\end{array}\!\right]\succeq0
	\bigg\}.
\end{align}

Leveraging this representation, the following theorem present a data-driven STS. 

\begin{theorem}{[\cite[Theorem 1]{WangCDC}]}\label{data:self:scheme}
	For a given set of parameters, including non-negative scalars {$\sigma_{1}\geq0$ and $\sigma_{2}\geq0$}, a positive definite matrix $\Omega\succ0$, controller gain $K$, and state vector $x(t_k)$ derived from the system presented in \eqref{sys:switch}, the function $\mathcal{Q}(x(t_k),s)$, which is defined in \eqref{sys:QMI}, conforms to the following condition 
	\begin{equation}\label{th3:Q}
		\mathcal{Q}(x(t_k),s)\geq0.
	\end{equation}
	This condition holds true for any $[\bar{A}^{s}~\underline{\bar{B}}^{s}]\in \Sigma_{AB}^{s}$, provided that a positive scalar $\gamma>0$ can be identified, satisfying the subsequent LMI for a certain value of $s\in\mathbb{N}_+$ 
	\begin{equation}\label{Th3:LMI}
		\tilde{\mathcal{Q}}(x(t_k))-\gamma \tilde{\mathcal{G}}^{s}(x(t_k)) \succeq 0,
	\end{equation}
	where the matrices $\tilde{\mathcal{Q}}(x(t_k))$ and $\tilde{\mathcal{G}}^{s}(x(t_k))$ are defined as
	\begin{align}
		\tilde{\mathcal{Q}}(x(t_k))&:=\left[\begin{array}{cc} I & 0\\0 &x^{\top}(t_k) \\\end{array}\right]
		\!\left[\!\!\begin{array}{cc}(\sigma_1-1)\Omega & \!\!\Omega\\\ast & \!\!(\sigma_2-1)\Omega \\\end{array}\!\!\right]\!\![\cdot]^{\top}\notag,\\
		\tilde{\mathcal{G}}^{s}(x(t_k))&:= \left[\begin{array}{ccc} I & 0 & 0\\0 &x^{\top}(t_k)& x^{\top}(t_k)\underline{K}^{s\top}\\\end{array}\right] \tilde{\Theta}_{AB}^{s} [\cdot]^{\top}\notag,\\
		\tilde{\Theta}_{AB}^{s}&:=\left[\!\!\begin{array}{cc}-\tilde{R}_c^{s} & {\tilde{S}_c^{s\top}}\\\ast &  -\tilde{Q}_c^{s} \\\end{array}\!\!\right],\left[\!\begin{array}{cc}\tilde{Q}_c^s & \tilde{S}_c^{s}\\\ast & \tilde{R}_c^{s} \\\end{array}\!\right] :=
		\left[\!\begin{array}{cc}\!Q_c^{s} & S_c^{s}\!\\ \!\ast & R_c^{s}\! \\\end{array}\!\right]^{-1}\notag.
	\end{align}
\end{theorem}

Following this theorem, a data-based version of the function $\mathcal{Q}(x(t_k),s_k)$ in \eqref{sys:self:function} can be derived as 
\begin{equation}\label{sys:self:function:data}
	t_{k+1}=t_k+\max \big\{s_k \in \mathbb{N}_+|\tilde{\mathcal{Q}}(x(t_k))-\gamma \tilde{\mathcal{G}}^{s_k}(x(t_k))\succeq 0\big\}.
\end{equation}


Under the data-based STS in \eqref{sys:self:function:data}, the following theorem presents a data-driven STC method designing the controller gain matrix $K$ and the triggering matrix $\Omega$ in $\tilde{\mathcal{Q}}(x(t_k))$.

\begin{theorem}[{ \cite[Theorem 3]{WangCDC}}]\label{Th6}
Given positive scalars $\sigma_{1}\geq0$, $\sigma_{2}\geq0$, and $\alpha$, it is possible to find a suitable controller gain $K$ such that the system described in \eqref{sec1:sys:disLTI} attains asymptotic stability under the triggering condition delineated in \eqref{sys:self:function:data}. This holds for any $[\bar{A} ~\bar{B}]\in \Sigma_{AB}^1$, contingent upon the existence of a positive scalar $\varepsilon>0$ and matrices $P\succ0$, $\Omega_z\succ0$, $G$, and $K_c$. The key requirement for this design is the fulfillment of the subsequent LMI
	\begin{equation}{\label {Th6:LMI1}}
		\left[\begin{array}{cc}\mathcal{Y}_1 & \mathcal{Y}_2+\mathcal{K}\\\ast & \mathcal{H}+\bar{\mathcal{J}}+\mathcal{Y}_3 \\\end{array}\right]
		\prec0.
	\end{equation}
	Here, the matrices and expressions involved in this theorem are defined as follows
	\begin{align*}
		\mathcal{H}&:=E_2^{\top} P E_2 - E_1^{\top} P E_1,\\
		\bar{\mathcal{J}}&:={\rm Sym}\big\{-\mathcal{L}GE_{2})\big\}+\sigma_1 E_1^{\top} \Omega_z E_1+\sigma_2 E_{3}^{\top}\Omega_z E_{3}- (E_1-E_{3})^{\top} \Omega_z(E_1-E_{3}),\\
		\mathcal{L}&:=(E_1+\alpha E_2)^{\top},~\mathcal{K}:=\Big[E_1^{\top}G^{\top},\,E_{3}^{\top}K_c^{\top} \Big]^{\top},\\
		\mathcal{Y}_1&:=\varepsilon\mathcal{Z}_1\Theta_{AB}^1\mathcal{Z}_1^{\top},
		\mathcal{Y}_2:=\varepsilon\mathcal{Z}_1\Theta_{AB}^1\mathcal{Z}_2^{\top},\\
		\mathcal{Y}_3&:=\varepsilon\mathcal{Z}_2\Theta_{AB}^1\mathcal{Z}_2^{\top},
		\mathcal{Z}_1:=
		[\begin{array}{ccc}I & 0\\\end{array}],~
		\mathcal{Z}_2:=
		[\begin{array}{ccc} 0& \mathcal{L},\\\end{array}]\\
		E_i&:=\left[0_{n\times (i-1)n}, \,I_n, \,0_{n\times (3-i)n} \right], ~(i=1, 2,3).
	\end{align*}
	Furthermore, the controller gain $K$ and triggering matrix $\Omega$ are given by $K=K_cG^{-1}$ and $\Omega={G^{-1}}^{\top} \Omega_z G^{-1}$.
\end{theorem}
 {\begin{remark}
The proposed data-driven ETC/STC methodologies offer three salient advantages over existing aperiodic control strategies:  
$1)$ While model-based event-triggered approaches in \cite{Ding2020,Hu2016}, require system identification that fundamentally limits their theoretical guarantees under noisy conditions, our method eliminates the need for system identification while systematically addressing noise effects through convex hull uncertainty modeling.
$2)$ In contrast to the model-free adaptive event-triggered strategies in \cite{Bu2022Adaptive,Bu2022Power} requiring real-time iterative computations of control inputs and parameters using streaming data, our approach utilizes finite historical data to pre-design fixed control parameters, significantly reducing online computational overhead.
$3)$ Differing from the time-triggered data-driven sampling schemes in \cite{wildhagen2021datadriven} (which determine transmission sequences by optimizing maximum sampling intervals), our method employs a signal filtering mechanism where preconfigured triggering rules dynamically adjust transmission instants based on real-time state trajectories, achieving superior transmission efficiency.
\end{remark}
}

\section{Network security}
{As discussed in Section \ref{sec:intro:security}, DoS attacks and FDI attacks are two of the most notorious attacks.
In addition, resilient control plays an important role in maintaining system performances against these attacks.
Hence, we consider in this section data-driven resilient control against DoS attacks and FDI attacks.}

\subsection{Data-driven resilient control under DoS}

This section reviews a recent data-driven resilient control method against DoS attacks, which is summarized from \cite{Liu2021data}.

\subsubsection{DoS attack modeling}\label{sec:dos}

To start with, a general model that characterizes DoS attacks by only constraining its frequency and duration is introduced.
This model, initially introduced by \cite{depersis2015input} for continuous-time systems, was extended to discrete-time systems in \cite{wakaiki2019stabilization}.
To be specific, for each discrete time instant $t \in \mathbb{N}$, a binary DoS indicator $k(t)$ is introduced as follows
\begin{equation}
	k(t) :=
	\left\{
	\begin{array}{ll}
		0,& {\text{no DoS attack happens at}}~t,\\
		1,&{\text{a DoS attack happens at}}~t.
	\end{array}
	\right.
\end{equation}
The DoS duration during the interval $\mathbb{N}_{[t_1, t_2 -1]}$ is defined as $\Phi_d(t_1, t_2) = \sum_{t = t_1}^{t_2 - 1} k(t)$. Additionally, another binary variable $d(t)$ is defined for each $t \in \mathbb{N}$:
\begin{equation}
	d(t) := \left\{
	\begin{array}{ll}
		1, &k(t) = 1~ {\text{and}}~k(t-1) = 0,\\
		0, &{\text{otherwise}}.
	\end{array}
	\right.
\end{equation}
The DoS frequency during the interval $\mathbb{N}_{[t_1, t_2 -1]}$ is then expressed as $\Phi_f(t_1, t_2) = \sum_{t = t_1}^{t_2 - 1} d(t)$.

Assumptions on DoS frequency and duration adapted from \cite[Assumptions 2.1, 2.2]{wakaiki2019stabilization} are given as follows.
\begin{assumption}[{DoS frequency}]\label{as:dosfre}
	
 There exist constants {$\kappa_f \in \mathbb{R}_{\ge 0}$ and $\nu_f \in \mathbb{R}_{\ge 2}$},  such that the DoS frequency satisfies
	\begin{equation}\label{eq:dosfre}
		\Phi_f(t_1, t_2) \le \kappa_f + {(t_2 - t_1)}/{\nu_f},
	\end{equation}
	over every time interval $\mathbb{N}_{[t_1, t_2 -1]}$, where $t_1\le t_2 \in \mathbb{N}$.
\end{assumption}

\begin{assumption}[{DoS duration}]\label{as:dosdur}
	There exist constants {$\kappa_d \in \mathbb{R}_{\ge 0}$ and $\nu_d \in \mathbb{R}_{\ge 1}$},  such that the DoS duration satisfies
	\begin{equation}\label{eq:dosdur}
		\Phi_d(t_1, t_2) \le \kappa_d + {(t_2 - t_1)}/{\nu_d},
	\end{equation}
	over every time interval $\mathbb{N}_{[t_1, t_2 -1]}$, where $t_1\le t_2 \in \mathbb{N}$.
\end{assumption}


Let $\{s_r\}_{r\in \mathbb{N}}$ denote the time instants of successful transmission, i.e., $k(s_r) = 0$.
A direct consequence of Assumptions \ref{as:dosfre} and \ref{as:dosdur} is that the number of time steps between two successful transmissions is upper-bounded. 
This property plays an important role in resilient controller design and stability analysis, and is introduced as follows. 
\begin{lemma}[\!\!
	Maximum resilience {\cite[Lemma 3]{FengResilient}}]
	\label{lem:dos}
	
	Suppose  the DoS attacks satisfy Assumptions \ref{as:dosfre} and \ref{as:dosdur} with
	\begin{equation}\label{eq:doscondition}
		\setlength{\abovedisplayskip}{3pt}
		\setlength{\belowdisplayskip}{3pt}
		{1}/{\nu_f} + {1}/{\nu_d} < 1.
	\end{equation}
	Then it holds that $s_0 \le T_0 - 1$, and $s_{r + 1} - s_{r} \le T_0$ for all $r\in \mathbb{N}$ with $T_0 := (\kappa_d + \kappa_f)(1 - {1}/{\nu_d} - {1}/{\nu_f})^{-1} + 1$.
\end{lemma}


Condition \eqref{eq:doscondition} is referred to as the maximum resilience against DoS attacks one can achieve
for an open-loop unstable system.

\subsubsection{Data-driven resilient predictive control}\label{sec:ddmpc:state:d}

Consider system \eqref{sys:data:perturbed} with $B_w = I$ and $w(t)\in \mathbb{R}^{n}$, i.e.,
\begin{equation}\label{eq:sys:state}
	x(t+1) = A x(t) + B u(t) + w(t),~~ t \in \mathbb{N}.
\end{equation}
Let $\bar{w} := \max_{t \in \mathbb{N}}\{\Vert w(t)\Vert\}$.
At each time $t$, sensors located at the plant side measure the state $x(t)$ and transmit it to the controller, which then computes a control input and transmits it back to the plant.
The sensor-to-controller (S-C) channel is subject to DoS attacks causing transmission failures of the state $x(t)$ while the controller-to-plant (C-P) is ideal such that the plant can receive input $u(t)$ at every time instant $t$.

The objective here is that, given noisy input-state data $(u^{pre},x^{pre})$ generated from \eqref{eq:sys:state},  design $u(t)$ rendering the closed-loop ISS and achieving maximum resilience against DoS attacks.
To this aim, a data-driven resilient predictive controller is designed below, which is adapted from \cite{Liu2021data}.

For a given prediction horizon $L \in \mathbb
N_+$, the following data-driven MPC is solved at each successful transmission time instant (i.e., $t = s_r$).


\begin{subequations}\label{eq:mpc:state}
	\begin{align}
		J^*_L(u(t), x(t)) := 
		\underset{g(t), h(t)\atop
			\bar{u}_i(t), \bar{x}_i(t)}{\min}
		~&\sum_{i = 0}^{L - 1} c(\bar{u}_i(t), \bar{x}_i(t)) \!+\! \lambda_{g}\bar{w} \Vert g(t)\Vert^2 \!+\! \frac{\lambda_h}{\bar{w}}\Vert h(t)\Vert^2 \nonumber\\
		{\rm s.t.}\quad \;&
		\left[
		\begin{matrix}
			\bar{u}(t)\\
			\bar{x}(t) + h(t)
		\end{matrix}
		\right] =
		\left[
		\begin{matrix}
			{H}_{L+1}(u^{pre})\\
			{H}_{L+1}(x^{pre})
		\end{matrix}
		\right] g(t)\label{eq:mpc:state1},\\
		&	\left[
		\begin{matrix}
			\bar{u}_{-1}(t)\\
			\bar{x}_{-1}(t)
		\end{matrix}
		\right] =
		\left[
		\begin{matrix}
			u(t)\\
			x(t)
		\end{matrix}
		\right] ,\label{eq:mpc:state2}\\
		&	\left[
		\begin{matrix}
			\bar{u}_{L - 1}(t)\\
			\bar{x}_{L - 1}(t)
		\end{matrix}
		\right] =
		\left[
		\begin{matrix}
			0\\
			0
		\end{matrix}
		\right],\label{eq:mpc:state3}\\
		&~ \bar{u}_i \in \mathbb{U},~~ i \in \mathbb{N}_{[0,L - 1]}\label{eq:mpc:state4}.
	\end{align}
\end{subequations}


It can be observed that, when no DoS attack is present (i.e., $t = s_r$), solving problem \eqref{eq:mpc:state} predicts a input-state trajectory of $L$ steps into the future, i.e., from $t$ to $t + L - 1$.
Therefore, during a DoS attack (i.e., $t \neq s_r$), for $t \in \mathbb{N}_{[s_r, s_r + L]}$, we can sequentially use the first $t - s_r$ computed inputs $\bar{u}_i(s_r) \in \mathbb{N}_{[0, t - s_r]}$, and for $t \in \mathbb{N}_{s_r + L + 1}$, we simply use zero inputs until the next successful transmission takes place.

For reference, this scheme is summarized in Algorithm \ref{alg:mpc:state}.
\begin{algorithm}[h]
	\caption{Data-driven resilient control via input-state data}
 \small
	\label{alg:mpc:state}
	\begin{algorithmic}[1]
		\STATE {\bfseries Input:} Prediction horizon $L \ge  2$; parameters of the cost function $R_1 \succ 0$, $R_2 \succ 0$, $\lambda_g > 0$ and $\lambda_h > 0$;
		noise bound $\bar{w}$;
		input-state trajectories $(u^{pre}, x^{pre})$ of system \eqref{eq:sys:state} with $\left[
		\begin{matrix}
			{H}_{L+1}(u^{pre})\\
			{H}_{L+1}(x^{pre})
		\end{matrix}
		\right]$ having full row rank.
		\STATE {\bfseries Construct} Hankel matrix for the input-state trajectory, i.e., $H\! =\! [H^\top_{L}(u^{pre}), H^\top_{L}(x^{pre})]^\top$.
		\STATE {\bfseries If} $t = s_r$, do \label{alg:mpc1}				
		\STATE \quad Use the state $x(t-1)$ and input $u(t-1)$ to solve problem \eqref{eq:mpc:state}.
		Set $u(t) = \bar{u}_0(t)$.
		\STATE {\bfseries Else if} $t \ne s_r$
		\STATE {\bfseries \quad if} $t - s_r \le L -1$
		\STATE \quad \quad Set $u(t) = \bar{u}_{t - s_r}(s_r)$.\label{alg:mpcu1}
		\STATE {\bfseries \quad else if} $t - s_r > L -1$ \label{alg:mpcL}
		\STATE \quad \quad Set $u(t) = 0$.\label{alg:mpcu2}
		\STATE {\bfseries \quad End if}
		\STATE {\bfseries End if}
		\STATE {\bfseries Set} $t = t + 1$ and go back to \ref{alg:mpc1}.
	\end{algorithmic}
\end{algorithm}
Under this algorithm, both maximum resilience and ISS can be achieved, which is consistent with the model-based results presented in \cite{FengNetworked}.
Recursive feasibility of problem \eqref{eq:mpc:state} and the
stability analysis of the closed-loop system \eqref{eq:sys:state} can be found in \cite[Theorem 1]{Liu2021data}.

When only input-output data are available, collect a sequence of input-output data $(u^{pre}, y^{pre}) := (u_{[0,T-1]}, y_{[0,T-1]})$ generated from the following system
\begin{subequations}\label{eq:sys:output}
	\begin{align}
		x(t+1) &= A x(t) + B u(t) + w(t),\\
		y(t) &= Cx(t) + v(t).
	\end{align}
\end{subequations}
Similar to Section \ref{sec:sts:mpc}, let $\eta$ denote the observability index of the pair $(A,C)$.
Additionally, we have bounded process noise $w(t) \in \mathbb{R}^{n}$ and measurement noise $v(t) \in \mathbb{R}^{p}$, both satisfying $\bar{w} := \max_{t \in \mathbb{N}}\{\Vert w(t)\Vert, \Vert v(t)\Vert\}$.

In this setting, several changes are made with respect to the input-state case.
Specifically, the prediction horizon should satisfy $L \in \mathbb{N}_{\eta}$.
In addition, at each successful transmission time instant (i.e., $t = s_r$), instead of transmitting the state $x(s_r)$, an output packet containing the past $\eta$ measurements, i.e., $y_{[s_r - \eta, s_r - 1]}$ is transmitted.
To accommodate these changes, the data-driven MPC scheme in \eqref{eq:mpc:state} is replaced by the following formulation.
\begin{subequations}\label{eq:mpc}
	\begin{align}
		J^*_L(u_{[t - \eta, t - 1]},  y_{[t - \eta, t - 1]}) := 
		\underset{g(t), h(t)\atop
			\bar{u}_i(t), \bar{y}_i(t)}{\min}
		~&\sum_{i = 0}^{L - 1} c(\bar{u}_i(t), \bar{y}_i(t)) \!+\! \lambda_{g}\bar{w} \Vert g(t)\Vert^2 \!+\! \frac{\lambda_h}{\bar{w}}\Vert h(t)\Vert^2 \nonumber\\
		{\rm s.t.}\quad \;&
		\left[
		\begin{matrix}
			\bar{u}(t)\\
			\bar{y}(t) + h(t)
		\end{matrix}
		\right] =
		\left[
		\begin{matrix}
			{H}_{L + \eta}(u^{pre})\\
			{H}_{L + \eta}(y^{pre})
		\end{matrix}
		\right] g(t), \label{eq:mpc1}\\
		&	\left[
		\begin{matrix}
			\bar{u}_{[-\eta, -1]}(t)\\
			\bar{y}_{[-\eta, -1]}(t)
		\end{matrix}
		\right] =
		\left[
		\begin{matrix}
			u_{[t - \eta, t - 1]}\\
			y_{[t - \eta, t - 1]}
		\end{matrix}
		\right]\label{eq:mpc2},\\
		&	\left[
		\begin{matrix}
			\bar{u}_{[L -\eta, L-1]}(t)\\
			\bar{y}_{[L-\eta, L-1]}(t)
		\end{matrix}
		\right] =
		\left[
		\begin{matrix}
			0\\
			0
		\end{matrix}
		\right],\label{eq:mpc3}\\
		&~ \bar{u}_i \in \mathbb{U},~~ i \in \mathbb{N}_{[0,L - 1]}\label{eq:mpc4}.
	\end{align}
\end{subequations}


The resilient control scheme and the stability result remain consistent with Algorithm \ref{alg:mpc:state} and  \cite[Theorem 1]{Liu2021data}.

\subsection{Data-driven Resilient Control under FDI}


To design data-driven resilient controllers against FDI attacks, there are three emerging challenges: i) how to model unknown FDI attacks; 
	ii) how to design a resilient controller against FDI attacks based only on input-state data; and, iii) the associated stability analysis and robustness guarantees.
 This section reviews the main results in \cite{liu2024robustfdi} to address these challenges.

\subsubsection{Healthy system}
For the design of a resilient controller against FDI attacks, we consider system \eqref{eq:sys:state} implementing a time-varying state-feedback controller, i.e.,
\begin{subequations}\label{eq:sys:fdi:ideal}
	\begin{align}
		x(t + 1) &= Ax(t) + Bu(t) + w(t),\label{eq:sys:fdi:ideal:x}\\ 
		u(t) &= K(t)x(t)\label{eq:sys:fdi:ideal:u}.
	\end{align}
\end{subequations}
During online operation, the state and input signals are transmitted through a vulnerable communication channel. These signals are subject to FDI attacks, compromising the integrity of both input and state data, leading to changes in the dynamics described in \eqref{eq:sys:fdi:ideal}.
In this context, the objective is to devise a stabilizing controller in the form of \eqref{eq:sys:fdi:ideal:u} to mitigate the impact of FDI attacks and ensure the closed-loop stability of the resultant FDI-corrupted system.

\subsubsection{Switched FDI Modeling}
Before moving forward, we introduce a series of assumptions to characterize the behavior of an attacker in the context of FDI attacks. 
To distinguish between the pristine offline data denoted as $x(t)$ and $u(t)$ and the data tainted by attacks, we employ the labels $x_p(t)$ and $u_p(t)$ to represent the online polluted states and control inputs, respectively. We posit the following assumption.

\begin{assumption}
	[FDI attack]
	\label{as:3:FDI}
	The FDI attack is governed by the following conditions:
	\begin{itemize}
		\item [i)]
		The attacker possesses access to clean input-state data $U$, $X$, and $X_+$, based on which an attack strategy is devised offline.
		\item [ii)]
		Upon launching an FDI attack, the attacker intercepts $x_p(t)$ as it is transmitted from the plant to the controller, and simultaneously eavesdrops on $u_o = K(t) x_p(t)$, the command sent from the controller to the actuators.
		\item [iii)]
		Based on the attack strategy and the state $x_p(t)$, the attacker computes the injection signal $u_a(t)$ and adds it to the non-corrupted input $u_o(t)$, so that the actuator implements $u_o(t) + u_a(t)$ instead.
	\end{itemize}	
\end{assumption}

{\begin{remark}
    Assumption \ref{as:3:FDI}  is consistent with the model-based setting in \cite{wu2018optimal,wu2019optimalswitching}.
    This is realistic since the attacker can inject zero attack inputs, i.e., $u_a(t) = 0$, for some times to eavesdrop the clean input-state data of the actual system before designing and launching the attack.
    By this assumption, we mean that the considered attack model involves the worst case, i.e., the attacker can design an optimal attack strategy that maximizes the performance degradation.
\end{remark}}


Under this premise, the system corrupted by FDI attacks can be formally stated as
\begin{subequations}\label{eq:sys:c}
	\begin{align}
		x_p(t + 1) &= A x_p(t) + B u_p(t) + w(t),\label{eq:sys:c:x}\\ 
		u_p(t) &= u_o(t) +  u_a(t),\label{eq:sys:c:u}\\
		u_o(t) &= K(t) x_p(t).
	\end{align}
\end{subequations}

Additionally, consider a potent attacker with the capability to compromise a maximum of $m$ actuator channels, where $m$ is the dimension of the input, i.e., $u_p(t) \in \mathbb{R}^{m}$. 
Specifically, at each time step $t$, the attacker selects a maximum of $m$ actuator channels from a predetermined strategy within a finite set $\mathcal{M} := \{0, 1, \cdots,M\}$, where $M$ is defined as $M := \sum_{i = 0}^{m} C_{m}^i$ {with $C_{m}^i$ the combination formula defined in Table \ref{tab:notation}}. 
Furthermore, the attacker employs distinct state feedback matrices for each actuator channel compromise. In other words, for the $j$-th channel combination with $j = 0, \cdots, M$, the attacker deploys the feedback matrix $K_a^j$, resulting in $u_a(t) = D_a^jK^j_a x_p(t)$, where $D_a^j$ characterizes the attack direction.
The following example provides an illustration of $j$-th channel combination and matrix $D_a^j$.

\begin{example}
		Consider a system as in \eqref{eq:sys:fdi:ideal} with $m = 3$ actuator channels, resulting in $M = \sum_{i = 0}^{3} C_{3}^i = 8$ different channel combinations, i.e., $\{\emptyset\}, \{1\}, \{2\}, \{3\}, \{1, 2\}, \{1, 3\}, \{2, 3\}$, $\{1, 2, 3\}$, which are sequentially indexed by the elements in $\mathcal{M}=\{0,1, 2, 3, 4, 5, 6, 7\}$. Suppose that at time $t$ the first and third channels (which therefore corresponds to the $6$-th channel combination $\{1,3\}$) are attacked, i.e., $\sigma(t) = 6$, and the associated channel-selection matrix $D_a^{6}$ is given by $D_a^{6}
			=\left[\begin{smallmatrix}
				1&0&0\\
				0&0&0\\
				0&0&1
			\end{smallmatrix}\right]$.
	\end{example}

Under this configuration, we introduce a piece-wise constant function $\sigma(t)$, which varies with time and takes values from the set $\mathcal{M}$, thus facilitating the switching between subsystems. Consequently, the system under attacks in \eqref{eq:sys:c}  can be reformulated as a switched system with $M$ subsystems that switch based on the signal $\sigma(t)$, as follows 
\begin{subequations}\label{eq:sys:s}
	\begin{align}
		x_p(t + 1) &= A_{\sigma(t)}x_p(t) + B u_o(t) + w(t)\label{eq:sys:s:x},\\ 
		u_o(t) &= K(t) x_p(t)\label{eq:sys:s:u},
	\end{align}
\end{subequations}
where $A_{\sigma(t)} := A  + B  
D_a^{\sigma(t)}K^{\sigma(t)}_a$ is derived as a combination of matrices to capture the attack effect.

We introduce the concept of time instances $t_s$ to denote the moments when an attack is initiated. Specifically, $t_s$ marks the time when the $s$-th attack occurs, defined as $t_{s} = \min\{t >t_{s - 1}:\sigma(t) \ne \sigma(t_{s - 1}) \}$ for $s \in \mathbb{N}_+$, with $t_0=0$. Furthermore, we assume that the system is in mode $j$ at time $t_s$, implying that $\sigma(t) = j$ holds for all $t \in [t_s, t_{s + 1} - 1]$.

To maintain a level of stealth and considering the attacker's limited energy resources, we introduce two crucial assumptions that restrict the switching frequency and the magnitude of the injection signal.
\begin{assumption}[Switching frequency]\label{as:dwell}
	For any $t_1 \le t_2 \in \mathbb{N}$, the function $N_{\sigma}(t_1, t_2)$ counts the number of discontinuities in the signal $\sigma$ over the time interval $[t_1, t_2)$. We assume the existence of constants $\kappa \in \mathbb{N}$ and $\tau \in \mathbb{N}_2$,  such that the following condition is satisfied
	\begin{equation}\label{eq:dwell}
		N_\sigma(t_1,t_2) \le \kappa + (t_2 - t_1)/\tau.
	\end{equation}
\end{assumption}

\begin{assumption}[Attacking power] \label{as:4:delta}
	There exists a constant $\phi >0$ such that $\Vert D_a^{j}K_a^{j}\Vert \le \phi$ holds for $j \in \mathcal{M}$.
\end{assumption}

\begin{assumption}[Attacked system]\label{as:5:ctrl}
	The pairs $(A_j, B)$ for all $j \in \mathcal{M}$ are unknown to the defender but are assumed to be controllable.
\end{assumption}
Note that $C_m^0$ indicating that no attacks occurs, which is also one of the channel combinations. This further implies the controllability of the pair $(A,B)$.

Building on the preliminaries above, the objective becomes that, given noisy input-state data $(u^{pre}, x^{pre})$ generated from \eqref{eq:sys:fdi:ideal:x}, design a time-varying state feedback controller in the form of \eqref{eq:sys:s:u} such that the unknown switched system in \eqref{eq:sys:s} is ISS.

\subsubsection{Data-driven resilient control}\label{sec:ctrl:ctrl}
To begin with, let $\Phi := [X^\top~U^\top]^\top$.
It follows from Lemma \ref{lem:sys_rep:qmi} that
	 the healthy system $Z := [A~B]^\top \in \mathcal{E}^{\bar{w}}$ where
	\begin{equation}
		\mathcal{E}^{\bar{w}} := \Big\{\bar{Z} : \bar{Z}^\top \mathbf{A}_{\bar{w}}\bar{Z} + \bar{Z}^\top \mathbf{{B}}_{\bar{w}} + \mathbf{{B}}_{\bar{w}}^\top \bar{Z} + \mathbf{{C}}_{\bar{w}} \preceq 0\Big\},\label{eq:ellipsoid:bard}
	\end{equation} 
	with 
	\begin{align}\label{eq:ellipsoid:bard:abc}
		\small
		&\mathbf{A}_{\bar{w}} := 
		\Phi \Phi^\top,~\mathbf{B}_{\bar{w}}: = -\Phi X_+^\top, ~\mathbf{C}_{\bar{w}} := X_+X_+^\top \!\!-\! T\bar{w}^2I.
	\end{align}

Let  $\phi_1 := \phi\Vert  B\Vert$.
Under Assumptions \ref{as:4:delta} and \ref{as:5:ctrl}, it is evident that each switching subsystem $Z_j := [A_j~B]^\top$ is at most $\phi_1$-far from the healthy system $[A~B]$, and therefore, for all $j\in\mathcal{M}$, 
\begin{equation}
	\label{eq:set:z_delta}
	Z_j\in 	\mathcal{B}^{\phi_1} :=\!\left \{\bar{Z} = [\bar{A}~\bar{B}]^\top\!\!: \Vert \bar{Z} - Z \Vert  \le \phi_1\right\}.
\end{equation}
A set $\mathcal{B}^{\delta}$ that contains all matrices within set $\mathcal{B}^{\phi_1}$ can be constructed as follows
\begin{equation}\label{eq:ellipsoid:delta}
		\mathcal{B}^{\delta} :=\Big \{\bar{Z}:\bar{Z}^\top\bar{Z} - \bar{Z}^\top \tilde{Z} -\tilde{Z}^\top \bar{Z} + \mathbf{C}^{\delta}  \preceq 0\Big\},
	\end{equation}
	where the matrix $\tilde{Z} = - \mathbf{A}_{\bar{w}}^{-1}\mathbf{B}_{\bar{w}}$ and the constant $\mathbf{C}^{\delta} := \tilde{Z}^\top\tilde{Z} - \delta^2 I$ with 
	\begin{equation*}
		\delta := \underline{\lambda}_{\mathbf{A}_{\bar{w}}}^{-1/2} \Vert (\mathbf{B}_{\bar{w}}^\top\mathbf{A}_{\bar{w}}^{-1}\mathbf{B}_{\bar{w}} - \mathbf{A}_{\bar{w}})^{1/2} \Vert + \phi_1.
	\end{equation*}
\begin{remark}
     {Set \eqref{eq:ellipsoid:delta} defines a matrix ellipsoid constructed based on an upper bound of the attacking power (cf. Assumption \ref{as:4:delta}). While this set is guaranteed to contain all systems consistent with attack-corrupted data, it may also include redundant matrices that do not correspond to any physically realizable attack, which introduces a degree of conservatism in the subsequent controller design.
It is worth highlighting that, since no assumptions are imposed on the attack strategy, the level of conservatism cannot be explicitly quantified in the general case. However, if prior knowledge about the attacker’s behavior, such as the distribution of attack power, the dynamics of the attack, or the frequency of occurrence, becomes available, the conservatism of the ellipsoidal set $\mathcal{B}^\delta$ could be more accurately evaluated.
In such cases, a robustness margin analysis for the designed controller could also be systematically derived, offering a more precise assessment of its performance under attack. Investigating this direction represents an interesting and valuable avenue for future work.}
\end{remark}

During online operation, we assume that the initial conditions $u_p(0)$ and $x_p(0)$ are arbitrary. 
	At time $t \in \mathbb{N}_+$, 
	the key idea is to combine only the most fresh data $(x_p(t-1),u_o(t-1),x_p(t))$ from the subsystem $(A_{\sigma(t-1)},B)$ and the set of offline data from the health system $(A, B)$ to design the controller $K(t)$ on the fly. 
	
	To this end, consider the switched system \eqref{eq:sys:s} at any time $t \in \mathbb{N}_+$, for which we have observed the online data $(x_p(t-1),u_o(t-1),x_p(t))$. 
 The set of matrices ${Z}_t = [{A}_t~{B}_t]^\top$ that can generate $(x_p(t-1),u_o(t-1),x_p(t))$ is given by  
	\begin{equation}
		\mathcal{E}_t = \Big\{Z_t : Z_t^\top \mathbf{A}_{t}Z_t + Z_t^\top \mathbf{{B}}_{t} + \mathbf{{B}}_{t}^\top Z_t + \mathbf{{C}}_{t} \preceq 0\Big\},\label{eq:ellipsoid:t}
	\end{equation}
	where
	\begin{subequations}\label{eq:ellipsoid:t:abc}
		\begin{align}
			&\mathbf{A}_{t} := 
			\left[
			\begin{matrix}
				x_p(t -1)\\
				u_o(t-1 )
			\end{matrix}
			\right]
			\left[
			\begin{matrix}
				x_p(t -1)\\
				u_o(t -1)
			\end{matrix}
			\right]^\top,~~\mathbf{B}_{t}: = -\left[
			\begin{matrix}
				x_p(t -1)\\
				u_o(t -1)
			\end{matrix}
			\right]x_p^\top(t), ~~\mathbf{C}_{t} := x_p(t)x_p^\top(t) - \bar{w}^2 I\label{eq:ellipsoid:t:bc}.
		\end{align}
	\end{subequations}
Set $\mathcal{E}_t$ is also a matrix ellipsoid. 
It is evident that  the active subsystem $[A_{\sigma(t-1)} ~B]^\top \in \mathcal{E}_t$.

Note from  \eqref{eq:ellipsoid:delta} that all subsystems are contained in $\mathcal{B}^{\delta}$, which implies $Z_{\sigma(t-1)} = [A_{\sigma(t-1)}~B]^\top \in \mathcal{B}^{\delta}$ for all $t$. Combining this with \eqref{eq:ellipsoid:t}, we conclude that
	\begin{equation}\label{eq:set:int}
		Z_{\sigma(t-1)} \in \mathcal{E}_{t} \cap \mathcal{B}^{\delta},\quad \forall t \in \mathbb{N}_+,
	\end{equation} 
	where the intersection set $\mathcal{E}_{t} \cap \mathcal{B}^{\delta}$ is bounded.
	The problem thus reduces to finding a controller $K(t)$ to stabilize all the systems in the set $\mathcal{E}_{t} \cap \mathcal{B}^{\delta}$ for all $t$.
 This can be achieved by the following SDP

	\begin{subequations}\label{eq:lqr:on}
	
		\begin{align}
			 \min_{\gamma,\beta,  \tau_1, \tau_2,\atop P, Y, L, Q} & ~~\gamma \label{eq:lqr:on:a}\\
			{\rm s.t. }  ~~~ & \left[
			\begin{matrix}
				P - \beta I\!\!&0 \!\!\!& 0\!\!\!& 0\\
				\star \!\!& -P \!\!\!& -Y^\top \!\!\!& 0\\
				\star \!\!& \star \!\!\!& 0 \!\!\!& Y\\
				\star \!\!&\star \!\!\!& \star \!\!\!& P
			\end{matrix}
			\right] 
			- \tau_1 
			\left[
			\begin{matrix}
				-\mathbf{C}_t \!&\! -\mathbf{B}_t^\top \!&\!\! 0\\
			\star \!&\! -\mathbf{A}_t \!&\!\! 0\\
				\star \!&\! \star \!&\!\! 0
			\end{matrix}
			\right] 
			- 
		\tau_2 
			\left[
			\begin{matrix}
				-\mathbf{C}^{\delta} & -Z_{\rm tr}^\top & 0\\
				\star & -I & 0\\
				0 & 0 & 0
			\end{matrix}
			\right] \succeq 0, \label{eq:lqr:on:b}\\
			& \beta >0,~~\tau_1 \ge 0,~~ \tau_2 \ge 0, ~~P \succ 0,\\
			&\left[
			\begin{matrix}
				L & Y\\
				Y^\top & P
			\end{matrix}
			\right] \succeq 0,\label{eq:lqr:on:d}\\
			&\left[
			\begin{matrix}
				Q & I\\
				I & P
			\end{matrix}
			\right] \succeq 0, \label{eq:lqr:on:e}\\
			&{\rm Tr}(P) + {\rm Tr}(L) + \epsilon \Vert Q\Vert\le \gamma. \label{eq:lqr:on:f}
		\end{align}
	\end{subequations}

The feasibility of this SDP at each time $t$ can be guaranteed following a similar step as in \cite[Theorem 3.1]{Liu2023fdi}.
	Let $(\gamma^*(t), P^*(t), Y^*(t), L^*(t),Q^\ast(t))$ denote any optimal solution. The stabilizing controller $u_o(t) = K(t)x_p(t)$ for all systems in $\mathcal{I}_t^\ast$ 
	can be designed as follows
	\begin{equation}\label{eq:Kk}
		K(t) = {Y}^*(t) ({P}^{*}(t))^{-1}.
	\end{equation}
	
	Based on \eqref{eq:lqr:on}--\eqref{eq:Kk}, the online data-driven controller is summarized in Algorithm \ref{alg:ctrl}.
Recursive feasibility of the SDP \eqref{eq:set:int} and stability analysis under this data-based controller can be found in \cite[Theorem 3.1]{liu2024robustfdi}.
 {\begin{remark}
It is worth highlighting several notable advantages of the reviewed data-driven method in \cite{Liu2023fdi} compared to previous works such as \cite{wu2019optimalswitching,rotulo2021online}.
\begin{itemize}
    \item [1)]
    Although \cite{wu2019optimalswitching} designed a resilient controller under FDI attacks, their method relies on pre-collected injected data and requires exact knowledge of the system matrices, neither of which is needed in Algorithm \ref{alg:ctrl}.
    Moreover, instead of computing a time-invariant controller gain matrix, the gain \eqref{eq:Kk} is time-varying and hence can automatically adapt to the unknown FDI-induced changes in system dynamics.
    
    \item [2)]
    Although an online data-driven LQR method for stabilizing unknown switched systems was considered in \cite{rotulo2021online}, the method in Algorithm \ref{alg:ctrl} offers four main differences and advantages:
    \begin{itemize}
        \item[i)] The core idea in Algorithm \ref{alg:ctrl} is the online implementation of a modified SDP formulation from \cite{bisoffi2022data}, where the LQR structure is adopted mainly to ensure recursive stability; therefore, the associated proofs deviate significantly from those in \cite{rotulo2021online}.
        \item[ii)] Algorithm \ref{alg:ctrl} leverages Petersen's lemma, which enables direct extension to scenarios involving process noise, a feature not addressed in \cite{rotulo2021online}.
        \item[iii)] Algorithm \ref{alg:ctrl} reduces the computational complexity from $\mathcal{O}(n_x^3(n_x(n_u + 1) + n_u)^3)$ to $\mathcal{O}(n_x^3(n_x + n_u)^3)$.
        \item[iv)] Algorithm \ref{alg:ctrl} does not impose a requirement that the minimum switching interval must be larger than $2(n_x(n_u + 1) + n_u) - 1$, thereby allowing for stronger and more frequent attacker behaviors.
    \end{itemize}
\end{itemize}
\end{remark}
}


\begin{algorithm}[t]
	\caption{Online data-driven control.}
 \small
	\label{alg:ctrl}
	\begin{algorithmic}[1]
		\STATE {\bfseries Offline:} 
		Collect input-state data $(u_{[0,T\!-1]}, x_{[0,T]})$ and
			form matrices $X,X_+,U$, and $\Phi$.
			Compute $\mathcal{B}^{\delta}$ as in \eqref{eq:ellipsoid:delta}.
			\STATE {\bfseries Online:} 
			Given initial conditions $x_p(0)$, $u_o(0)$ and 
			constant $\epsilon >0$, for $t =1,2,3,\ldots$, do 
			\STATE 
			\begin{itemize}
				\item[1)]
				Compute the matrix ellipsoid $\mathcal{E}_t$ in \eqref{eq:ellipsoid:t} based on $u_o(t - 1)$, $x_p(t - 1)$, and $x_p(t)$.
				\item[2)]
				Solve the SDP in \eqref{eq:lqr:on} and denote the solution as $(\gamma^*(t), P^*(t), Y^*(t), L^*(t),Q^*(t))$.
				\item[4)]
				Compute the control input $u_o(t) = K(t)x_p(t)$, where $K(t) = Y^*(t)(P^*(t))^{-1}$ is given in \eqref{eq:Kk}.
				\item[5)]
				Set $t = t+1$ and go back to 1).
			\end{itemize}
	\end{algorithmic}
\end{algorithm}


\section{Distributed configuration}

In this section, our focus shifts from centralized network systems to the distributed network systems, i.e., MASs. 
We survey results in \cite{wang2023event,li2023selftriggered,li2023poly}, including distributed data-driven methods for MASs to design event-triggered, self-triggered, and output synchronization control protocol.
To avoid notations burden, we use ETC/STC also to represent event-/self-triggered consensus control in the following subsequent sections.


Before delving further, some fundamental graph theory concepts frequently employed in the analysis and synthesis of MASs are recalled.

\emph{Notation  (graph theory).}
In the context of our communication network, we employ a weighted graph ${\mathcal{G}}=({\mathcal{V}}, {\mathcal{E}})$ to depict the communication topology among our agents. Here, ${\mathcal{V}}=\{v_1,\ldots, v_N \}$ signifies the set of nodes, while ${\mathcal{E}} \subseteq {\mathcal{V}}\times {\mathcal{V}}$ represents the set of edges. Each element in ${\mathcal{E}}$, denoted as $(v_i, v_j)$, symbolizes the link from node $v_j$ to node $v_i$. The construction of the adjacency matrix $\mathcal{A}=[a_{ij}]\in \mathbb{R}^{N\times N}$ involves setting $a_{ij}>0$ only if $(v_j, v_i)\in \mathcal{E}$; otherwise, $a_{ij}=0$. It is pertinent to note that self-loops are not considered, meaning that $a_{ii}=0$ holds for all $i\in\mathbb{N}_{[1,N]}$. The in-degree of node $v_i$ is described by $d_i=\sum_{j=1}^{N}a_{ij}$. The degree matrix $\mathcal{D}$ is introduced as ${\rm{diag}}\{d_i\}_{i=1}^{N}$. Furthermore, the Laplacian matrix $\mathcal{L}=[l_{ij}]\in \mathbb{R}^{N\times N}$ associated with ${\mathcal{G}}$ is formally defined as $\mathcal{L}=\mathcal{D}-\mathcal{A}$. The neighbor set of node $v_i$ is denoted by $\mathcal{N}_i=\{j\in {\mathcal{V}}|(i,j)\in {\mathcal{E}}\}$.

We extend our graph to encompass additional elements by introducing $\bar{\mathcal{G}}=(\bar{\mathcal{V}}, \bar{\mathcal{E}})$, where $\bar{\mathcal{V}}={\mathcal{V}}\cup v_0$, with $v_0$ representing the node associated with the leader. In this extension, $\bar{\mathcal{E}}$ includes all the arcs present in ${\mathcal{E}}$ as well as those spanning from $v_0$ to ${\mathcal{E}}$. Specifically, a graph $\bar{\mathcal{G}}$ is deemed to contain a directed spanning tree if a particular node, known as the root, exists such that every other node within $\bar{\mathcal{V}}$ can be reached via a directed path originating from this root node.
The pinning matrix $\mathcal{P} = {\rm{diag}}\{p_i\}_{i=1}^{N}$ serves as a valuable tool for describing the accessibility of the leader node $v_0$ to the remaining nodes $v_i\in \mathcal{V}$. It is worth noting that the pinning gain $p_{i} > 0$ if $(v_0,v_i)\in \bar{\mathcal{E}}$, while $p_{i} = 0$ otherwise.
In addition, define the matrix $\Lambda := \mathcal{L} + \mathcal{P}$. 
{It should be noted that a topology $\bar{\mathcal{G}}$ that contains a directed spanning tree is discussed over this section} (see Assumption \ref{as:graph1} in Section \ref{sec:etc:mas} for details).

\subsection{Distributed data-driven ETC}
\label{sec:etc:mas}
{Generally, MASs can be categorized into two types depending on whether all agents have identical systems, namely, homogeneous (i.e., identical systems $(A,B)$) and heterogeneous (i.e., non-identical systems $(A_i,B_i)$).
The property of having identical systems simplifies many challenges in MASs; therefore, data-driven ETC design and stability analysis can be directly extended from Section \ref{sec:etc}; see \cite{Li2023data} for details.
Therefore, this section reviews a data-driven design for heterogeneous MASs, which is summarized from \cite{wang2023event}.
}
Consider a heterogeneous leader-following MAS, comprising a leader (indexed as $0$) and $N$ followers (indexed from $1$ to $N$).
The communication network among these agents is represented by a directed graph $\bar{\mathcal{G}}$, in accordance with the following assumption.
\begin{assumption}[Communication topology]
	\label{as:graph1}
	Assume that only a subset of followers has direct access to the leader's information, and the graph $\bar{\mathcal{G}}$ contains a directed spanning tree with the leader serving as the root.
\end{assumption}

For all $i\in \mathbb{N}_{[0,N]}$, the dynamics are described by the following linear discrete-time recursions
\begin{subequations}\label{eq:mas2:all}
    \begin{align}
	x_i(t+1)&=A_i x_i(t)+B_i u_i(t), \label{eq:mas2}\\
u_i(t)& =\left\{\begin{array}{lll}
		K_{i}x_i(t_k^i), ~~~~~~~~~~~~i=0\\
		\sum_{j \in \mathcal{N}} K_{ij}e_{ij}(t_k^i),~~i>0\\
	\end{array}
	\right., t \in \mathbb{N}_{[t_k^i, t_{k+1}^i -1]} ,\label{eq:mas2:controller}
\end{align}
\end{subequations}
 where $e_{ij}(t_k^i):=x_i(t_k^i)- x_j(t_{k'(t)}^j)$ represents the state measurement error between agents $i$ and $j$, and $k'(t):={\rm arg} \min_{l\in \mathbb{N}:t\geq t_l^j}$ $ \{t-t_l^j\}$ (where $t_l^j$ is the transmitted instant of agent $j$),
and therefore, for each $t\in \mathbb{N}_{[t_k^i, t_{k+1}^i-1]}$, $t_{k'(t)}^j$ is the last transmitted time of agent $j$; $t_k^i \in \mathbb{N}$ is the sampling time generated by the distributed version of the periodic
dynamic ETS in Section \ref{sec:etc}, i.e., 
\begin{equation}\label{eq:mas2:trigger}
	t_{k+1}^i=t_k^i+h\min_{v\in \mathbb{N}}\big\{v>0\big|\eta_i({\tau_{k,v}^i})+\theta_i\rho_i({\tau_{k,v}^i}) < 0\big\},
\end{equation}
where
$\tau_{k,v}^i:=t_k^i+vh$,
$v\in\mathbb{N}_+$, $h$ shares the same definition as in Section \ref{sec:etc}, constant $\theta_i>0$ is to be designed,
and functions
\begin{align*}
	\rho_0(\tau_v^0)&:=\sigma_0 x_0^\top (t_k^0)\Omega_0 x_0(t_k^0)-e_0^\top (\tau_v^0)\Omega_0 e_0(\tau_v^0),\\
	\rho_i({\tau_{k,v}^i})&:=\sum_{j\neq i}^N\sigma_{ij} e_{ij}^\top (t_k^i)\Omega_i e_{ij}(t_k^i)-e_i^\top ({\tau_{k,v}^i})\Omega_i e_i({\tau_{k,v}^i}),
\end{align*}
where $\Omega_i\succ0$ is a weight matrix, and $\sigma_0$, $\{\sigma_{ij}\}_{j\in\mathcal{N}}$ are parameters, all to be designed; {$e_i({\tau_{k,v}^i}):=x_i({\tau_{k,v}^i})-x_i(t_k^i)$ denotes the error of agent $i$ between the latest transmitted signal $x_i(t_k^i)$ and the current sampled signal $x_i({\tau_{k,v}^i})$; and, $\eta_i({\tau_{k,v}^i})$ satisfies}
\begin{equation}\label{eq:mas2:dynamic}
	\eta_i({\tau_{k,v+1}^i})-\eta_i({\tau_{k,v}^i})=-\lambda_i\eta_i({\tau_{k,v}^i})+\rho_i({\tau_{k,v}^i}),
\end{equation}
where $\eta_i(0)\geq0$ and $\lambda_i>0$ are given parameters. 
\begin{remark}
    {In addition to the MAS introduced in \eqref{eq:mas2}, most existing works consider a MAS, where the leader's input is zero, i.e., $u_0(t) = 0$ for all $t\in \mathbb{N}$; see e.g., \cite{Hong2013,Zhang2022self, Zegers2022}.
Although incorporating a nonzero, time-varying control input for the leader in \eqref{eq:mas2} complicates the design procedure, it brings several advantages. 
Specifically, this model is more realistic and applicable in practice since it allows the leader to adjust its behavior accordingly and optimize its performance.
For example, in vehicle platoon control systems, the leader vehicle may need to adjust its speed, acceleration, or trajectory based on traffic conditions, road inclines, or other dynamic factors.
To capture these changes, modeling the leader by involving a control input is more appropriate.}
\end{remark}

For data collection, we consider the perturbed version of \eqref{eq:mas2} as follows
\begin{equation}\label{eq:mas1:perturbed}
	x_i(t+1)=A_i x_i(t)+B_i u_i(t)+B_{w,i} w_i(t),\quad \forall i\in\mathbb{N}_{[0,N]},
\end{equation}
where  $B_{w,i} \in \mathbb{R}^{n\times n_{w}}$ is known and has full column rank.
The objective of distributed data-driven ETC is that, given input-state data $(u_i^{pre}, x_i^{pre}):=\{u_i(t), x_i(t)\}_{t = 0}^T$ of each agent $i$ generated from system \eqref{eq:mas1:perturbed}, design controller gain matrices $K_0$, $K_{ij}$, and matrices $\Omega_0$, $\Omega_{ij}$ in the ETS \eqref{eq:mas2:trigger} ensuring the state consensus of the MAS \eqref{eq:mas2:all}.

Since the objective for each agent $i$ is to follow the leader, we shifts our focus from the state of each agent to the error between the leader and each follower.
Let $\epsilon_i(t):=x_i(t)-x_0(t)$ denote the tracking error between the leader and follower $i$ with $i = 1, \cdots, N$. Upon collecting the errors of all agents along with the state of the leader to form $\epsilon(t):=[\epsilon_1^{\top}(t)~\cdots~\epsilon_N^{\top}(t)~x_0^{\top}(t)]^{\top}$, we establish the following closed-loop system expression
\begin{equation}\label{eq:mas2:close}
	\epsilon(t+1)=A \epsilon(t)+BK\epsilon(t_k), ~t\in \mathbb{N}_{[t_k^i, t_{k+1}^i-1]}.
\end{equation}
Here, $\epsilon(t_k):=[\epsilon_1^{\top}(t_k^1)~\cdots~\epsilon_N^{\top}(t_k^N)~ x_0^{\top}(t_{k'(t)}^0)]^{\top}$, $\epsilon_i(t_k^i):=x_i(t_k^i)-x_0(t_{k'(t)}^0)$, and
\begin{align*}
	A&:=\left[ \begin{array}{c|ccc}
		{\rm diag}\{A_i\}^N_{i=1} &\mathbf{1}_N {\rm diag}\{A_{i}-A_0\}^N_{i=1}\\
		\hline
		0&A_0 \\
	\end{array} \right],\\
	B&:=\left[ \begin{array}{c|cccc}
		{\rm diag}\{B_i\}^N_{i=1}&{\mathbf{1}_N} \otimes (-B_0) \\ \hline
		0&B_0 \\
	\end{array} \right],\\
	K&:=\left[ \begin{matrix}
		\sum_{j \in \mathcal{N}} K_{1j}& -K_{12} & \cdots & -K_{1N}& 0 \\
		\vdots   & \vdots   &\ddots & \vdots &\vdots\\
		-K_{N1}  & -K_{N2} & \cdots & \sum_{j \in \mathcal{N}} K_{Nj} & 0\\
		0  & 0  &\cdots & 0 & K_0\\
	\end{matrix} \right],
\end{align*}
where {$\mathbf{1}_N$ denotes an $N$-dimensional column vector whose elements are $I$.}

According to the definition of the tracking error $\epsilon_i(t)$, for each agent $i$, we compute from the state data $x_i^{pre}$ the tracking error data $\epsilon_i^{pre} :=\{\epsilon_i(t)\}^{T}_{t=0}$. 
Denote  $u(t):=[u_1^{\top}(t)~\cdots~u_N^{\top}(t)~ u_0^{\top}(t)]^{\top}$ and construct the following data matrices
\begin{align*}
	U & :=\left[\begin{matrix}
			u(0) & u(1) & \cdots & u(T-1)
		\end{matrix} \right],\\		
		E & :=\left[\begin{matrix}
			\epsilon(0) & \epsilon(1) & \cdots & \epsilon(T-1)
		\end{matrix}\right], \\
	 E_{+} & :=\left[\begin{matrix}
			\epsilon(1) & \epsilon(2) & \cdots & \epsilon(T)
		\end{matrix}\right].
\end{align*}

In accordance with Lemma \ref{lem:sys_rep:qmi} and leveraging these data, a data-based representation can be derived for the error system \eqref{eq:mas2:close}.

\begin{lemma}[{\cite[Lemma 1]{wang2023event}}]\label{lem:leadermas:noise}
	Suppose that the unknown noise matrix $W_i$ for each agent satisfies Assumption \ref{Ass:disturbance}. In this context, the set of system matrices $[\bar{A}~\bar{B}]$ capable of explaining the data $(U,E,E_{+})$ can be represented as
	\begin{align*}
		\Sigma_{AB}=
		\left\{[\bar{A}~\bar{B}] \Big |
		\left[\begin{smallmatrix}[\bar{A}~\bar{B}]^\top \\I \\\end{smallmatrix}\right]^\top
		\Theta_{AB}
		\left[\begin{smallmatrix}[\bar{A}~\bar{B}]^\top \\I \\\end{smallmatrix}\right]\succeq0
		\right\},
	\end{align*}
	where  $\Theta_{AB}:=
	\left[\begin{smallmatrix}-E & 0 \\ -U & 0 \\ \hline E_+ & B_w \\\end{smallmatrix}\right]
	Q_d
	\left[\begin{smallmatrix}-E & 0 \\ -U & 0 \\ \hline E_+ & B_w \\\end{smallmatrix}\right]^\top $ and
 	$B_{w}:=\left[\begin{smallmatrix}\begin{array}{c|c}
		{\rm diag}\{B_{w,i}\}^N_{i=1} & \mathbf{1}_N \otimes (-B_{w,0})\\ \hline
		0 & B_{w,0} \\ 
	\end{array} \end{smallmatrix}\right]$.
\end{lemma}
Building on the data-based system representation, the subsequent design follows a similar approach as in the model-based case in \cite[Theorem 2]{Wang2023Distributed}, and the main result is presented below.


\begin{theorem}
	[{\cite[Theorem 3]{wang2023event}}]\label{thm:mas2}
	Consider the heterogeneous MAS \eqref{eq:mas2} under the event-triggered state-feedback consensus control law \eqref{eq:mas2:controller}-\eqref{eq:mas2:trigger}. 
	Given positive scalars $\sigma_{0}$, $\sigma_{ij}$,  $\bar{h}$, $\underline{h}$, and $\lambda_i$, $\theta_i$ satisfying $1-\lambda_i-\frac{1}{\theta_i}\geq0$ for all $i\in\mathbb{N}_{[0,N]}$ and $j\in \mathcal{N}$,
	there exists a block controller gain $K$ such that asymptotic consensus of the system is achieved for any $[\bar{A} ~\bar{B}]\in \Sigma_{AB}$, and $\eta_i({\tau_{k,v}^i})$ tends to zero for any $\eta_i(0)\ge0$,
	if there exist matrices $R_1\succ0$, $R_2\succ0$, $P\succ0$,
	$S=S^\top$, $M_1$, $M_2$, $G$, $K_c$, and $\bar{\Omega}_i\succ0$ for all $i\in\mathbb{N}_{[0,N]}$, satisfying LMIs $\forall h\in\{\underline{h}, \bar{h}\}$, $\varsigma=1,2$,
	\begin{equation}{\label {Th3:LMI1}}
		\left[
		\begin{matrix}
			\mathcal{T}_1& \mathcal{F}+\mathcal{T}_2& 0\\
			\ast &  \Xi_0+h\Xi_\varsigma+{\hat{\Psi}}+\bar{\mathcal{Q}}+\mathcal{T}_3  & hM_\varsigma\\
			\ast &  \ast & -hR_\varsigma
		\end{matrix}
		\right]\prec0,
	\end{equation}
	where
	\begin{align*}
		\hat{\Psi}&:={\rm Sym}\big\{\!-\mathcal{D}GH_{2}\big\},~\mathcal{F}:=\left[H_1^\top G^\top,H_{5}^\top K_c^\top  \right]^\top, \\
		\mathcal{D}&:=(H_1+\epsilon H_2)^\top ,\mathcal{V}_1:=
		\left[\begin{matrix}I & 0\\\end{matrix}\right],~
		\mathcal{V}_2:=
		\left[\begin{matrix} 0& \mathcal{D}\\\end{matrix}\right],\\
		\mathcal{T}_1&:=\mathcal{V}_1\Theta_{AB}\mathcal{V}_1^\top ,~
		\mathcal{T}_2:=\mathcal{V}_1\Theta_{AB}\mathcal{V}_2^\top ,~
		\mathcal{T}_3:=\mathcal{V}_2\Theta_{AB}\mathcal{V}_2^\top,
	\end{align*}
 and other matrices are outlined in \cite[Theorem 3]{wang2023event}. 
	Moreover, the desired block controller matrix $K$ is given by $K=K_cG^{-1}$, and the triggering matrices are designed as  $\Omega_a={G^{-1}}^\top\bar{\Omega}_a{G^{-1}}$, and $\Omega_b={G^{-1}}^\top\bar{\Omega}_b{G^{-1}}$.
\end{theorem}

Having derived a distributed data-driven ETC, we now take a step further by considering noise also in the online operation, i.e., replace \eqref{eq:mas2} with
\begin{equation}\label{eq:mas2:noise}
	x_i(t+1)=A_i x_i(t)+B_i u_i(t) + B_{w,i} w_i(t), ~t\in \mathbb{N}.
\end{equation}
which is more general in practice.
In this case, an $\mathcal{H}_{\infty}$ should be designed from noisy data. 

The closed-loop error dynamics become 
\begin{equation}\label{sys:sampling:entire:noise}
	\epsilon(t+1)=A \epsilon(t)+BK\epsilon(t_k)+B_{w}w(t), ~t\in \mathbb{N}_{[t_k^i, t_{k+1}^i-1]},
\end{equation}
where $w(t):=[w_1^{\top}(t)~\cdots ~w_N^{\top}(t)~ w_0^{\top}(t)]^{\top} \in \mathbb{R}^{(N+1)n_w}$.


{Difference from \eqref{eq:mas2:close}, due to the presence of the noise $w(t)$, zero tracking error cannot be achieved.
To improve the performance of system \eqref{sys:sampling:entire:noise}, we seek for $\mathcal{H}_{\infty}$ control method.
For clarity, the definition of $\mathcal{H}_{\infty}$ stabilization for the closed-loop system \eqref{sys:sampling:entire:noise} is introduced as follows.
\begin{definition}\label{def:H}
Given a scalar $\gamma>0$, the system \eqref{sys:sampling:entire:noise} achieves $\mathcal{H}_{\infty}$ consensus with the disturbance attenuation $\gamma$ if the following conditions hold.
\begin{enumerate}
		\item [1)] The system \eqref{sys:sampling:entire:noise} with the controller \eqref{eq:mas2:controller} is asymptotically stable with zero disturbance $w(t)=0$;
		\item [2)] The following bounded $\mathcal{L}_2$-gain condition is satisfied under zero initial condition for all nonzero {$w_i(t)\in \mathcal{L}_2[0,\infty]$}
\begin{equation}
\sum_{t=0}^{+\infty} \epsilon^\top(t)\epsilon(t)\leq \sum_{t=0}^{+\infty} \gamma^2 w^\top(t)w(t).
\end{equation}
	\end{enumerate}
\end{definition}
}

Building upon Theorem \ref{thm:mas2} and Definition \ref{def:H}, a data-driven $\mathcal{H}_{\infty}$ ETC can be designed as follows.

\begin{theorem}
	[{\cite[Theorem 4]{wang2023event}}]\label{thm:mas2:H}
	Consider the system \eqref{sys:sampling:entire:noise} under the event-triggered state-feedback consensus control law \eqref{eq:mas2:controller}-\eqref{eq:mas2:trigger} over the graph $\bar{\mathcal{G}}$. Given the same scalars as in Theorem \ref{thm:mas2},
	there exists a controller gain $K$ such that $\mathcal{H}_{\infty}$ consensus of the system is achieved with a given disturbance attenuation $\gamma>0$ for any $[\bar{A} ~\bar{B}]\in \Sigma_{AB}$, and $\eta_i({\tau_{k,v}^i})$ tends to zero for any $\eta_i(0)\ge0$,
	if there exist matrices $R_1\succ0$, $R_2\succ0$, $P\succ0$,
	$S=S^\top$, $M_1$, $M_2$, $G$, $K_c$, and $\bar{\Omega}_i\succ0$ for all $i\in\mathbb{N}_{[0,N]}$, satisfying the following LMIs for all $ h\in\{\underline{h}, \bar{h}\}$
	\begin{align}{\label {eq:thm:mas2:H}}
		&\left[
		\begin{matrix}
			\mathcal{T}_1& \mathcal{F}+\mathcal{T}_2& 0& 0 \\
			\ast &  \Xi_0+h\Xi_\varsigma+{\tilde{\Psi}}+\bar{\mathcal{Q}}+\mathcal{T}_3  & hM_\varsigma&\mathcal{D}B_dG\\
			\ast &  \ast & -hR_\varsigma& 0\\
			\ast &  \ast &\ast &  -\gamma^2 G^\top G
		\end{matrix}\!
		\right]\!\prec0,~~~\varsigma=1,2,
	\end{align}
	where ${\tilde{\Psi}=\hat{\Psi}}+H_1^{\top}G^\top G H_1$.
	Furthermore, the desired controller matrix  can be computed as $K=K_cG^{-1}$, and the triggering matrices are designed as $\Omega_a={G^{-1}}^\top\bar{\Omega}_a{G^{-1}}$ and $\Omega_b={G^{-1}}^\top\bar{\Omega}_b{G^{-1}}$.
\end{theorem}


\subsection{Distributed data-driven STC}

Similar to the centralized case, predicting future states from noisy historical data and the state from the last triggering time plays an important role in the design of distributed data-driven STC.
However, this procedure is rather challenging even for homogeneous MASs, and there are few existing results, with the exception of \cite{li2023selftriggered}.
That work extended the switched systems approach in Section \ref{sec:sts:swithced} and proposed a data-driven STC for homogeneous MASs with a zero-input leader (i.e., $u(t) =0$ for all $t \in \mathbb{N}$).
In the following, we review the main results from \cite{li2023selftriggered}, which is only a starting point for distributed data-driven STC and suggests two possible research directions: i) data-driven STC for heterogeneous MASs, and ii) data-driven STC using an MPC-based method.




Consider a homogeneous leader-following MAS as follows
\begin{subequations}\label{eq:mas}
	\begin{align}
 {x}_{0}(t+1)&={A} x_{0}(t)\label{eq:leader},\\
		{x}_{i}(t+1)&={A} x_{i}(t)+{B} u_{i}(t), \quad  i\in\mathbb{N}_{[1,N]} \label{eq:follower},
		\\ 
  u_{i}(t)&=K {z_i(t_k^i)},\quad t\in \mathbb{N}_{[t_k^i,\,t_{k+1}^{i}-1]},\label{eq:self:controller}
	\end{align}
\end{subequations}
where $z_i(t)$ is the combined measurement variable given by
\begin{equation}
\label{z}
z_i(t):=\sum_{j=1}^{N} a_{i j}\left({x}_{i}(t)-{x}_{j}(t)\right)+p_{i}\left({x}_{i}(t)-{x}_{0}(t)\right).
\end{equation}
The triggering time $t_k^i$ is generated by the STS below
\begin{equation}
	\label{eq:self:trigger}
	t_{k+1}^i=t_k^i+\inf\left\{s_k^i \in \mathbb{N}_{[1,\bar{s}]}\big|f_i\left(x_i(t_k^i),\,z_i(t_k^i),\, s_k^i\right)\geq0 \right\},
\end{equation}
where the triggering function $f_i\left(x_i(t_k^i),z_i(t_k^i),s_k^i\right)$ is defined as $f_i\left(x_i(t_k^i),\,z_i(t_k^i),\,s_k^i\right)=e_{i}^{\top} (s_k^i) \Phi e_{i}(s_k^i)-\sigma z_i^{\top}(t_k^i) \Phi z_{i}(t_k^i)$.
Here, $\Phi\in \mathbb{R}^{n\times n}$ represents a positive definite matrix to be designed, $\sigma$ is a positive constant, $s_k^i=t_{k+1}^i-t_k^i$ represents the $k$th inter-event triggering interval for agent $i$, and $e_{i}(s_k^i)=x_i(t_k^i+s_k^i)-x_i(t_k^i)$ denotes the error between the last broadcast state at $t_k^i$ and the state at $t_{k+1}^i$. 

The objective here is that, given input-state data $(u_i^{pre},x_i^{pre})$ of leader and a follower $i$ generated from the perturbed version of system \eqref{eq:leader}--\eqref{eq:follower}, design controller gain matrix $K$, and matrix $\Phi$ in the STS \eqref{eq:self:trigger} ensuring the state consensus of the MAS \eqref{eq:mas}.
The rest of the design extend the  data-driven switched systems approach-based STC in Section \ref{sec:sts:swithced} from centralized systems to MASs.
Similar to Section \ref{sec:etc:mas}, controller design and stability analysis in the distributed setting focus on the dynamics of the state error between the $i$th agent and the leader, i.e., $\epsilon_i(t):=x_i(t)-x_0(t)$.
It follows from \eqref{eq:mas} that
\begin{equation}
	\label{eq:self:delta}
	\epsilon_i(t+1)={A}\epsilon_i(t)+{B}u_i(t), \quad  i\in\mathbb{N}_{[1,N]}.
\end{equation}
Interpreting the dynamics of the system \eqref{eq:self:delta}  as a switched system yields
\begin{equation}
	\label{eq:lift}
	\epsilon_i(t_k^i+s_k^i)={A}^{s_k^i}\epsilon_i(t_k^i)+\underline{B}^{s_k^i}\underline{K}^{s_k^i}z_i(t_k^i).
\end{equation}
Here, $\underline{K}^{s_k^i}:=[ {K^\top ~ K^\top ~ \cdots ~ K^\top}]^\top$ containing $s_k^i$ copies of $K$, and ${A}^{s_k^i}$ and $\underline{B}^{s_k^i}:= [ A^{s_k^i-1}B ~ A^{s_k^i-2}B ~ \cdots ~ B]$ are system matrices with $s_k^i\in \mathbb{N}_{[1,\,\bar{s}]}$.
For any $i \in \mathbb{N}_{[1,N]}$, calculating  $\{\epsilon_i(t)\}_{t=0}^{T+s-1}$ for all $s\in \mathbb{N}_{[1,\bar{s}]}$ using $x^{pre}_i$.
Similar to Section \ref{sec:sts:swithced}, construct the following data matrices
\begin{equation*}
	\begin{split}
E_i^1 &:=\left[\begin{matrix}\epsilon_i(0) \quad \epsilon_i(1) \quad \dots \quad \epsilon_i(T-1)\end{matrix}\right],\\		
  E_{i+}^s&:=\left[\begin{matrix}\epsilon_i(s) \quad \epsilon_i(s+1) \quad \dots \quad \epsilon_i(T+s-1)\end{matrix}\right],\\
		U_i^s&:=\left[\begin{matrix}
			u_i(0) & u_i(1) & \cdots &u_i(T-1) \\ \vdots & \vdots &\ddots &\vdots \\  u_i(s-1) & u_i(s) & \cdots &u_i(T+s-2)
		\end{matrix}\right],\\
		W_i^1 &:=W_i, \\
		W^s_i& :=\left[ A^{s-1}B_w ~ A^{s-2}B_w ~ \cdots ~ B_w \right]\underline{W}_i^s,\\
		\underline{W}_i^s&:=\left[\begin{matrix}
			w_i(0) & w_i(1) & \cdots &w_i(T-1) \\ \vdots & \vdots &\ddots &\vdots \\  w_i(s-1) & w_i(s) & \cdots &w_i(T+s-2)
		\end{matrix}\right],
	\end{split}
\end{equation*}
where $s\in \mathbb{N}_{[2,\bar{s}]}$ and matrices $W_i^1$, $W_i^s$, and $\underline{W}_i^s$ are unknown.

Consequently, for all $s\in \mathbb{N}_{[1,\bar{s}]}$ and $i \in \mathbb{N}_{[1,N]}$, the following set contains all matrices $\bar{A}^s$ and $\underline{\bar{B}}^s$ consistent with the data $(U_i^s,E_i^1,E_{i+}^s,  W_i^s)$
\begin{equation}
	\label{eq:perturb}
	\Sigma_i^s:=\left\{[\bar{A}^s\; \underline{\bar{B}}^s] \mid E_{i+}^s=\bar{A}^s E_i^1+\underline{\bar{B}}^s U_i^s+B_w^s W_i^s\right\},
\end{equation}
where $B_w^1:=B_w$ for $s=1$ and $B_w^s:=I$ for all $s\in \mathbb{N}_{[2,\bar{s}]}$.

\begin{figure*}[t]
	\begin{equation}
		\label{long}
		\tag{FQ}
		\begin{split}
			\mathcal{F}_i(\epsilon_i(t_k^i),\, z_i(t_k^i))&:=\left[\begin{matrix}
				I & {0} & {0} \\ {0} & \epsilon_i(t_k^i) &{0} \\   {0} & {0} & z_i(t_k^i)
			\end{matrix}\right]^\top
			\left[\begin{matrix}
				-\Phi & \Phi & {0} \\ \Phi & -\Phi &{0} \\   {0} & {0} & \sigma\Phi
			\end{matrix}\right]\left[\begin{matrix}\cdot\end{matrix}\right]\\
			\mathcal{Q}_i^s(\epsilon_i(t_k^i),\,z_i(t_k^i))&:=\left[\begin{matrix}
				I & {0} & {0} &{0}&{0}\\ {0} & \epsilon_i^\top(t_k^i) &z_i^\top(t_k^i) (\underline{K}^s)^\top &{0}&{0}  \\  {0}& {0} & {0} & \epsilon_i^\top(t_k^i) & z_i^\top(t_k^i) (\underline{K}^s)^\top	\end{matrix}\right] \left[\begin{matrix}\hat{\Theta}_i^s & {0}\\ {0} & M\end{matrix}\right]
			\left[\begin{matrix} \cdot \end{matrix}  \right]^\top.
		\end{split}
	\end{equation}
	\rule[-10pt]{17.9cm}{0.08em}
\end{figure*}

The MASs version of Assumption \ref{as:Theta:single} in given as follows.

\begin{assumption}[Requirement of data]
	\label{as:Theta}
	The matrix 	
	$$\Theta_{i}^s:=\left[\begin{array}{cc}-E_i^1 & {0} \\ -U_i^s & {0} \\ \hline E_{i+}^s & B_w^s\end{array}\right]\left[\begin{array}{cc}Q_{d}^s & S_{d}^s \\ * & R_{d}^s\end{array}\right]\left[\begin{matrix} \cdot\end{matrix} \right]^\top, $$
	has full column rank.
\end{assumption}

Building upon Assumption \ref{as:Theta}, 
the following lemma provides a data-based representation of the MAS \eqref{eq:lift}, which is the MASs version of the representation  \eqref{data:represent:self}.

\begin{lemma}[{\cite[Lemma 2]{li2023selftriggered}}]\label{lem:self:lift}
	Suppose Assumptions \ref{Ass:disturbance:self} and \ref{as:Theta} hold. The set $\bar{\Sigma}_i^s$ of agent $i$ can be represented in the form of a QMI
	\begin{equation}
		\label{eq:lift:Theta}
		\begin{split}
			\bar{\Sigma}_i^s:=&\bigg\{[\bar{A}^s\; \underline{\bar{B}}^s] \in\,   \mathbb{R}^{n \times(n+sp)} \Big| \left[\begin{smallmatrix}{[\bar{A}^s\; \underline{\bar{B}}^s]} \\ I\\I\end{smallmatrix}\right]^\top  \left[\begin{smallmatrix}\hat{\Theta}_i^s & {0}\\ {0} & M\end{smallmatrix}\right]\left[\begin{smallmatrix}{[\bar{A}^s\; \underline{\bar{B}}^s]} \\ I\\I\end{smallmatrix}\right]\succeq 0\bigg\},
		\end{split}
	\end{equation}
	where $M\succ0$, $	\hat{\Theta}_i^s:=\left[\begin{smallmatrix}-\hat{R}_{d}^s & \hat{S}_{d}^{s\top} \\ * & -\hat{Q}_{d}^s\end{smallmatrix}\right] $, and	$\left[\begin{smallmatrix}\hat{Q}_{d}^s & \hat{S}_{d}^{s\top} \\ * & \hat{R}_{d}^s\end{smallmatrix}\right]:=\left[\begin{smallmatrix}Q_{d}^s & S_{d}^s \\ * & R_{d}^s\end{smallmatrix}\right]^{-1}$.
\end{lemma}

To implement the data-driven design, STS \eqref{eq:self:trigger} should be transformed into a QMI form, given as follows.

\begin{lemma}[{\cite[Lemma 3]{li2023selftriggered}}]
	\label{lem:model:tri}
	The model-based STS \eqref{eq:self:trigger} is satisfied if the following condition holds for all $[\bar{A}^s\; \underline{\bar{B}}^s]\in \bar{\Sigma}_i^s~(s\in\mathbb{N}_{[1,\bar{s}]})$
	\begin{equation}
		\label{eq:self:trigger_AB}
		\left[\begin{matrix}{\bar{A}^{s}\epsilon_i(t_k^i)+\underline{\bar{B}}^{s}\underline{K}^{s}z_i(t_k^i)}\\ \epsilon_i(t_k^i) \\z_i(t_k^i)\end{matrix}\right]^\top  
		\left[\begin{matrix}
			-\Phi & \Phi & {0} \\ \Phi & -\Phi &{0} \\   {0} & {0} & \sigma\Phi
		\end{matrix}\right]
		\left[\begin{matrix}\cdot\end{matrix}\right]\geq0.
	\end{equation}
\end{lemma}
\begin{remark}
    It can be observed from Lemmas~\ref{lem:self:lift}, \ref{lem:model:tri} that although each agent $i$ has the same dynamics, both the switched system representation of MASs and the STS are different from that of centralized systems in Section \ref{sec:sts:swithced}.
    Such differences come from the distributed setting, which introduces local interaction between agents (cf. $z_i(t)$ in \eqref{z}).
    Moreover, although each agent has identical matrix, different triggering time $t_k^i$ renders set $\bar{\Sigma}_i^s$ different.
    It is evident that this data-based representation can become more complex when a heterogeneous MAS is considered.
    This can be an interesting future direction.
\end{remark}

Building on Lemmas~\ref{lem:self:lift} and  \ref{lem:model:tri}, the following theorem designs a distributed data-driven STS.

\begin{theorem}[{\cite[Theorem 1]{li2023selftriggered}}]
	\label{thm:self:trigger}
	Consider the MAS \eqref{eq:leader}-\eqref{eq:follower} and the state-feedback controller \eqref{eq:self:controller} under the graph $\bar{\mathcal{G}}$. 
	Under Assumptions \ref{Ass:disturbance:self}, \ref{as:graph1} and \ref{as:Theta} and given a scalar $\sigma>0$, a controller gain $K$, a triggering matrix $\Phi\succ 0$, and the latest transmitted state $x_i(t_k^i)$ of agent $i$, the triggering condition \eqref{eq:self:trigger_AB} is satisfied for any $[\bar{A}^s\; \underline{\bar{B}}^s]\in \bar{\Sigma}_i^s$, if and only if there exists a scalar $\alpha>0$ such that the following LMI holds for $i\in\mathbb{N}_{[1,N]}$ and some $s\in\mathbb{N}_{[1,\bar{s}]}$
	\begin{equation}
		\label{pf1}
		\mathcal{F}_i(\epsilon_i(t_k^i),\,z_i(t_k^i))-\alpha \mathcal{Q}_i^s(\epsilon_i(t_k^i),\,z_i(t_k^i))\succeq0,
	\end{equation}
	where $\mathcal{F}_i(\epsilon_i(t_k^i),\,z_i(t_k^i))$ and $\mathcal{Q}_i^s(\epsilon_i(t_k^i),\,z_i(t_k^i))$ are defined in \eqref{long}.
\end{theorem}

{In light of Theorem~\ref{thm:self:trigger}, we can formulate a data-driven STS as follows}
\begin{equation}
	\label{eq:trigger:data}
	t_{k+1}^i=t_k^i+\max  \left\{s_k^i\in \mathbb{N}_+\big|\hat{f}_i\left(x_i(t_k^i),\,z_i(t_k^i),\, s_k^i\right)\succeq 0
	\right\},
\end{equation}
where
\begin{align}
	\hat{f}_i\left(x_i(t_k^i),\,z_i(t_k^i),\, s_k^i\right)&=\mathcal{F}_i(\epsilon_i(t_k^i),\,z_i(t_k^i))-\alpha \mathcal{Q}_i^s(\epsilon_i(t_k^i),\,z_i(t_k^i)). \label{eq:function:data}
\end{align}

{Leveraging the data-driven representation of MASs in Lemma~\ref{lem:self:lift} and the unknown MAS \eqref{eq:leader}-\eqref{eq:follower} under the controller \eqref{eq:self:controller} and the STS \eqref{eq:trigger:data},}
we now ready to establish a data-driven condition that ensures the stability of the system \eqref{eq:self:delta} in closed-loop with the controller \eqref{eq:self:controller} for all $[\bar{A}\; \bar{B}]\in \Sigma_i$.

\begin{theorem}[{\cite[Theorem 2]{li2023selftriggered}}]
	\label{thm:self:stability}
	Consider the MAS \eqref{eq:leader}-\eqref{eq:follower} operating under the graph $\bar{\mathcal{G}}$. Under Assumptions~\ref{Ass:disturbance:self}, \ref{as:graph1}, and \ref{as:Theta} and given scalars $\sigma>0, \epsilon$, the leader-following consensus is achieved asymptotically for any initial state under the state-feedback controller \eqref{eq:self:controller} and the data-driven STS \eqref{eq:trigger:data} for any $[\bar{A}\; \bar{B}]\in \Sigma_i$, if there exist a scalar $\beta>0$ and matrices $P\succ0$, $\bar{\Phi}\succ0$, $G$, $K_G$ such that the following LMIs are satisfied for $i\in\mathbb{N}_{[1,N]}$
	\begin{equation}
		\label{eq:thm:self}
		\left[\begin{matrix} {0}&\mathcal{T}\\ \ast &\Omega+{\Psi}\end{matrix}\right]+\beta (I_N\otimes\widetilde{\Theta}_i)\prec 0 ,
	\end{equation}
	where
	\begin{align*}
		\mathcal{R}&:=(L_1+\epsilon L_2)^\top,\\
		\mathcal{T}&:=[(I_N\otimes G L_1)^\top~~~(\mathcal{H}\otimes K_G L_3 )^\top]^\top,\\
		\Omega &:=L_2^\top(I_N\otimes P)L_2-L_1^\top(I_N\otimes P)L_1,\\
		\Psi &:={\rm Sym}\{-(I_N\otimes \mathcal{R} GL_2)\}+\sigma(L_3^\top (\mathcal{H}^2\otimes \bar{\Phi} )L_3)-(L_3-L_1)^\top(I_N\otimes \bar{\Phi} )(L_3-L_1),\\
		L_\kappa&:=\big[{0}_{n\times (\kappa-1)n},\, I_n,\, {0}_{n\times (3-\kappa)n}\big],\;\kappa=1,\,2,\,3,\\
		\widetilde{\Theta}_i&:=\left[\begin{smallmatrix} I&{0} \\ {0}&\mathcal{R}\end{smallmatrix}\right]\Theta_i\left[\begin{smallmatrix} I&{0} \\ {0}&\mathcal{R}\end{smallmatrix}\right]^\top.
	\end{align*}
	Moreover, the controller gain is given by $K=K_{G}G^{-1}$ and the triggering matrix is designed as $\Phi=({G^{-1}})^\top\bar{\Phi}G^{-1}$.
\end{theorem}

It is evident that both the controller \eqref{eq:self:controller} and the data-driven STS \eqref{eq:self:trigger} are implemented online in a distributed fashion, where only local (neighbor-to-neighbor) communications are performed. Nonetheless, our design procedure (cf. Theorem \ref{thm:self:stability}) is static and relies on certain global information of the MAS, in terms of the Laplacian matrix associated with the communication graph. 
In this sense, our control paradigm involves centralized design and distributed execution. It is reasonable when considering a scenario, in which minimal information is available online and local data can be acquired during the system operation online, which has been widely studied in model-based STC works \cite{Yi2019,Zegers2022,Zhang2022self}. Exploring methods to eliminate the dependence on such global information in data-driven STC design is an option for future work.

\subsection{Data-driven output synchronization}\label{sec:output}
In practical scenarios, the presence of inherent variations among agents and uncertainties arising from the diverse physical characteristics of systems introduces heterogeneity in the dynamics of MASs \cite{yan2024cooperative}. Consequently, departing from previous research that primarily focused on homogeneous agents or heterogeneous MASs with agents of the same dimension, this section delves into a more comprehensive framework. Herein, agents exhibit variations in both their dimensions and dynamics.
In such a diverse setting, achieving state consensus among agents is often an impractical goal. 
Instead, the emphasis shifts toward the challenging problem of output synchronization. 

We begin by introducing the problem setup for output synchronization.
Consider a leader-following heterogeneous MAS, composed by a leader (indexed as 0) and $N$ followers (indexed from $1$ to $N$).
The dynamics of each follower $i$ and the leader are described as
\begin{subequations}\label{eq:hetero:mas}
    \begin{align}
		\label{eq:hetero:follower}
				{x}_{i}(t+1)&={A}_{i} x_{i}(t)+{B}_{i} u_{i}(t),\\
				y_i(t)&={C}_ix_{i}(t),\\
				{x}_{0}(t+1)&={S} x_{0}(t)	,\label{eq:hetero:leader}\\
				{y}_{0}(t)&={H} x_{0}(t).
		\end{align}
\end{subequations}
The followers and the leader in the MAS \eqref{eq:hetero:mas} interact via a 
graph $\bar{\mathcal{G}}$, which satisfies Assumption~\ref{as:graph1}.
To proceed with, a standard assumption is imposed.
		
		
		\begin{assumption}
			\label{as:pole}
			The leader dynamics matrix $S$ has all its poles on the unit circle and they are non-repeated.
		\end{assumption}

Based on this assumption, we employ a distributed observer $\eta_i(t)\in \mathbb{R}^{n_0}$ to estimate the state of the leader, governed by the following dynamics
\begin{align}
    	\label{eq:hetero:observer}
	\eta_i(t+1)&=S\eta_i(t)+(1+d_i+p_i)^{-1}F \Big(\sum_{j=1}^{N}a_{ij}(\eta_j(t)-\eta_i(t))+p_i(x_0(t)-\eta_i(t))\Big),
\end{align}
where $F\in \mathbb{R}^{n_0\times n_0}$ is a gain matrix. 
Now, we consider a distributed state feedback controller for each follower in \eqref{eq:hetero:follower} as
\begin{equation}
	\label{eq:hetero:controller}
	u_{i}(t)=K_i(x_i(t)-\Pi_i\eta_i(t))+\Gamma_i\eta_i(t),
\end{equation}
where $K_i\in \mathbb{R}^{p_i\times n_i}$ is the feedback gain matrix to be designed, and $\Pi_i\in \mathbb{R}^{n_i\times n_0}$ and $\Gamma_i\in \mathbb{R}^{p_i\times n_0}$ are the solutions to the OREs given by
\begin{equation}
	\label{eq:regulator}
	\begin{split}
		{A}_i\Pi_i+{B}_i\Gamma_i&=\Pi_i S,\\
		{C}_i\Pi_i&=H.
	\end{split}
\end{equation}

The objective here is that, given state-input-output data $\{x_i(t),u_i(t),y_i(t)\}_{t=0}^{T}$ for each follower $i$ generated from \eqref{masnoisy}, design matrices $K_i$, $\Pi_i$, and $\Gamma_i$,  ensuring $m_{t\rightarrow\infty}e_i(t)=m_{t\rightarrow\infty}(y_i(t)-y_0(t))=0$ for all $i \in \mathbb{N}_{[1,N]}$. 

Due to the presence of noises $w_i(t)$ and $v_i(t)$, a set of systems consistent with the data exists, i.e., 
$(\bar{A}_i, \bar{B}_i, \bar{C}_i) \in \mathcal{M}_i$.
{This consequently renders the infeasibility of the initial OREs \eqref{eq:regulator}.}
Specifically, it is impossible to seek a solution $(\Pi_i, \Gamma_i)$ to 
    \begin{align*}
        \left[ \begin{matrix}
            (S^\top \otimes I) - (I \otimes \bar{A}_i) & -(I \otimes \bar{B}_i)\\
        I \otimes \bar{C}_i & 0
        \end{matrix}
        \right] 
        \left[\begin{matrix}
            {\rm vec}(\Pi_i)\\
            {\rm vec}(\Gamma_i)
        \end{matrix}\right] = \left[\begin{matrix}
            0\\
            {\rm vec}({H})
        \end{matrix}\right],
    \end{align*}
since there are infinitely many systems 
$(\bar{A}_i, \bar{B}_i, \bar{C}_i)$ contained in $\mathcal{M}_i$ compared to the finite number of variables in $(\Pi_i, \Gamma_i)$.
This renders data-driven output synchronization under noisy data challenging.
To tackle this issue, a data-driven polytopic reachability analysis approach was proposed in \cite{li2023poly}.
Instead of the exact solution of the OREs \eqref{eq:regulator}, they seek an approximate solution $({\Pi}_i,{\Gamma}_i)$ minimizing the error of the OREs for all possible matrices $(\bar{A}_i, \bar{B}_i, \bar{C}_i)$ and achieve near-optimal output synchronization, i.e., ensuring $\lim_{t \rightarrow \infty} \Vert y_i(t) - y_0(t)\Vert \le \delta_i$ for all $i \in \mathbb{N}_{[1,N]}$ where $\delta_i$ relates to the noise.
In the following, we briefly review the main results in \cite{li2023poly}.




Let $\Delta_{i1}$ and $\Delta_{i2}$ represent the errors of OREs in \eqref{eq:regulator} induced by noisy data. Building upon this foundation, the OREs \eqref{eq:regulator} for all $(\bar{A}_i,\bar{B}_i,\bar{C}_i)\in\mathcal{M}_i$ satisfy
\begin{equation}
\label{eq:regu:noise}
\begin{split}
\Delta_{i1}&=\bar{A}_i\Pi_i+\bar{B}_i\Gamma_i-\Pi_i S,\\
\Delta_{i2}&=\bar{C}_i\Pi_i-H.
\end{split}
\end{equation}

As has been discussed in the above paragraph, neither finding the same gains $\bar{\Pi}_i$ and $\bar{\Gamma}_i$ for the actual system $(A_i,B_i,C_i)$ as with the model-based case nor seeking a single solution satisfying \eqref{eq:regulator} for all $(\bar{A}_i,\bar{B}_i,\bar{C}_i)\in\mathcal{M}_i$ are possible from noisy data.
Therefore, we instead seek a solution of the relaxed OREs \eqref{eq:regu:noise} that minimizes the errors $\Delta_{i1}$ and $\Delta_{i2}$ for all $(\bar{A}_i,\bar{B}_i,\bar{C}_i)\in\mathcal{M}_i$. 
Drawing upon the data-based polytopic representation unveiled in Lemma~\ref{lem:sys_rep:poly}, this idea can be formulated by the following optimization problem
\begin{equation}\label{eq:opt}
\underset{\Pi_i,\Gamma_i}{\min}
~\Big\|\mathcal
{M}_{Z_i}\left[\begin{matrix}\Gamma_i \\\Pi_i\end{matrix}\right]-\Pi_iS\Big\|_F+\|\mathcal
{M}_{C_i}\Pi_i-H\|_F,
\end{equation}
which is a convex optimization problem, amenable to efficient resolution via off-the-shelf solvers. 
With the optimal solution $(\Pi_i^*,\Gamma_i^*)$ at our disposal, $(\Delta_{i1}^*,\Delta_{i2}^*)$ denotes the resulting error of the relaxed OREs \eqref{eq:regu:noise}.
The following lemma provides upper bounds on $\Delta_{i1}^*$ and $\Delta_{i2}^*$.


\begin{lemma}[{\cite[Lemma 3]{li2023poly}}]\label{lem:bound}
Consider the MAS \eqref{eq:hetero:mas} with the relaxed OREs \eqref{eq:regu:noise}. Suppose that Assumptions \ref{as:noise:poly}, \ref{as:graph1}, and \ref{as:pole} hold.
For any $(\bar{A}_i,\bar{B}_i,\bar{C}_i)\in \mathcal{M}_{i}$ and $i\in \mathbb{N}_{[1,N]}$, 
there exist two bounded matrix polytopes {$\mathcal{M}_{\Delta_{i1}}$ and $\mathcal{M}_{\Delta_{i2}}$ such that 
the regulator equation errors $\Delta_{i1}\in \mathcal{M}_{\Delta_{i1}}$ and $\Delta_{i2}\in\mathcal{M}_{\Delta_{i2}}$. Here,} 
\begin{subequations}\label{eq:poly:delta}
\begin{align}
\mathcal{M}_{\Delta_{i1}}:=&\sum_{k=1}^{\gamma_{w_i}\rho}\big[(\hat{\beta}_{W,i}^{(k)}-{\beta}_{W,i}^{(k)})\hat{W}_{i}^{(k)}\big]\left[\begin{matrix}U_i \\X_i\end{matrix}\right]^{\dagger}\left[\begin{matrix}\Gamma_i^s \\\Pi_i^s\end{matrix}\right],\\
	\mathcal{M}_{\Delta_{i2}}:=&\sum_{k=1}^{\gamma_{v_i}\rho}\big[(\hat{\beta}_{V,i}^{(k)}-{\beta}_{V,i}^{(k)})\hat{V}_{i}^{(k)}\big] X_i^{\dagger}\Pi_i^s,
\end{align}
where $\hat{\beta}_{W,i}^{(k)}$ and $\hat{\beta}_{V,i}^{(k)}$ are polytopic parameters associated with the true system matrices $({A}_i,{B}_i,{C}_i)$, i.e.,
\begin{equation*}
	\begin{split}
	[{B}_i\;\, {A}_i]&=\Big(X_{i+}-\sum_{k=1}^{\gamma_{w_i}\rho}\hat{\beta}_{W,i}^{(k)}\hat{W}_{i}^{(k)}\Big)\left[\begin{matrix}U_i \\X_i\end{matrix}\right]^{\dagger},\\
	{C}_i&=\Big(Y_{i}-\sum_{k=1}^{\gamma_{v_i}\rho}\hat{\beta}_{V,i}^{(k)}\hat{V}_{i}^{(k)}\Big)X_i^{\dagger},
	\end{split}
\end{equation*}
and $(\Pi_i^s, \Gamma_i^s)$ are the exact solution of the initial OREs \eqref{eq:regulator}.
\end{subequations}
\end{lemma}
Having obtained $\Pi_i^*$ and $\Gamma_i^*$,
achieving output synchronization reduces to finding a matrix $K_i$ for each agent $i$ such that the matrix $A_i + B_i K_i$ is Schur-stable.
This can be achieved from noisy data by the following theorem.




\begin{theorem}[{\cite[Theorem 1]{li2023poly}}]
\label{thm:k}
Consider the MAS \eqref{eq:hetero:mas} under the distributed data-driven feedback protocol  {\eqref{eq:hetero:observer}--\eqref{eq:hetero:controller}} over the graph $\bar{\mathcal{G}}$. Let Assumptions \ref{as:graph1} and \ref{as:pole} hold and define $\Omega_i=X_{i+}-\sum_{k=1}^{\gamma_{w_i}\rho}\beta_{W,i}^{(k)}\hat{W}_{i}^{(k)}$.  Then, the following SDP is feasible, and the gain matrix $K_i=U_iM_i(X_iM_i)^{-1}$ with any $M_i\in\mathbb{R}^{\rho\times n_i}$ satisfying \eqref{sdp} renders $A_i+B_iK_i$ Schur-stable for all $(\bar A_i, \bar B_i)\in \mathcal{M}_{Z_i}$
\begin{equation}
\label{sdp}
\begin{split}
	X_iM_i-\Omega_iM_i(X_iM_i)^{-1}(\Omega_iM_i)^\top&\succ0,\\
	X_iM_i&\succ0.
\end{split}
\end{equation}
\end{theorem}

Leveraging previous results, including the optimization problem~\eqref{eq:opt}, Lemma \ref{lem:bound}, and Theorem~\ref{thm:k}, a distributed data-driven output synchronization solution for the unknown heterogeneous MAS \eqref{eq:hetero:mas} is established below.

\begin{theorem}[{\cite[Theorem 2]{li2023poly}}]
\label{thm:uub}
Consider the MAS \eqref{eq:hetero:mas} along with the graph $\bar{\mathcal{G}}$. Let Assumptions \ref{as:graph1} and \ref{as:pole} hold.
Given that the optimal solution of problem \eqref{eq:opt} is denoted as $(\Pi_i^*,\Gamma_i^*)$, the distributed data-driven feedback control protocol \eqref{eq:hetero:observer}--\eqref{eq:hetero:controller} ensures that the tracking errors $e_i(t)$ are ultimately uniformly bounded. This holds for any initial state and all $i\in \mathbb{N}_{[1,N]}$, provided that the following two conditions are met
\begin{enumerate}
\item The controller gain $K_i$ is designed in accordance with Theorem~\ref{thm:k}.
\item Matrix $F$ satisfies the condition that $I_N\otimes S- ((I_N+\mathcal{D}+\mathcal{P})^{-1}\mathcal{H})\otimes F$ is Schur-stable.
\end{enumerate}
\end{theorem}

 \begin{remark}
    It is important to highlight that the solution to problem~\eqref{eq:opt} and the SDP in~\eqref{sdp} need only examine the vertices of the matrix polytopes \(\mathcal{M}_{Z_i}\) and \(\mathcal{M}_{C_i}\) defined in~\eqref{M_AB}.  
Since the number of vertices grows combinatorially with the state dimension, explicit enumeration can become the dominant computational cost for large‐scale systems. 
Developing dedicated complexity-reduction or approximation techniques to mitigate this burden is an interesting direction for future research (see, e.g., \cite{Althoff2021set,alonso2021robust}), but it lies beyond the scope of the present paper.
\end{remark}

\section{Conclusions and future directions}
In this paper, we navigated the landscape of data-driven control for network systems, uncovering insights into four key themes: communication delay, aperiodic transmission, network security, and distributed configuration, while outlining potential future research avenues.

\subsection{Key insights}
 {Before outlining future research directions, we would like to emphasize the fundamental differences and advantages of the Willems' Fundamental Lemma-based data-driven control framework compared to traditional model-based approaches, particularly in the presence of uncertainties and network imperfections.

Traditional model-based control strategies fundamentally rely on the availability of accurate mathematical models to represent system dynamics. However, in real-world scenarios, obtaining such models is often extremely difficult due to factors such as unmodeled dynamics, measurement noise, time-varying environments, and incomplete knowledge of external disturbances. This inherent dependence on precise models can significantly limit both the robustness and generalizability of model-based controllers.

Over the years, various robust identification and control techniques have been developed to address uncertainties and disturbances. Nevertheless, applying these methods in practical settings remains challenging. One key difficulty lies in the fact that system models obtained through identification are often structurally inconsistent with the assumptions required by robust control design. This structural mismatch introduces a gap between the identification and control stages, which can undermine the effectiveness of model-based approaches when deployed in real-world systems.

In contrast, the data-driven control methods reviewed in this paper avoid explicit model identification by utilizing measured input-output data directly. These methods exploit the rich structure implicitly encoded in the data, thereby enabling a unified and coherent framework for both system representation and controller synthesis. This integrated approach eliminates the structural inconsistencies that typically arise when model identification and control design are handled separately.

Furthermore, the use of real-world data inherently captures the system's actual behavior, including latent nonlinearities and its response to external disturbances. As a result, data-driven controllers have the potential to outperform their model-based counterparts, particularly in situations where the latter depend on oversimplified or approximate models. This characteristic makes data-driven control strategies especially attractive for modern applications, including distributed control in unreliable networks and systems operating under dynamic and uncertain conditions.

Building on these insights, it is clear that there remain many open challenges and opportunities for future research. Addressing these challenges will be essential for the continued advancement of both the theoretical underpinnings and the practical adoption of data-driven control methods.}

\subsection{Future directions}

Looking ahead, several promising research directions and open challenges merit further attention.

 {\emph{d1) Nonlinear systems.}
One of the most important research frontiers for data-driven control is its extension to nonlinear network systems. These systems are common in real-world applications but are notoriously difficult to model and control due to their complex dynamics \cite{wang2025timedelay}. Recent studies have demonstrated encouraging progress by incorporating data-driven techniques into well-established nonlinear control frameworks. Examples include sum-of-squares optimization \cite{depersis2021sos}, kernel-based methods \cite{van2022kernel}, and linearization-based data-driven MPC \cite{berberich2022linear}. However, developing a unified framework that can address system uncertainty, enforce state and input constraints, and handle distributed nonlinear interactions remains an open challenge. Future research could explore the integration of machine learning and optimization methods \cite{jiang2025inexact,jiang2023distri-sdp} into data-driven control frameworks to enhance their ability to cope with the complex behaviors of large-scale nonlinear systems.

\emph{d2) Enhanced cybersecurity.}
The increasing reliance on networked control systems raises new security concerns, particularly with regard to cyberattacks such as false data injection (FDI) and denial-of-service (DoS) attacks. While existing solutions are predominantly model-based \cite{murguia2020security}, it remains an open question whether comparable detection and mitigation capabilities can be achieved in purely data-driven scenarios. This is especially challenging when only noisy or partially observed data are available. Combining real-time anomaly detection, resilient control synthesis, and secure data collection techniques in a data-driven framework could offer a promising direction for future research.

\emph{d3) Distributed control.}
The increasing scale and complexity of modern network systems make centralized control approaches impractical. Distributed data-driven control, where each agent makes decisions based on its local data and limited communication with neighboring agents, provides a scalable alternative. Initial efforts, such as the localized controller synthesis framework proposed in \cite{chang2024localized}, have demonstrated the potential of this approach. Nonetheless, the field is still at an early stage, and future research is expected to focus on distributed optimization, event-triggered communication schemes, and plug-and-play architectures. These developments will be critical for enabling robust operation in dynamic and uncertain network environments.

\emph{d4) Real-world validation.}
Although the theory of data-driven control has seen significant progress, its real-world application remains limited. Some early experimental studies, such as the implementation of data-driven model predictive control for quadcopters \cite{elokda2021data, ral2023zhou}, have demonstrated the feasibility of these methods. However, these examples also highlight the gap between theoretical assumptions and practical requirements. Future research should emphasize the validation of data-driven controllers under realistic conditions, including model uncertainties, time delays, and network-induced imperfections. Bridging this gap is essential for transitioning data-driven control methods from laboratory settings to industrial and safety-critical applications.}

In conclusion, the fusion of data-driven techniques with the challenges of networked systems offers a rich and exciting research landscape. Addressing these challenges will not only further the theoretical understanding of data-driven control but also facilitate the development of more adaptable, resilient, and efficient networked systems.





\Acknowledgements{This work was partially supported by the National Natural Science Foundation of China under Grants U23B2059, 62173034, 62495090, 62495095, and 62088101.}


\bibliographystyle{scis}

\bibliography{cas-refs}

@article{Althoff2021set,
   author = "Althoff, Matthias and Frehse, Goran and Girard, Antoine",
   title = "Set Propagation Techniques for Reachability Analysis", 
   journal= "Annu. Rev. Control Robot. Auton. Syst.",
   year = "2021",
   volume = "4",
   number = "",
   pages = "369-395",
  }

@misc{alonso2021robust,
      title={Robust Distributed and Localized Model Predictive Control}, 
      author={Carmen Amo Alonso and Jing Shuang Li and Nikolai Matni and James Anderson},
      year={2021},
      Journal={arXiv: 2103.14171},
    month={Mar.}
}

@article{kitano2002systems,
	title={Systems biology: {A} brief overview},
	author={Kitano, Hiroaki},
	journal={Science},
	volume={295},
	number={5560},
	pages={1662--1664},
	year={2002},
	month = {Mar.},
}

@article{zhang2023obstacle,
	title={Obstacle avoidance in human-robot cooperative transportation with force constraint},
	author={Zhang, Ying and Yang, Chenguang and Xu, Sheng and Ou, Yongsheng},
	journal={Sci. {China} Inf. Sci.},
	volume={66},
	number={1},
	pages={119205},
	year={2023},
}

@article{jiang2025inexact,
  title={Inexact proximal gradient algorithm with random reshuffling for nonsmooth optimization},
  author={Jiang, Xia and Fang, Yanyan and Zeng, Xianlin and Sun, Jian and Chen, Jie},
  journal={Sci. China Inf. Sci.},
  volume={68},
  number={1},
  pages={112201},
  year={2025},
  publisher={Springer}
}

@ARTICLE{jiang2023distri-sdp,
  author={Jiang, Xia and Zeng, Xianlin and Sun, Jian and Chen, Jie},
  journal={IEEE Trans. Autom. Control}, 
  title={Distributed Synchronous and Asynchronous Algorithms for Semidefinite Programming With Diagonal Constraints}, 
  year={2023},
  volume={68},
  number={2},
  pages={1007-1022},
}

@article{li2023topology,
  title={Topology Inference for Network Systems: Causality Perspective and Nonasymptotic Performance},
  author={Li, Yushan and He, Jianping and Chen, Cailian and Guan, Xinping},
  journal={IEEE Trans. Autom. Control},
  volume={69},
  number={6},
  pages={3483--3498},
  year={2023},
  publisher={IEEE}
}

@article{li2024preserving,
  title={Preserving Topology of Network Systems: Metric, Analysis, and Optimal Design},
  author={Li, Yushan and Wang, Zitong and He, Jianping and Chen, Cailian and Guan, Xinping},
  journal={IEEE Trans. Autom. Control},
  year={2024},
  publisher={IEEE}
}

@article{zhang2016unsupervised,
  title={Unsupervised learning of Dirichlet process mixture models with missing data},
  author={Zhang, Xunan and Song, Shiji and Zhu, Lei and You, Keyou and Wu, Cheng},
  journal={Sci. China Inf. Sci.},
  volume={59},
  number={1},
  pages={1--14},
  year={2016}
}

@article{zou2024leader,
  title={Leader--follower circumnavigation control of non-holonomic robots using distance-related information},
  author={Zou, Yao and Zhong, Liangyin and He, Wei and Silvestre, Carlos},
  journal={Automatica},
  volume={169},
  pages={111831},
  year={2024},
  publisher={Elsevier}
}

@article{jiang2024distributed,
  title={Distributed Adaptive Time-Varying Optimization with Global Asymptotic Convergence},
  author={Jiang, Liangze and Wu, Zheng-Guang and Wang, Lei},
  journal={IEEE Trans. Autom. Control},
page = {1-8},
  year={2024, doi: 10.1109/TAC.2024.3492952},
  publisher={IEEE}
}

@article{li2024controller,
	title={Controller Synthesis from Noisy-Input Noisy-Output Data},
	author={Li, Lidong and Bisoffi, Andrea and De Persis, Claudio and Monshizadeh, Nima},
	journal={arXiv:2402.02588},
	year={2024},
}

@article{van2023behavioral,
  title={A Behavioral Approach to Data-Driven Control With Noisy Input--Output Data},
  author={van Waarde, Henk J and Eising, Jaap and Camlibel, M Kanat and Trentelman, Harry L},
  journal={IEEE Trans. Autom. Control},
  volume={69},
  number={2},
  pages={813--827},
  year={2023},
  publisher={IEEE}
}

@article{Blanchini1999,
title = {Set invariance in control},
journal = {Automatica},
volume = {35},
number = {11},
pages = {1747-1767},
year = {1999},
author = {F. Blanchini},
year={Nov.}
}

@ARTICLE{ral2023zhou,
  author={Zhou, Ziyu and Wang, Gang and Sun, Jian and Wang, Jikai and Chen, Jie},
  journal={IEEE Robot. Autom. Lett.}, 
  title={Efficient and Robust Time-Optimal Trajectory Planning and Control for Agile Quadrotor Flight}, 
  year={2023},
  volume={8},
  number={12},
  pages={7913-7920},
}

@article{zhang2019networked,
  title={Networked control systems: {A} survey of trends and techniques},
  author={Zhang, Xian-Ming and Han, Qing-Long and Ge, Xiaohua and Ding, Derui and Ding, Lei and Yue, Dong and Peng, Chen},
  journal={IEEE/CAA J. Autom. Sin.},
  volume={7},
  number={1},
  pages={1--17},
  year={2020},
  month = {Jan.},
}

@book{wang2008networked,
  title={Networked control systems},
  author={Wang, Fei-Yue and Liu, Derong},
  publisher={Springer, 2008},
address = {London, UK},
}

@Book{FB-LNS,
  author =    {F. Bullo},
  title =     {{Lectures on Network Systems}},
  edition =   {{1.6}},
  publisher = {Kindle Direct Publishing, 2022},
  ISBN =      {978-1986425643},
  url =       {https://fbullo.github.io/lns},
}

@article{huo2022secure,
  title={Secure output synchronization of heterogeneous multi-agent systems against false data injection attacks},
  author={Huo, Shicheng and Huang, Dalin and Zhang, Ya},
  journal={Sci. China Inf. Sci.},
  volume={65},
  number={6},
  pages={162204},
  year={2022},
month = {May,},
  publisher={Springer}
}

@article{hou2022deep,
  title={Deep reinforcement learning for optimal denial-of-service attacks scheduling},
  author={Hou, Fangyuan and Sun, Jian and Yang, Qiuling and Pang, Zhonghua},
  journal={Sci. China Inf. Sci.},
  volume={65},
  number={6},
  pages={162201},
  year={2022},
month = {Apr.},
  publisher={Springer}
}

@article{wang2025timedelay,
  title={Time-delay effects on the dynamical behavior of switched nonlinear time-delay systems},
  author={Wang, Zhichuang and He, Wei and Sun, Jian and Wang, Gang },
  journal={Sci. China Inf. Sci.},
  year = {2025, doi: 10.1007/s11432-024-4392-0},
}

@ARTICLE{chen2025tac,
  author={Chen, Guoliang and Liu, Xinru and Ren, Xiaoqiang and Xia, Jianwei and Park, Ju H.},
  journal={IEEE Trans. Autom. Control}, 
  title={Asynchronous-Based Countermeasure for Stealthy Attack on Aperiodic Sampled-Data Control Systems}, 
  year={2025, doi: 10.1109/TAC.2025.3552745},
  volume={},
  number={},
  pages={1-8},
}

@ARTICLE{yang2025ddset,
  author={Yang, Te and Bu, Keqing and Chen, Guoliang and Xie, Xiang-Peng and Xia, Jianwei},
  journal={IEEE Trans. Syst. Man. Cybern.}, 
  title={Model-Driven and Data-Driven Reachable Set Estimation for Multirate Sampled-Data Truck-Trailer System}, 
  year={2024},
  volume={54},
  number={11},
  pages={7079-7091},
}

@article{Hou2013,
 author = {Z. Hou and Z. Wang},
 title = {From model-based control to data-driven control: {Survey}, classification and perspective},
 journal = {Inf. Sci.},
 volume={235},
 number={},
 pages={3-35},
 year = {2013},
 month = {Jun.},
 }

@article{Yang2021unmanned,
      title={An Overview of Recent Advances in Distributed Coordination of Multi-Agent Systems}, 
      author={Ruohan Yang and Lu Liu and Gang Feng},
      year={2021},
      journal={Unmanned Syst.},
      volume={10},
      number={3},
      month={Dec.},
      pages={307-325}
}

@article{Chen2020scis,
      title={How often should one update control and estimation: review of networked triggering techniques}, 
      author={Zhiyong Chen and Qing-Long Han and Yamin Yan and Zheng-Guang Wu},
      year={2020},
      journal={Sci. China Inf. Sci.},
      volume={63},
      number={},
      month={May,},
      pages={150201}
}

@article{dorfler2023data,
  title={Data-driven control: Part one of two: {A} special issue sampling from a vast and dynamic landscape},
  author={D{\"o}rfler, Florian},
  journal={IEEE Control Syst. Mag.},
  volume={43},
  number={5},
  pages={24--27},
  year={2023},
month = {Oct.},
}

@ARTICLE{Armanini2023robot,
  author={Armanini, Costanza and Boyer, Frédéric and Mathew, Anup Teejo and Duriez, Christian and Renda, Federico},
  journal={IEEE Trans. Robot.}, 
  title={Soft Robots Modeling: {A} Structured Overview}, 
  year={2023},
  volume={39},
  number={3},
  pages={1728-1748},
  }

@article{berberich2022linear,
  title={Linear tracking MPC for nonlinear systems—Part II: The data-driven case},
  author={Berberich, Julian and K{\"o}hler, Johannes and M{\"u}ller, Matthias A and Allg{\"o}wer, Frank},
  journal={IEEE Trans. Autom. Control},
  volume={67},
  number={9},
  pages={4406--4421},
  year={2022},
  month = {Sept.},
}

@ARTICLE{li2023selftriggered,
  author={Li, Yifei and Wang, Xin and Sun, Jian and Wang, Gang and Chen, Jie},
  journal={IEEE Trans. Autom. Control}, 
  title={Self-Triggered Consensus Control of Multi-Agent Systems From Data}, 
  year={2024},
  volume={},
  number={},
  pages={1-8},
  month={Jan.},
  doi={10.1109/TAC.2024.3351865},
}

@article{li2023poly,
      title={Data-driven Polytopic Output Synchronization of Heterogeneous Multi-agent Systems from Noisy Data}, 
      author={Yifei Li and Wenjie Liu and Jian Sun and Gang Wang and Lihua Xie and Jie Chen},
     journal={IEEE Trans. Autom. Control}, 
  year={2024},
month = {Dec.},
  volume={69},
  number={12},
  pages={8513-8525},
}

@article{annaswamy2023control,
  title={Control for Societal-Scale Challenges: Road Map 2030},
  author={Annaswamy, Anuradha M and Johansson, Karl H and Pappas, George J and others},
  journal={IEEE Control Systems Society Publication: Piscataway, NJ, USA},
  year={2023}
}

@ARTICLE{Su2012,
  author={Su, Youfeng and Huang, Jie},
  journal={IEEE Trans. Autom. Control}, 
  title={Cooperative Output Regulation of Linear Multi-Agent Systems}, 
  year={2012},
  volume={57},
  number={4},
  pages={1062-1066},
  month={Apr.}
}

@article{Hong2013,
author = {Hong, Yiguang and Wang, Xiaoli and Jiang, Zhong-Ping},
title = {Distributed output regulation of leader–follower multi-agent systems},
journal = {Int. J. Robust Nonlin. Control},
volume = {23},
number = {1},
pages = {48-66},
year = {2013},
month={Oct.}
}

@article{hashemi2022co,
  author={Hashemi, Navid and Ruths, Justin},
  journal={IEEE Trans. Control  Netw. Syst.}, 
  title={Codesign for Resilience and Performance}, 
  year={2023},
  volume={10},
  number={3},
  pages={1387-1399},
month={Sept.}
}

@article{wu2019optimalswitching,
	title={Optimal Switching Attacks and Countermeasures in Cyber-Physical Systems},
	author={Wu, Guangyu and Wang, Gang and Sun, Jian and Xiong, Lu},
	journal={IEEE Trans. Syst. Man Cybern. Syst.},
	year={2021},
	volume={51},
	number={8},
	pages={4825-4835},
	month={Aug.},}

@article{chang2024localized,
  title={Localized Data-driven Consensus Control},
  author={Chang, Zeze and Jiao, Junjie and Li, Zhongkui},
  journal={arXiv: 2401.12707},
  year={2024},
volume = {},
pages = {1-12},
}

@article{elokda2021data,
  title={Data-enabled predictive control for quadcopters},
  author={Elokda, Ezzat and Coulson, Jeremy and Beuchat, Paul N and Lygeros, John and D{\"o}rfler, Florian},
  journal={Int. J. Robust Nonlin.},
  volume={31},
  number={18},
  pages={8916--8936},
  year={2021},
month = {Jun.},
}

@article{murguia2020security,
	title={Security metrics and synthesis of secure control systems},
	author={Murguia, Carlos and Shames, Iman and Ruths, Justin and Ne{\v{s}}i{\'c}, Dragan},
	journal={Automatica},
	volume={115},
	pages={108757},
	year={2020},
month = {May,},
}

@article{Liu2023learning,
	title={Learning Robust Data-based {LQG} Controllers from Noisy Data},
	author={Liu, W. and Sun, J. and Wang, G. and Bullo, F. and Chen, J.},
	journal={IEEE Trans. Autom. Control},
	year={2024},
	 volume={69},
  number={12},
  pages={8526-8538},
	month = {Dec.},
	doi = {10.1109/TAC.2024.3409749},
}

@article{vanwaarde2023quadratic,
      title={Quadratic matrix inequalities with applications to data-based control},
  author={van Waarde, Henk J and Camlibel, M Kanat and Eising, Jaap and Trentelman, Harry L},
  journal={SIAM J. Control Optim.},
  volume={61},
  number={4},
  pages={2251--2281},
  year={2023},
month = {Feb.},
}

@ARTICLE{Anne2022diss,
  author={Koch, Anne and Berberich, Julian and Allgöwer, Frank},
  journal={IEEE Trans. Autom. Control}, 
  title={Provably Robust Verification of Dissipativity Properties from Data}, 
  year={2022},
  volume={67},
  number={8},
  pages={4248-4255},
  month={Aug.}}

@ARTICLE{van2022diss,
  author={van Waarde, Henk J. and Camlibel, M. Kanat and Rapisarda, Paolo and Trentelman, Harry L.},
  journal={IEEE Control Syst. Mag.}, 
  title={Data-Driven Dissipativity Analysis: {Application} of the Matrix {S}-Lemma}, 
  year={2022},
  volume={42},
  number={3},
  pages={140-149},
  month={June,}}

@article{wang2021data,
  title={Data-driven control of dynamic event-triggered systems with delays},
  author={Wang, Xin and Sun, Jian and Berberich, Julian and Wang, Gang and Allg{\"o}wer, Frank and Chen, Jie},
  journal={Int. J.  Robust  Nonlin. Control},
  volume={33},
  number={12},
  pages={7071--7093},
  year={2023},
month = {Aug.},
  publisher={Wiley Online Library}
}

@article{Liu2023data,
  title={Data-driven Self-triggering Mechanism for State Feedback Control},
  author={Liu, Wenjie and Li, Yifei and Sun, Jian and Wang, Gang and Chen, Jie},
  journal={IEEE Control Syst. Lett.},
  year={2023},
month = {June,},
pages = {1975 - 1980},
volume={7},
}

@ARTICLE{Liu2023self,
  author={Liu, Wenjie and Sun, Jian and Wang, Gang and Bullo, Francesco and Chen, Jie},
  journal={IEEE Trans. Autom. Control}, 
  title={Data-Driven Self-Triggered Control via Trajectory Prediction}, 
  year={2023},
month = {Nov.},
  volume={68},
  number={11},
  pages={6951-6958},
  }

@article{wolff2024robust,
  title={Robust data-driven moving horizon estimation for linear discrete-time systems},
  author={Wolff, Tobias M and Lopez, Victor G and M{\"u}ller, Matthias A},
  journal={IEEE Trans. Autom. Control},
  year={2024},
  volume={69},
  number={8},
  pages={5598-5604},
}

@ARTICLE{wu2018optimal,
	author={G. {Wu} and J. {Sun} and J. {Chen}},
	journal={IEEE Trans. Cybern.},
	title={Optimal Data Injection Attacks in Cyber-Physical Systems},
	year={2018},
	volume={48},
	number={12},
	pages={3302-3312},
	month = {June,},
}

@ARTICLE{Yi2019,
  author={Yi, Xinlei and Liu, Kun and Dimarogonas, Dimos V. and Johansson, Karl H.},
  journal={IEEE Trans. Autom. Control}, 
  title={Dynamic Event-Triggered and Self-Triggered Control for Multi-agent Systems}, 
  year={2019},
  volume={64},
  number={8},
  pages={3300-3307},
  month={Aug.}
  }

@ARTICLE{Zhang2022self,
  author={Zhang, Yifang and Wu, Zhengguang and Shi, Peng},
  journal={IEEE Trans. Ind. Inf.}, 
  title={Resilient Event-/Self-Triggering Leader-following Consensus Control of Multi-agent Systems Against {DoS} attacks}, 
  year={2023},
 volume={19},
  number={4},
  pages={5925-5934},
  month={Apr.},
  }

@ARTICLE{Zegers2022,
  author={Zegers, Federico M. and Deptula, Patryk and Shea, John M. and Dixon, Warren E.},
  journal={IEEE Trans. Autom. Control}, 
  title={Event/Self-Triggered Approximate Leader-Follower Consensus With Resilience to {B}yzantine Adversaries}, 
  year={2022},
  volume={67},
  number={3},
  pages={1356-1370},
  month={Mar.}
  }

@INPROCEEDINGS{Jiao2021,
  author={Jiao, Junjie and van Waarde, Henk J. and Trentelman, Harry L. and Camlibel, M. Kanat and Hirche, Sandra},
  booktitle={Proc. IEEE Conf. Decis. Control, Austin, TX, USA}, 
  title={Data-driven output synchronization of heterogeneous leader-follower multi-agent systems}, 
  month={Dec. 14-17,},
  year={2021},
  volume={},
  number={},
  pages={466-471},
}

@ARTICLE{Wang2023Distributed,
  author={Wang, Xin and Sun, Jian and Wang, Gang and Allg{\"o}wer, Frank and Chen, Jie},
  journal={IEEE/CAA J. Autom. Sin.},
  title={Data-Driven Control of Distributed Event-Triggered Network Systems},
  year={2023},
  month = {Feb.},
  volume={10},
  number={2},
  pages={351-364},
 }

@ARTICLE{Paulo2007,
  author={Tabuada, Paulo},
  journal={IEEE Trans. Autom. Control}, 
  title={Event-Triggered Real-Time Scheduling of Stabilizing Control Tasks}, 
  year={2007},
  volume={52},
  number={9},
  pages={1680-1685},
  month={Sept.}}

@article{persis2022event,
	title={Event-triggered Control From Data},
	author={De Persis, Claudio and Postoyan, Romain and Tesi, Pietro},
	journal={IEEE Trans. Autom. Control},
volume = {},
	year={2023},
pages = {1-16},
 doi={10.1109/TAC.2023.3335002}
}

@inproceedings{SminSecure,
	title={Secure control: {T}owords survivable cyber-physical systems},
	author = {C{\'{a}}rdenas, A. and Amin, S. and Sastry, Shankar},
	booktitle = {Proc. Int. Conf. Distrib. Comput. Syst. Workshops, Beijing, China},
	pages = {495-500},
	year = {2008},
	month = {June, 17-20,},
}

@article{depersis2015input,
	title={Input-to-State Stabilizing Control under {Denial-of-Service}},
	author={{De Persis}, C. and Tesi, P.},
	journal = {IEEE Trans. Autom. Control},
	volume={60},
	number={11},
	pages={2930-2944},
	year={2015},
	month = {Nov.},
}

@article{FengResilient,
	title={Resilient control under {Denial-of-Service}: {R}obust design},
	author={Feng, S. and Tesi, P.},
	journal={Automatica},
	volume={79},
	pages={42-51},
	year={2017},
	month = {Mar.},
}

@article{FengNetworked,
	title={Networked control systems under {Denial-of-Service}: {C}o-located vs. remote architectures},
	author={Feng, S. and Tesi, P.},
	journal={Syst. Control Lett.},
	volume={108},
	pages={40-47},
	year={2017},
	month = {Sept.},
	
}

@article{liu2025OutReg,
  title={Data-driven Internal Model Control for Output Regulation},
  author={Liu, Wenjie and Li, Yifei and Sun, Jian and Wang, Gang and You, Keyou and Xie, Lihua and Chen, Jie},
  journal={arXiv:2505.09255},
  year={2025}
}

@ARTICLE{Liu2021resilient,  author={Liu, Wenjie and Sun, Jian and Wang, Gang and Bullo, Francesco and Chen, Jie},  
	journal={IEEE Trans. Autom. Control},   
	title={Resilient Control Under Quantization and Denial-of-Service: {C}odesigning a Deadbeat Controller and Transmission Protocol},   year={2022},  
	month = {Aug.},
	volume={67},  
	number={8}, 
	pages={3879-3891}, 
}

@article{Liu2021data,
	author = {Liu, W.  and  Sun, J.  and  Wang, G.  and  Bullo, F.  and  Chen, J.},
	title = {Data-Driven Resilient Predictive Control under Denial-of-Service},
journal={IEEE Trans. Autom. Control},   
	year={2023},
	volume={68},
	number={8},
	pages={4722-4737},
	month = {Aug.},
	}

@inproceedings{liu2024robustfdi,
  title={Robust Data-Driven Control Against Actuator {FDI} for Unknown Linear Systems},
  author={Liu, Wenjie and Sun, Jian and Deng, Fang and Wang, Gang and Chen, Jie},
  booktitle={IEEE Int. Conf. Control Autom., Reykjavík, Iceland},
  pages={246--251},
  year={2024},
month = {June 18-24,},
}

@article{wakaiki2019stabilization,
	title={Stabilization of networked control systems under {DoS} attacks and output quantization},
	author={Wakaiki, Masashi and Cetinkaya, Ahmet and Ishii, Hideaki},
	journal={IEEE Trans. Autom. Control},
	volume={65},
	number={8},
	pages={3560--3575},
	year={2019},
	month = {Oct.},
}

@article{xu2023reinforcement,
  title={Reinforcement learning-based unknown reference tracking control of HMASs with nonidentical communication delays},
  author={Xu, Yong and Wu, Zheng-Guang and Che, Wei-Wei and Meng, Deyuan},
  journal={Sci. China Inf. Sci.},
  volume={66},
  number={7},
  pages={170203},
  year={2023},
  publisher={Springer}
}

@article{chen2024state,
  title={State estimation for delayed switched positive systems: delayed radius approach},
  author={Chen, Weizhong and Fei, Zhongyang and Zhao, Xudong and Wu, Zheng-Guang},
  journal={Sci. China Inf. Sci.},
  volume={67},
  number={8},
  pages={182202},
  year={2024},
  publisher={Springer}
}

@article{yan2024cooperative,
  title={Cooperative control for heterogeneous multi-agent systems: progress, applications, and challenges},
  author={Yan, Bing and Shi, Peng and Chambers, Jonathon},
  journal={Sci. China Inf. Sci.},
  volume={67},
  number={5},
  pages={156201},
  year={2024},
month = {Apr.},
  publisher={Springer}
}

@article{Bai2017Data,
	title={Data-Injection Attacks in Stochastic Control Systems: {D}etectability and Performance Tradeoffs},
	author={Bai, Cheng Zong and Pasqualetti, Fabio and Gupta, Vijay},
	journal={Automatica},
	volume={82},
	pages={251-260},
	year={2017},
	month  ={Apr.}
}

@misc{wang2025data,
  title={Data-Based Online Linear Quadratic Gaussian Control From Noisy Data},
  author={Wang, Linqi and Liu, Wenjie and Li, Yifei and Sun, Jian and Wang, Gang},
  year={2025, doi: https://doi-org.remotexs.ntu.edu.sg/10.1002/rnc.7886},
  publisher={John Wiley \& Sons, Inc. Hoboken, USA}
}

@article{2013Attack,
	title={Attack Detection and Identification in Cyber-Physical Systems},
	author={ Pasqualetti, F.  and  D\"{o}rfler, F.  and  Bullo, F. },
	journal={IEEE Trans. Autom. Control},
	volume={58},
	number={11},
	pages={2715-2729},
	year={2013},
	month = {June,},
}

@article{depersis2021sos,
	author    = {Meichen Guo and
	Claudio {De Persis} and
	Pietro Tesi},
	title     = {Data-Driven Stabilization of Nonlinear Polynomial Systems With Noisy
	Data},
	journal   = {{IEEE} Trans. Autom. Control},
	volume    = {67},
	number    = {8},
	pages     = {4210--4217},
	year      = {2022},
	month = {Sept.},
}

@article{zhao2022data,
	title={Data-driven Control of Unknown Linear Systems via Quantized Feedback},
	author={Zhao, Feiran and Li, Xingchen and You, Keyou},
	journal={Proc. Mach. Learn. Res.},
	volume={144},
	pages={1--12},
	year={2022}
}

@article{wang2024social,
  title={Social Power Games in Concatenated Opinion Dynamics},
  author={Wang, Lingfei and Chen, Guanpu and Bernardo, Carmela and Hong, Yiguang and Shi, Guodong and Altafini, Claudio},
  journal={IEEE Trans. Autom. Control},
  year={2024},
  volume={69},
  number={11},
  pages={7614-7629},
}

@article{berberich2019data,
	title={Data-driven model predictive control with stability and robustness guarantees},
	author={Berberich, Julian and K{\"o}hler, Johannes and M{\"u}ller, Matthias A and Allg{\"o}wer, Frank},
	journal={IEEE Trans. Autom. Control},
	year={2021},
	volume={66},
	number={4},
	pages={1702-1717},
	month = {June,},
}

@ARTICLE{persis2020data,
	author={C. {De Persis} and P. {Tesi}},
	journal={IEEE Trans. Autom. Control}, 
	title={Formulas for Data-Driven Control: {S}tabilization, Optimality, and Robustness}, 
	year={2020},
	month = {Mar.},
	volume={65},
	number={3},
	pages={909-924},
}

@article{Rotulo2021Online,
	title = {Online learning of data-driven controllers for unknown switched linear systems},
	journal = {Automatica},
	volume = {145},
	pages = {110519},
	year = {2022},
	month = {Nov.},
	issn = {0005-1098},
	author = {Monica Rotulo and Claudio {De Persis} and Pietro Tesi},
}

@ARTICLE{krishnan2021data,  
	author={Krishnan, Vishaal and Pasqualetti, Fabio},  
	journal={IEEE Control Syst. Lett.},   
	title={Data-Driven Attack Detection for Linear Systems},
	year={2021},  
	volume={5},  number={2},  pages={671-676},  month = {June,},
}

@article{bisoffi2022data,
	title={Data-driven control via {P}etersen’s lemma},
	author={Bisoffi, Andrea and De Persis, Claudio and Tesi, Pietro},
	journal={Automatica},
	volume={145},
	pages={110537},
	year={2022},
	month = {Nov.},
}

@article{van2020noisy,
	title={From noisy data to feedback controllers: non-conservative design via a matrix {S}-lemma},
	author={van Waarde, Henk J. and Camlibel, M. Kanat and Mesbahi, Mehran},
	journal={IEEE Trans. Autom. Control},
	year={2020},
	month = {Jan.},
	volume={67},
	number={1},
	pages={162-175},
	publisher={IEEE}
}

@article{Li2023data,
  author = "Yifei Li and Xin Wang and Jian Sun and Gang Wang and Jie Chen",
  title = "Data-driven consensus control of fully distributed event-triggered multi-agent systems",
  journal = "Sci. China Inf. Sci.",
  year = "2023",
  volume = "66",
  number = "5",
  pages = "152202",
month={May,},
  }

@inproceedings{chen2023data,
  title={Data-driven input-to-state stabilization with respect to measurement errors},
  author={Chen, Hailong and Bisoffi, Andrea and De Persis, Claudio},
  booktitle={Proc. IEEE Conf. Decis.  Control, Singapore},
  pages={1601--1606},
  year={2023},
}

@article{hespanha2007survey,
  title={A survey of recent results in networked control systems},
  author={Hespanha, Joo P and Naghshtabrizi, Payam and Xu, Yonggang},
  journal={Proc. IEEE},
  volume={95},
  number={1},
  pages={138--162},
  year={2007},
month = {Jan.},
  publisher={IEEE}
}

@article{wang2010event,
  title={Event-triggering in distributed networked control systems},
  author={Wang, Xiaofeng and Lemmon, Michael D},
  journal={IEEE Trans. Autom. Control},
  volume={56},
  number={3},
  pages={586--601},
  year={2010},
month = {July,},
  publisher={IEEE}
}

@article{sandberg2015cyberphysical,
  title={Cyberphysical security in networked control systems: {An} introduction to the issue},
  author={Sandberg, Henrik and Amin, Saurabh and Johansson, Karl Henrik},
  journal={IEEE Control Syst. Mag.},
  volume={35},
  number={1},
  pages={20--23},
  year={2015},
month = {Jan.},
  publisher={IEEE}
}

@inproceedings{mo2009secure,
  title={Secure control against replay attacks},
  author={Mo, Yilin and Sinopoli, Bruno},
  booktitle={Proc. Allerton Conf. Commun. Control Comput., Monticello, IL, USA},
  pages={911--918},
  year={Sept. 30-Oct. 2, 2009},
}

@inproceedings{aaarzen1999simple,
	title={A simple event-based {PID} controller},
	author={{\AA}arz{\'e}n, Karl-Erik},
	booktitle={Proc. IFAC World Congress, Beijing, China},
	volume={32},
	number={2},
	pages={8687--8692},
	year={1999},
	month = {June, 7-9,},
}

@ARTICLE{anta2010to,
	author={Anta, Adolfo and Tabuada, Paulo},
	journal={IEEE Trans. Autom. Control}, 
	title={To Sample or not to Sample: {S}elf-Triggered Control for Nonlinear Systems}, 
	year={2010},
	volume={55},
	number={9},
	pages={2030-2042},
	month = {Sept.},
}

@article{heemels2013model,
	title={Model-based periodic event-triggered control for linear systems},
	author={Heemels, W. P. M. H. and Donkers, M C F},
	journal={Automatica},
	volume={49},
	number={3},
	pages={698--711},
	year={2013},
	month = {Mar.},
	publisher={Elsevier}
}

@article{gommans2015resource,
	title={Resource-aware {MPC} for constrained nonlinear systems: A self-triggered control approach},
	author={Gommans, T. M. P. and Heemels, W. P. M. H.},
	journal={Syst. Control Lett.},
	volume={79},
	pages={59--67},
	year={2015},
	month = {May,},
	publisher={Elsevier},
}

@inproceedings{heemels2012introduction,
	title={An introduction to event-triggered and self-triggered control},
	author={Heemels, W. P. M. H. and Johansson, Karl Henrik and Tabuada, Paulo},
	booktitle={Proc. IEEE Conf. Decis. Control, Maui, HI, USA},
	pages={3270--3285},
	year={2012},
	month = {Dec. 10-13,},
}

@inproceedings{matsume2020resilient,
	title={Resilient Self/Event-Triggered Consensus Based on Ternary Control},
	author={Matsume, Hiroki and Wang, Yuan and Ishii, Hideaki},
	booktitle={Proc. IEEE Conf. Decis. Control, Jeju, Korea},
	pages={6240--6245},
	year={2020},
	month = {Dec. 14-18},
}

@article{wildhagen2021datadriven,
      title={Data-Driven Analysis and Controller Design for Discrete-Time Systems Under Aperiodic Sampling},
      author={S. Wildhagen and J. Berberich and M. Hertneck and F. Allg{\"o}wer},
      year={2023},
      journal = {IEEE Trans. Autom. Control},
      month = {June,},
        volume={68},
    number={6},
    pages={3210-3225},
}

@article{martin2023guarantees,
  title={Guarantees for data-driven control of nonlinear systems using semidefinite programming: A survey},
  author={Martin, Tim and Sch{\"o}n, Thomas B and Allg{\"o}wer, Frank},
  journal={Annu. Rev. Control},
  volume={56},
  pages={100911},
  year={2023},
  publisher={Elsevier}
}

@article{chen2021mean,
  title={Mean square exponential stability analysis for It{\^o} stochastic systems with aperiodic sampling and multiple time-delays},
  author={Chen, Guoliang and Fan, Chenchen and Sun, Jian and Xia, Jianwei},
  journal={IEEE Trans. Autom. Control},
  volume={67},
  number={5},
  pages={2473--2480},
  year={2021},
  month = {May,},
}

@article{WILLEMS2005,
title = {A note on persistency of excitation},
author = {Willems, Jan C. and Rapisarda, Paolo and Markovsky, Ivan and De Moor, Bart L.M.},
journal = {Syst. Control Lett.},
volume = {54},
number = {4},
pages = {325-329},
year = {2005},
month={Apr.},
}

@inproceedings{hu2024hinfty,
  title={Data-Driven $\mathcal{H}_{\infty}$ Control for Unknown Piecewise Affine Systems with Bounded Disturbances},
  author={Hu, Kaijian and Liu, Tao},
  booktitle={Proc. Europ. Control Conf.},
  pages={2630--2635},
  year={2024},
}

@article{hu2025robust,
  title={Robust Data-Driven Predictive Control for Unknown Linear Systems with Bounded Disturbances},
  author={Hu, Kaijian and Liu, Tao},
  journal={IEEE Trans. Autom. Control},
  year={2025, doi: 10.1109/TAC.2025.3560697},
pages = {1-16},
  publisher={IEEE}
}

@ARTICLE{liu2025switchmpc,
  author={Wang, Zhi-Min and Liu, Kun-Zhi and Cheng, Xiao-Lin and Sun, Xi-Ming},
  journal={IEEE Trans. Autom. Control}, 
  title={Online Data-Driven Model Predictive Control for Switched Linear Systems}, 
  year={2025, doi: 10.1109/TAC.2025.3556322},
  volume={},
  number={},
  pages={1-8},
}

@ARTICLE{wang2024ddmpc,
  author={Wang, Zhi-Min and Liu, Kun-Zhi and Wen, Si-Xin and Sun, Xi-Ming},
  journal={IEEE Trans. Autom. Sci. Eng.}, 
  title={Data-Driven Switched Model Predictive Control Without Terminal Ingredients}, 
  year={2024},
  volume={21},
  number={3},
  pages={4247-4260},
  }

@ARTICLE{liu2023ddevent,
  author={Qi, Wan-Ling and Liu, Kun-Zhi and Wang, Rui and Sun, Xi-Ming},
  journal={IEEE Trans. Ind. Electron.}, 
  title={Data-Driven $\mathcal{L}_{2}$-Stability Analysis for Dynamic Event-Triggered Networked Control Systems: {A} Hybrid System Approach}, 
  year={2023},
  volume={70},
  number={6},
  pages={6151-6158},
}

@article{Liu2023fdi,
    author = {Liu, Wenjie and  Li, Lidong and Sun, Jian and Deng, Fang and Wang, Gang and Chen, Jie},
    title = {Data-driven control against false data injection attacks},
journal = {Automatica},
volume = {179},
pages = {112399},
year = {2025},
}

@inproceedings{Coulson2019,
title = {Data-enabled predictive control: {I}n the shallows of the {D}ee{PC}},
booktitle = {Proc. Eur. Control Conf., Naples, Italy},
author = {J. {Coulson} and J. {Lygeros} and F. {Dörfler}},
year = {2019},
month={Jun.},
pages = {307-312},

}

@article{Sche2001,
title = {{L}{P}{V} control and full block multipliers},
journal = {Automatica},
author = {Scherer, C. W.},
volume = {37},
number = {3},
year = {2001},
month = {Mar.},
pages = {361-375},
}

@article{vlacic2024automation,
  title={Automation 5.0: {T}he key to systems intelligence and Industry 5.0},
  author={Vlacic, Ljubo and Huang, Hailong and Dotoli, Mariagrazia and Wang, Yutong and Ioannou, Petros A and Fan, Lili and Wang, Xingxia and Carli, Raffaele and Lv, Chen and Li, Lingxi and others},
  journal={IEEE/CAA J. Autom. Sin.},
  volume={11},
  number={8},
  pages={1723--1727},
  year={2024},
month = {Aug.},
  publisher={IEEE}
}

@article{pang2022comparison,
  title={Comparison of three data-driven networked predictive control methods for a class of nonlinear systems},
  author={Pang, Zhong-Hua and Zhao, Xue-Ying and Sun, Jian and Shi, Yuntao and Liu, Guo-Ping},
  journal={IEEE/CAA J. Autom. Sin.},
  volume={9},
  number={9},
  pages={1714--1716},
  year={2022},
month = {Sept.},
  publisher={IEEE}
}

@article{he2024analysis,
  title={Analysis and Control of Frequency Stability in Low-Inertia Power Systems: {A} Review},
  author={He, Changjun and Geng, Hua and Rajashekara, Kaushik and Chandra, Ambrish},
  journal={IEEE/CAA J. Autom. Sin.},
  volume={11},
  number={12},
  pages={2363--2383},
  year={2024},
month = {Dec.},
  publisher={IEEE}
}

@ARTICLE{Yue2013,
  author={D. {Yue} and E. {Tian} and Q. {Han}},
  journal={IEEE Trans. Autom. Control},
  title={A Delay System Method for Designing Event-Triggered Controllers of Networked Control Systems},
  year={2013},
  month = {Feb.},
  volume={58},
  number={2},
  pages={475-481},}

@ARTICLE{Peng2013,
  author={C. {Peng} and Q. {Han}},
  journal={IEEE Trans. Autom. Control},
  title={A Novel Event-Triggered Transmission Scheme and $\mathcal{L}_{2}$ Control Co-Design for Sampled-Data Control Systems},
  year={2013},
  month = {Oct.},
  volume={58},
  number={10},
  pages={2620-2626},}

@CONFERENCE{Fridman2014,
  author  = {Fridman, E.},
  title   = {Introduction to time-delay systems: {A}nalysis and control},
  booktitle = {Boston: Birkhouser},
  year    = {2014},
  pages   = {}
}

@article{Ding2020,
language = {English},
copyright = {Compilation and indexing terms, Copyright 2021 Elsevier Inc.},
copyright = {Compendex},
title = {Event-triggered static/dynamic feedback control for discrete-time linear systems},
journal = {Inf. Sci.},
author = {Ding, S. and Xie, X. and Liu, Y.},
volume = {524},
year = {2020},
month = {Jul.},
pages = {33-45}
}

@article{Hu2016,
author = {Hu, S. and Yue, D. and Yin, X. and Xie, X. and Ma, Y.},
title = {Adaptive event-triggered control for nonlinear discrete-time systems},
journal = {Int. J. Robust Nonlin. Control},
volume = {26},
number = {18},
pages = {4104-4125},
year = {2016},
month = {Apr.}
}

@ARTICLE{Bu2022Adaptive,
  author={Bu, Xuhui and Yu, Wei and Yu, Qiongxia and Hou, Zhongsheng and Yang, Junqi},
  journal={IEEE Trans. Cybern.},
  title={Event-Triggered Model-Free Adaptive Iterative Learning Control for a Class of Nonlinear Systems Over Fading Channels},
  year={2022},
  volume={52},
  number={9},
  pages={9597-9608},
  month = {Sep.},}

@ARTICLE{Bu2022Power,
  author={Bu, Xuhui and Yu, Wei and Cui, Lizhi and Hou, Zhongsheng and Chen, Zongyao},
  journal={IEEE Trans. Ind. Informat.},
  title={Event-Triggered Data-Driven Load Frequency Control for Multiarea Power Systems},
  year={2022},
  volume={18},
  number={9},
  pages={5982-5991},
  month = {Sep.},}

@ARTICLE{Rueda2022datadelay,
  author={Rueda-Escobedo, Juan G. and Fridman, Emilia and Schiffer, Johannes},
  journal={IEEE Trans. Autom. Control}, 
  title={Data-Driven Control for Linear Discrete-Time Delay Systems}, 
  year={2022},
  month = {July,},
  volume={67},
  number={7},
  pages={3321-3336},
  }

@ARTICLE{Jad2003,
  author={Jadbabaie, A. and Jie Lin and Morse, A.S.},
  journal={IEEE Trans. Autom. Control}, 
  title={Coordination of groups of mobile autonomous agents using nearest neighbor rules}, 
  year={2003},
  volume={48},
  number={6},
  pages={988-1001},
  month={June,}}

@ARTICLE{Olfati2004,
  author={Olfati-Saber, R. and Murray, R.M.},
  journal={IEEE Trans. Autom. Control}, 
  title={Consensus problems in networks of agents with switching topology and time-delays}, 
  year={2004},
  volume={49},
  number={9},
  pages={1520-1533},
  month={Sept.}}

@ARTICLE{Ren2005,
  author={Wei Ren and Beard, R.W.},
  journal={IEEE Trans. Autom. Control}, 
  title={Consensus seeking in multiagent systems under dynamically changing interaction topologies}, 
  year={2005},
  volume={50},
  number={5},
  pages={655-661},
  month={May,}}

@ARTICLE{Lizk2010,
  author={Li, Zhongkui and Duan, Zhisheng and Chen, Guanrong and Huang, Lin},
  journal={IEEE Trans. Circuits and Syst. I, Reg. Papers}, 
  title={Consensus of Multiagent Systems and Synchronization of Complex Networks: {A} Unified Viewpoint}, 
  year={2010},
  volume={57},
  number={1},
  pages={213-224},
  month={Jan.}}

@article{Dima2012,
  title={Distributed Event-Triggered Control for Multi-Agent Systems},
  author={D. V. Dimarogonas and E. Frazzoli and K. H. Johansson},
  journal={IEEE Trans. Autom. Control},
  volume={57},
  number={5},
  pages={1291-1297},
  year={2012},
  month={May,}
}

@article{Li2021,
  title={Distributed Dynamic Event-triggered Consensus Control for Multi-agent
Systems under Fixed and Switching Topologies},
author={Y. Li and X. Liu and H. Liu and C. Du and P. Lu},
  journal={J. Franklin. Inst.},
  volume={358},
  number={8},
  pages={4348-4372},
  year={2021},
  month={May,},
}

@article{wang2023event,
  title={Event-triggered consensus control of heterogeneous multi-agent systems: {M}odel-and data-based approaches},
  author={Wang, Xin and Sun, Jian and Deng, Fang and Wang, Gang and Chen, Jie},
  journal={Sci. China Inf. Sci.},
  volume={66},
  number={9},
  pages={192201},
  year={2023},
  month={Aug.},
  publisher={Springer}
}

@INPROCEEDINGS{Matsuda2022event,
  author={Matsuda, Yuma and Kato, Shuichi and Wakasa, Yuji and Adachi, Ryosuke},
  booktitle={Proc. Annual Conf. Soc. Instrum. Control Eng., Kumamoto, Japan}, 
  title={State-Feedback Event-Triggered Control Using Data-Driven Methods}, 
  year={2022},
month = {Sept. 6-9,},
  volume={},
  number={},
  pages={1287-1292},
  }

@ARTICLE{Wang2023Tc,
  author={Wang, Xin and Berberich, Julian and Sun, Jian and Wang, Gang and Allg{\"o}wer, Frank and Chen, Jie},
  journal={IEEE Trans. Cybern.},
  title={Model-Based and Data-Driven Control of Event- and Self-Triggered Discrete-Time Linear Systems},
  year={2023},
  month = {Sept.},
  volume={53},
  number={9},
  pages={6066-6079},
 }

@ARTICLE{Qi2023event,
  author={Qi, WanLing and Liu, KunZhi and Wang, Rui and Sun, XiMing},
  journal={IEEE Trans. Ind. Electron.}, 
  title={Data-Driven $\mathcal {L}_{2}$-Stability Analysis for Dynamic Event-Triggered Networked Control Systems: {A} Hybrid System Approach}, 
  year={2023},
month = {June,},
  volume={70},
  number={6},
  pages={6151-6158},
  }

@article{van2022kernel,
  title={Kernel-based models for system analysis},
  author={Van Waarde, Henk J and Sepulchre, Rodolphe},
  journal={IEEE Trans. Autom. Control},
  year={2022},
pages = {5317--5332},
volume={68},
  number={9},
month = {Sept.},
}

@article{LIU2017BL,
title = {Stability analysis of systems with time-varying delays via the second-order Bessel-{L}egendre inequality},
author = {Kun Liu and Alexandre Seuret and Yuanqing Xia},
journal = {Automatica},
volume = {76},
pages = {138-142},
year = {2017},
month = {Feb.},
}

@ARTICLE{tcns2022wangxin,
  author={Wang, X. and Sun, J. and Wang, G. and Dou, L.},
  journal={IEEE Trans. Control Netw. Syst.},
  title={A Mixed Switching Event-Triggered Transmission Scheme for Networked Control Systems},
  year={2022},
  month = {Mar.},
  volume={9},
  number={1},
  pages={390-402},}

@ARTICLE{tcb2023wzc,
  author={Wang, Zhichuang and He, Wei and Sun, Jian and Wang, Gang and Chen, Jie},
  journal={IEEE Trans. Cybern.}, 
  title={Event-Triggered Control of Switched Nonlinear Time-Delay Systems With Asynchronous Switching}, 
  year={2023},
month = {Sept.},
  volume={},
  number={},
  pages={1-11},
  doi={10.1109/TCYB.2023.3312491}
}

@INPROCEEDINGS{WangCDC,
  author={Wang, Xin and Li, Yifei and Sun, Jian and Wang, Gang and Chen, Jie and Dou, Lihua},
  booktitle={Proc. IEEE Conf. Decis. Control}, 
  title={Data-Driven Self-Triggered Control for Linear Networked Control Systems}, 
  year={2023},
  pages={6869-6874},
  adddress = {Singapore},
}

@article{CHEN2025TS,
title = {Stability analysis for {T}akagi-{S}ugeno fuzzy systems with a periodically varying delay via a generalized allowable delay set partitioning approach},
journal = {Fuzzy Set. Syst.},
volume = {518},
pages = {109502},
year = {2025},
month = {Oct.},
author = {Yun Chen and Xin Wang and Yaqi Li and Yunfei Qiu and Shuangcheng Sun},
}

@ARTICLE{Tan2025Data,
  author={Tan, Haoran and Zhang, Xueming and Wang, Yaonan and Wu, You and Feng, Yun and Hou, Zhongsheng},
  journal={IEEE/CAA J. Autom. Sin.}, 
  title={Data-Driven Bipartite Consensus Control for Large Workpieces Rotation of Nonlinear Multi-Robot Systems}, 
  year={2025},
  month = {Jun.},
  volume={12},
  number={6},
  pages={1144-1158},
  }

@article{CHAO2025Data,
title = {Data-driven adaptive formation control based on preview mechanism for networked multi-robot systems with communication delays},
journal = {Neurocomputing},
volume = {620},
pages = {129151},
year = {2025},
author = {Chenzhuolei Chao and Haoran Tan and Xueming Zhang and Gang Wang and You Wu and Yaonan Wang},
month = {Mar.},
}







\end{document}